\newcommand{\myfig}[5]{
\begin{figure}[{#5}]
\includegraphics[keepaspectratio,width=#4,angle=0]{#1}%
\caption{#2}\label{#3}%
\end{figure}}

\newcommand{\figref}[1]{Fig.\ \ref{#1}}
\newcommand{\secref}[1]{Sec.\ \ref{#1}}
\newcommand{\eqnref}[1]{Eq.\ (\ref{#1})}

\documentclass[aps,pre,twocolumn,showpacs,amsmath,amssymb]{revtex4}
\usepackage{graphicx,dcolumn,bm,verbatim,color,hyperref,subfigure}

\begin{document}

\title{Phase transitions in random Potts systems
and the community detection problem: spin-glass type
and dynamic perspectives}
\begin{abstract}

Phase transitions in spin glass type systems and, more recently, in
related computational problems have gained broad interest in
disparate arenas. In the current work, we focus on the ``community
detection'' problem when cast in terms of a general Potts spin glass
type problem. As such, our results apply to rather broad Potts spin
glass type systems. Community detection describes the general
problem of partitioning a complex system involving many elements
into optimally decoupled  ``communities''  of such elements. We
report on phase transitions between {\em solvable} and {\em
unsolvable} regimes. Solvable region may further split into {\em
``easy''} and {\em ``hard''} phases. Spin glass type phase
transitions appear at both low and high temperatures (or noise). Low
temperature transitions correspond to an ``order by disorder''  type
effect wherein fluctuations render the system ordered or solvable.
Separate transitions appear at higher temperatures into a disordered
(or an unsolvable) phase. Different sorts of randomness lead to
disparate behaviors. We illustrate the spin glass character of both
transitions and report on memory effects. We further relate Potts
type spin systems to mechanical analogs and suggest how {\em
chaotic-type} behavior in general thermodynamic systems can indeed
naturally arise in hard-computational problems and spin-glasses. The
correspondence between the two types of transitions (spin glass and
dynamic) is likely to extend across a larger spectrum of spin glass
type systems and hard computational problems. We briefly discuss
potential implications of these transitions in complex many body
physical systems.
\end{abstract}

\author{Dandan Hu}
\affiliation{Department of Physics, Washington University in St.
Louis, Campus Box 1105, 1 Brookings Drive, St. Louis, MO 63130, USA}
\author{Peter Ronhovde}
\affiliation{Department of Physics, Washington University in St.
Louis, Campus Box 1105, 1 Brookings Drive, St. Louis, MO 63130, USA}
\author{Zohar Nussinov}
\affiliation{Department of Physics, Washington University in St.
Louis, Campus Box 1105, 1 Brookings Drive, St. Louis, MO 63130, USA}

 \maketitle{} \vskip 0.1in

\section{Introduction}

One of the highly significant recent applications of statistical
mechanics concerns a topic of broad interest-- that of community
detection\cite{fortunato,rosvall,
radicchi,ramasco,blondel,gregory,duch,peter2,peter1,mod,RB,gudkov,Hastings} in
complex networks \cite{fortunato,duch,jsima} and related
computational problems \cite{bayati,jie,hidetoshi}. In this article,
we address further development in the challenging quest of studying
these difficult computational problems by bringing additional tools
from physics into the fore. Our aim is not only to study the
community detection problem itself. Rather, we use the community
detection problem as a platform for a detailed investigation of
phase transitions \cite{mezard,florent,zecchina,taghogg} associated
with complex computational problems and, generally,
Potts spin glass type systems. Various applications of physics
to computational problems have enabled significant advances in the design of
new algorithms and the identification and understanding of various
{\em ``phases''} of computational problems in way that has dramatically
advanced previous approaches.

In this article, we provide direct evidence for earlier indications
of \emph{two} phase transitions in the community detection problem
and more generally in Potts type spin glass systems. These Potts
type spin glass transitions occur at \emph{both} low and high
temperatures (or, similarly, at low and high levels of randomness or
noise). These transitions reflect different underlying physics.
Earlier reports of such transitions were afforded by information
theory measures (as in Appendix E of \cite{peter2}) and a
computational ``computational susceptibility'' to be defined in
later sections of the current work that monitors the onset of a
large number of local minima, or large computational complexity (as
in Appendix B of \cite{peter1}). As was earlier shown (e.g., Fig. 11
in \cite{peter1}), overlap parameters (to be defined herein) such as
the normalized mutual information $I_N$ exhibit progressively
sharper changes as the system size $N$ increases. This suggests the
existence of bona fide thermodynamic transitions. In this article,
we will investigate ``fixed'' spin glass type Potts Hamiltonian. By
``fixed'', we allude new spin glass systems with fixed parameters
which are not dependent on the problem itself. Thus, this fixed
approach contrasts with, e.g., ``modularity''
\cite{fortunato,mod,goodMC} or other models that involved
comparisons to random case systems- so called ``null models''
\cite{fortunato,mod,RB} that have been earlier invoked on in the
community detection problem. When cast in terms of canonical fixed
Potts spin Hamiltonian,  the system exhibits sharper phase transitions
\cite{peter1}.  By applying our model to a general random graph, we
can locate phase transitions between {\em solvable} and {\em
unsolvable} regions. Solvable regions may further splinter into {\em
``easy''} and {\em ``hard''} phases. We further elaborate on
disparate phase transitions (at low and high temperatures) in these
rather general Potts spin glass type systems.It is noteworthy that a similar analysis
can be done for any other method for detecting communities.
Within most of the easy phase, all of the known methods agree
on the solutions. The results of our analysis are not relevant to only
one specific method.

Insofar as the classification of computational problems, the main
tools of analysis to date were of a static nature and further invoke
various forms of ``cavity'' type approximations
\cite{cavity1,cavity2} and extremely powerful related approaches
such as ``belief propagation'' \cite{Hastings,gallager, bp}. Cavity
type methods were of immense success early on in studying mean-field
type theories in spin-glasses and, in the last decade, have seen a
rapid resurgence in enabling new and very potent algorithms and in
better enabling an understanding of complex problems.

In this article, we directly study the phase transition in
computational problems such as community detection from both static
(i.e., thermodynamic) and dynamic aspects. We directly numerically
investigate, sans any analytical approximations, thermodynamic
quantities characterizing the transition augmented by further direct
measures of the energy landscape of these systems by use of a
``computational susceptibility'' that we will introduce later on
that monitors the increase number of local minima and convergence
with local minima. In the dynamic approach, in order to relate hard
computational problems to classical dynamics, we will dualize, via a
Hubbard-Stratonovich transformation, the original (discrete) system
to be optimized by a continuous theory for which equations of motion
can be written down and the dynamics investigated. By employing
these two complimentary approaches ((i) static thermodynamic of
information measures of the energy landscape and (ii) classical
dynamics), we correspondingly report on the existence of (i) static
spin-glass-type transitions as well as (ii) dynamical transitions,
i.e., the transition of nodes from stable orbits to ``chaos''.  The
transitions as ascertained by both approaches occur at precisely the
same set of parameters describing the problem. As far as we are
aware, earlier studies have not investigated the general phase
diagram of this important problem. To date, links between dynamical
mechanical transitions and spin-glass type transitions in
computational problems such as this have, furthermore, not been
discovered.

\section{Outline}

The rest of the paper is organized as follows.
In Section \ref{sec:potts}, we introduce
the (general) Potts model that will form
the focus of our attention and its relation
to the community detection problem.
In Section \ref{def}, we introduce the
basic definitions of trials and replicas that are imperative to our
approach. These allow us to directly explore the energy landscape
of the system without the aid of approximations. This is followed, in Section \ref{information},
by a review of information theoretic quantities as they
pertain to our method. We then proceed to present
our findings. In Section \ref{sec:transition} we present
evidence for the existence of spin-glass type transitions
that may generally occur at both high and low temperatures.
We discuss the physical origin of these transitions
and the relation between the phase diagram of the
community detection problem (and, more generally,
that of the Potts model) to other important computational
problems. In Section \ref{dee}, we relate the Potts model
system to a continuous mechanical system. By examining
its dynamics, we note that, in this mechanical system,
the transition to chaos onsets exactly at the same set of parameters
at which the Potts model displays spin-glass type transitions.
Various technical details and further physical aspects
have been relegated to the appendices.

\section{The Potts model}
\label{sec:potts}
We will employ a, rather general, spin-glass type Potts model
Hamiltonian (denoted, henceforth, as the ``Absolute Potts Model''
(APM)) \cite{peter2} for solving the community detection problem.
The Hamiltonian reads
\begin{eqnarray}
H({\sigma}) = -\frac{1}{2}\sum_{i\neq j}(A_{ij}-\gamma
(1-A_{ij}))\delta(\sigma_i,\sigma_j).
 \label{eq:ourpotts}
\end{eqnarray}
Here, $A_{ij}$ is an adjacency matrix element which assumes a value of $1$
if nodes $i$ and $j$ are connected and a value of $0$ otherwise.
The spins $\{\sigma_{i} \}_{i=1}^{N}$ attain integer values: $1\leq\sigma_{i}\leq q$.
Their values reflect the community membership. That is, if
$\sigma_{i} =a$ then node $i$ belongs to community
number $a$. The parameter $q$ denotes the total
number of communities. To simplify the analysis, we will,
unless stated otherwise, set (the so-called resolution parameter \cite{peter1}) $\gamma=1$. (In recent work
\cite{image}, we reported on similar results for general $\gamma$ and
weighted version of \eqnref{eq:ourpotts}. In particular, in physics related
applications for many particle systems, the weights $A_{ij}$ were determined
by the two-body interactions \cite{peter3,peter4}.)

Although, as we elaborate on in Appendix A, we can achieve analytic
solutions for certain cases of graphs (e.g., employing the cavity
method \cite{gallager, parisi,RB2,jorgbook,Allah} when all of the nodes
are of a fixed degree of $k=3$ or Ising systems (i.e., systems with $q=2$ communities)),
most {\em general graphs} (with arbitrary degree and cluster size
distributions) require computer simulation. To this end, we will
undertake a direct numerical investigation of the system at hand
without the need to invoke analytical approximations or assumptions.
Our (``zero-temperature'') community detection algorithm for
minimizing \eqnref{eq:ourpotts} was discussed at length in Refs.
\cite{peter2,peter1,peter3,peter4,image}. In the current work, we
investigate the above Hamiltonian of \eqnref{eq:ourpotts} at zero
temperature \cite{peter2} and also at finite temperatures ($T>0$)
with the use of a heat bath algorithm (HBA) (Appendix B).  In brief,
within the HBA, we will sequentially allow each node an opportunity
to change the community membership during each time step with
probabilities determined by a Boltzman weight $e^{-\beta\Delta E}$
($\beta=\frac{1}{T}$) at a specified temperature $T$ and the energy
change ($\Delta E$) as the node were moved to each connected cluster
(or to a new cluster). Similarly, as elaborated on in more detail in
Appendix B, following each step, we further allow the possibility of
community merges based on a Boltzman weight.

\myfig{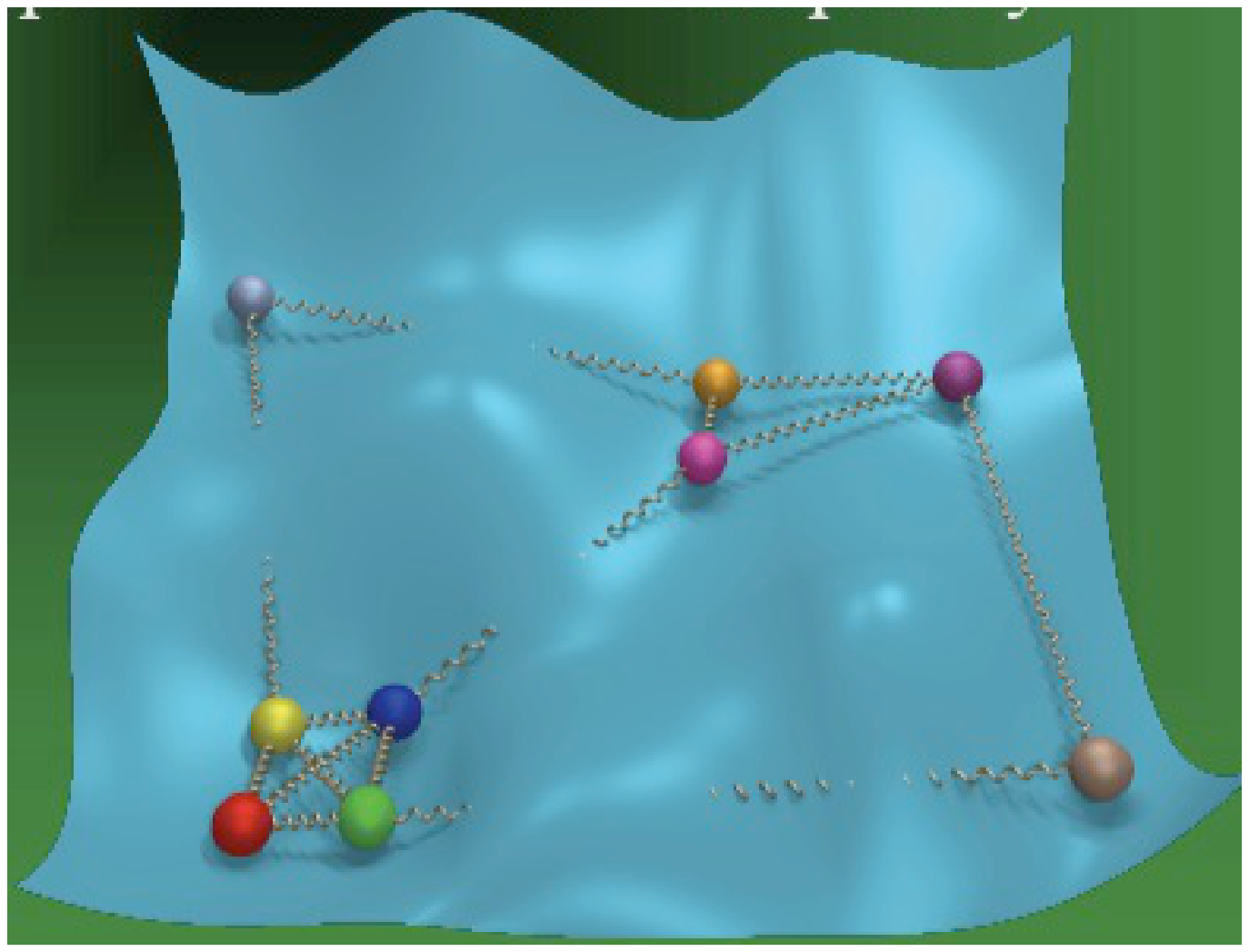}{A caricature of the information theory
correlations (springs) between ``replicas'' (denoted symbolically by balls) in a high dimensional energy landscape (in our case, graph partitions or
Potts spin configurations). ``Replicas'' are obtained from multiple
solutions of the same problem (in this case, the minimization of the Potts model Hamiltonian of Eq. (\ref{eq:ourpotts})). The information theory correlations
measure the agreement or overlap between the candidate solutions (``replicas'').
In earlier works and in the current work, we use such correlations to ascertain
system parameters (e.g., $\gamma$ of Eq. (\ref{eq:ourpotts})) for which
clearly defined solutions appear. Throughout most of the current
work, we will not employ inter-replica correlations but rather
the average of the correlations between all of the replica
and a known (or ``planted'') solution to the community detection
problem (a minimum of the Hamiltonian).
For detailed definitions of replicas and information theory
correlations, see \secref{def} and \secref{information},
respectively.
}{fig:replica}{0.65\linewidth}{}

\myfig{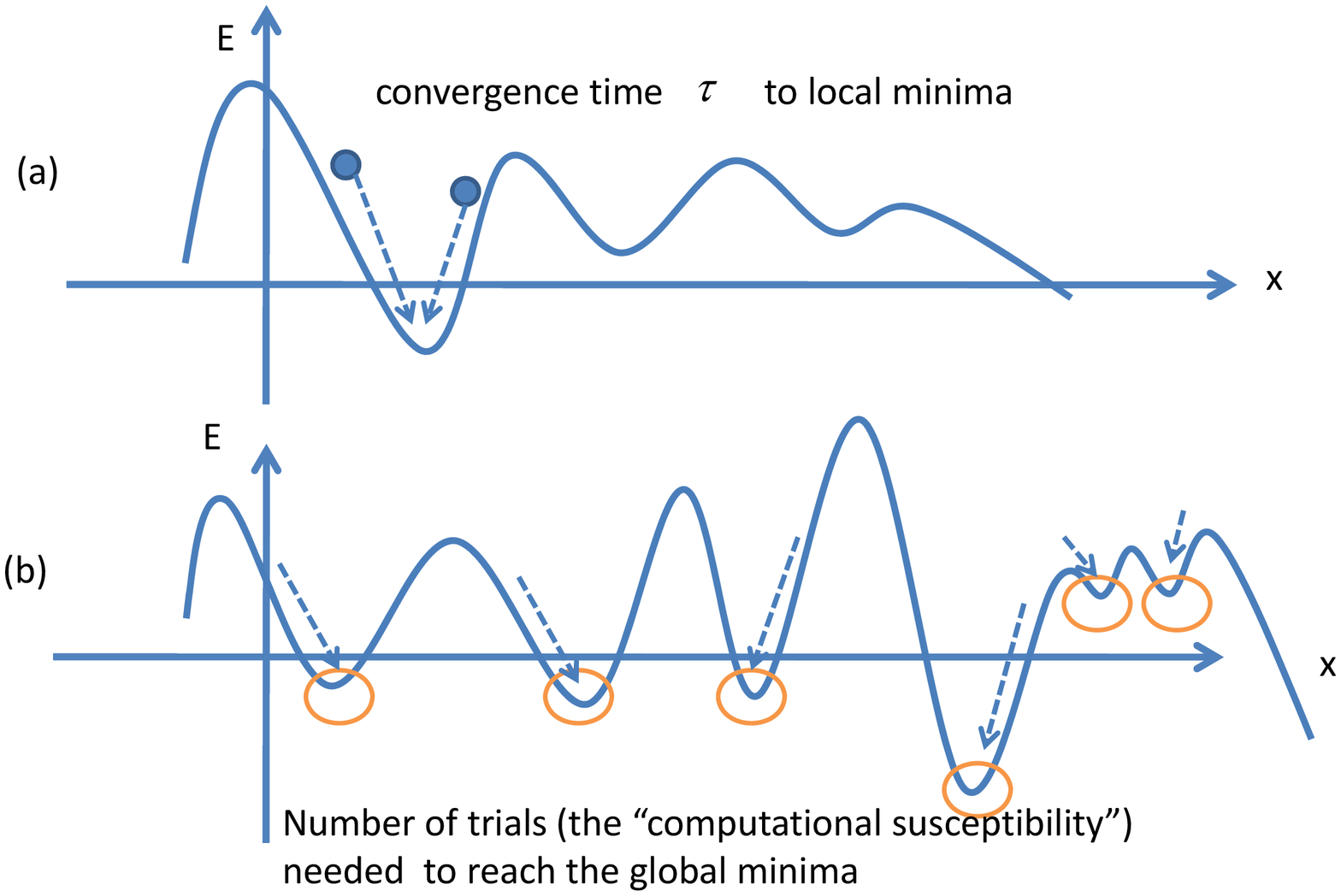}{A schematic of the physical content of
parameters that we employ: (a) The convergence time $\tau$ is the
number of steps the algorithm needed to reach local minima, (b) When
the energy landscape becomes complex, more ``trials'' are needed in
order to veer towards the global minimum (or minima). This requisite
number of trials relates to the ``computational susceptibility''
$\chi$ of Eq.(\ref{eq:susceptiblity}) that as will be explained
later records the improvement in the quality of the solutions (as
seen by the normalized mutual information $I_{N}$) as the number of
trials $s$ (different trajectories in panel (b)) is increased.
 }{fig:schematicplot2}{1\linewidth}{}

\myfig{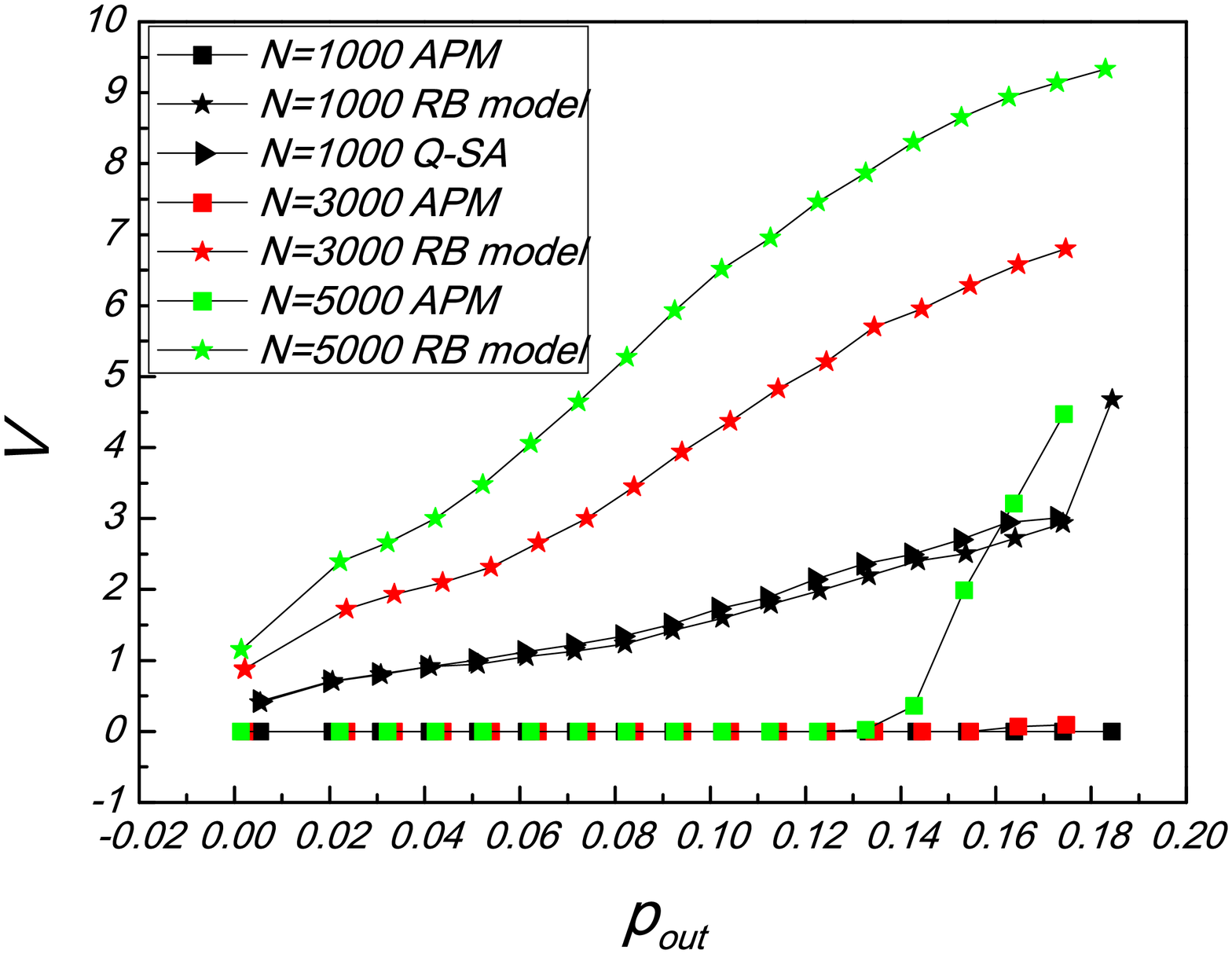}{The average variation of information $V$ of the
``noise'' $p_{out}$ (the density of links connecting different
communities). $V$ is calculated between the proposed solution and
the embedded constructed sample graph whose solution is known. (The
graph has a power-law distribution of community sizes with a minimum
$n_{min}=8$, maximal $n_{max}=40$, and with the exponent determining
the community size distribution set equal to $-1$). We show results
obtained by using our absolute Potts model (denoted as ``APM'' in
\eqnref{eq:ourpotts}). For comparison, we also plot the results
determined by ``RB Potts'' \cite{RB,peter2} model and modularity
optimization (``Q-opt'') \cite{mod} using simulated annealing. With
the ``APM'', our algorithm demonstrates extremely high accuracy for
the small and large systems shown above.
 }{fig:compare}{0.8\linewidth}{t}

\section{Definitions: Trials and Replicas} \label{def}

Before turning to the specifics of our results, we need to introduce
several basic notions. We start by discussing
two concepts which underlie our approach. Both concepts pertain to the
use of multiple identical copies of the same system which differ
from one another by a permutation of the site indices. In the
definitions of  {\em``trials''} and {\em ``replicas''} given below, we build on
the existence of a given algorithm (any algorithm) that may minimize
a given energy or cost function. In our particular case, we minimize
the Hamiltonian of Eq. (\ref{eq:ourpotts}). However, these ideas
and concepts are more general.
\bigskip

$\bullet$ {\underline{{\em Trials}.} We use trials alone in our bare
community detection algorithm \cite{peter1,peter2}. We run the
algorithm on the same problem ``$s$'' independent times. This may
generally lead to different contending states that minimize Eq.
(\ref{eq:ourpotts}). Out of these $s$ trials, we will pick the
lowest energy state and use that state as the solution.
In the current work, $ 4 \le s \le 20$.
We will canonically employ $s=4$ trials. We will use $s>4$ trials in
the calculation of the computational
susceptibility of Eq.(\ref{eq:susceptiblity}).

\bigskip

$\bullet$ {\underline{{\em Replicas}.}   Each sequence of the above
described $s$ trials is termed a {\em replica} (see the schematic
plot \figref{fig:replica} of {\em replicas} ). When using
``replicas'' in the current context, we run the aforementioned $s$
trials (and pick the lowest solution) ``$r$'' independent times. By
examining information theory correlations between the ``$r$'' {\em
replicas} and the known (or ``planted'') solution, we can assess the
quality of candidate solutions. In this work, we set $r=100$.

In this work, we will briefly remark on the determination
of optimal parameters of the system. To this end, we
will compute the average inter-replica information theory correlations within the ensemble
of $r$ replicas. Specifically, {\em information theory extrema} as a
function of the scale parameters, generally correspond to more
pertinent solutions that are locally stable to a continuous change
of scale.  It is in this way that we will detect the important
physical scales and parameters in the system.
\bigskip

In this work, we will compute the  {\em average} information
measures between the disparate candidate solutions found by
different {\em ``replicas''} and the known (or ``planted'') solution
to the problem which we label below as ``$K$''. In general, with $A$
denoting graph partitions in different ``replicas'' and $Q(A,K)$
denoting the information theory overlap between replica A and
the known solution $K$, the average for a general
quantity $Q$ that we will employ are, rather explicitly,
\begin{eqnarray}
\langle Q \rangle = \frac{1}{r} \sum_{A} Q(A,K).
\end{eqnarray}
In earlier works \cite{peter2,peter3,peter4,image}
we employed the average inter-replica information theory overlaps.
We will invoke this method once when discussing the optimal value of
the resolution parameter $\gamma$ of Eq. (\ref{eq:ourpotts}).
Apart from that single case, will generally not use these
average inter-replica measures here but rather their comparison
to a known solution $K$.

In the context of the Potts model Hamiltonian
of Eq. (\ref{eq:ourpotts}),
by ``replicas'', we allude \cite{peter1} to systems
that initially constitute identical copies of the system that differ
only by a permutation the Potts spin label. Different replicas will,
generally, lead to disparate
final contending solutions. By the use of an {\em
ensemble} of such replicas, we can attain accurate result
and determine information theory correlations between candidate
solutions and infer from these a detailed picture of the system.

These definitions might seem fairly abstract for the moment. We will
flesh these out and re-iterate their definition anew when detailing
our specific results  and invoked information theory based
correlations to which we turn next.

\section{Information theory and complexity
measures}\label{information}

In this section, we introduce and review information theory measures
(see the schematic plot \figref{fig:replica} depicting the
information theory correlations) (as they pertain to the community
detection problem) that we will employ in our analysis.

$\bullet$ {\em Shannon Entropy.}  If there are $q$
communities in a partition $A$, then the Shannon entropy is

\begin{eqnarray}
H_A=-\sum_{a=1}^q\frac{n_a}{N}\log_2\frac{n_a}{N}. \label{eq:HA}
\end{eqnarray}

The ratio $\frac{n_a}{N}$ is the probability for a randomly selected
node to be in a community $a$ with $n_a$ the number of nodes in
community $a$ and $N$ the total number of nodes. With
the aid of this probability distribution the Shannon entropy of Eq. (\ref{eq:HA})
follows.

$\bullet$ {\em The mutual
information.}
The mutual information $I(A,B)$ between candidate
partitions ($A$ and $B$) that are found by two replicas is

\begin{eqnarray}
I(A,B)=\sum_{a=1}^{q_A}\sum_{b=1}^{q_B}\frac{n_{ab}}{N}\log_2\frac{n_{ab}N}{n_an_b}. \label{eq:IAB}
\end{eqnarray}

Here, $n_{ab}$ is the number of nodes of community $a$ of partition
$A$ that are shared with community $b$ of partition $B$, $q_A/q_B$
is the number of communities in partition $A$ (or $B$), and (as earlier) $n_a$
(or $n_b$) is the number of nodes
in community $a$ (or $b$).

$\bullet$ {\em The variation of information.}

The variation of information $0\leq
V(A,B)\leq\log_2N$ between two partitions $A$ and $B$ is given by

\begin{eqnarray}
V(A,B)=H_A+H_B-2I(A,B).
\end{eqnarray}

$\bullet$ {\em The normalized mutual information.}
The normalized mutual information $0\leq I_N(A,B)\leq1$ is

\begin{eqnarray}
I_N(A,B)=\frac{2I(A,B)}{H_A+H_B}.
\end{eqnarray}

High $I_N$ and low $V$ values generally indicate high agreement
between different the partitions (or general Potts spin
configurations) $A$ and $B$.

The physical significance of two of the following concepts is sketched in Fig. \ref{fig:schematicplot2}.

$\bullet$ {\em The convergence time.}  The convergence time $\tau$ is the number of the algorithm
steps needed to find the local minimum following a greedy algorithm.
As just noted above, a schematic plot explaining the physical meaning of the
convergence time $\tau$ is shown in \figref{fig:schematicplot2}.

$\bullet$ {\em The complexity.} The complexity customarily denoted
as $\Sigma(e)$, can be derived from the number of states
${\cal{N}}(E)$ with energy $E$. Specifically, ${\cal{N}}(E) \sim
\exp[N\Sigma(e)]$, \cite{mezard} where the energy density $e = E/N$.
In this work, we will numerically determine the onset of the high
complexity (which probes the number of local minima) without any
prior assumptions or approximations by directly computing the
``computational susceptibility'' (\cite{peter1}) that we will
briefly define next.

$\bullet$ {\em The ``computational susceptibility''.}

A ``computational susceptibility'' monitoring the onset of high
complexity can be defined as:

\begin{eqnarray}
\chi_n=I_N(s=n)-I_N(s=4).
 \label{eq:susceptiblity}
\end{eqnarray}
That is, $\chi$ is  the increase in the normalized multure
information $I_N$ as the number of trials (number of initial
starting points in the energy landscape) $s=n$ is increased.
Physically, we ask how many different initial starting points in the
energy landscape (i.e., how many different initial ``trials'') are
required to achieve a certain desired threshold accuracy as measured
by information theory
measures. 

\begin{figure}[]
\begin{center}
\subfigure[The Variation of  information $V$ as a function of the
inter-community link density $p_{out}$.  Note that $V$ for the
$q=140$ system rapidly increases from zero at precisely $p_{out}
=p_{1} =0.2$.  At a value of $p_{out} = p_{2} = 0.24$, $V$ exhibits
a much more gradual increase (whence curves for different values of
$q$ cross).]{\includegraphics[width= 2.45in]{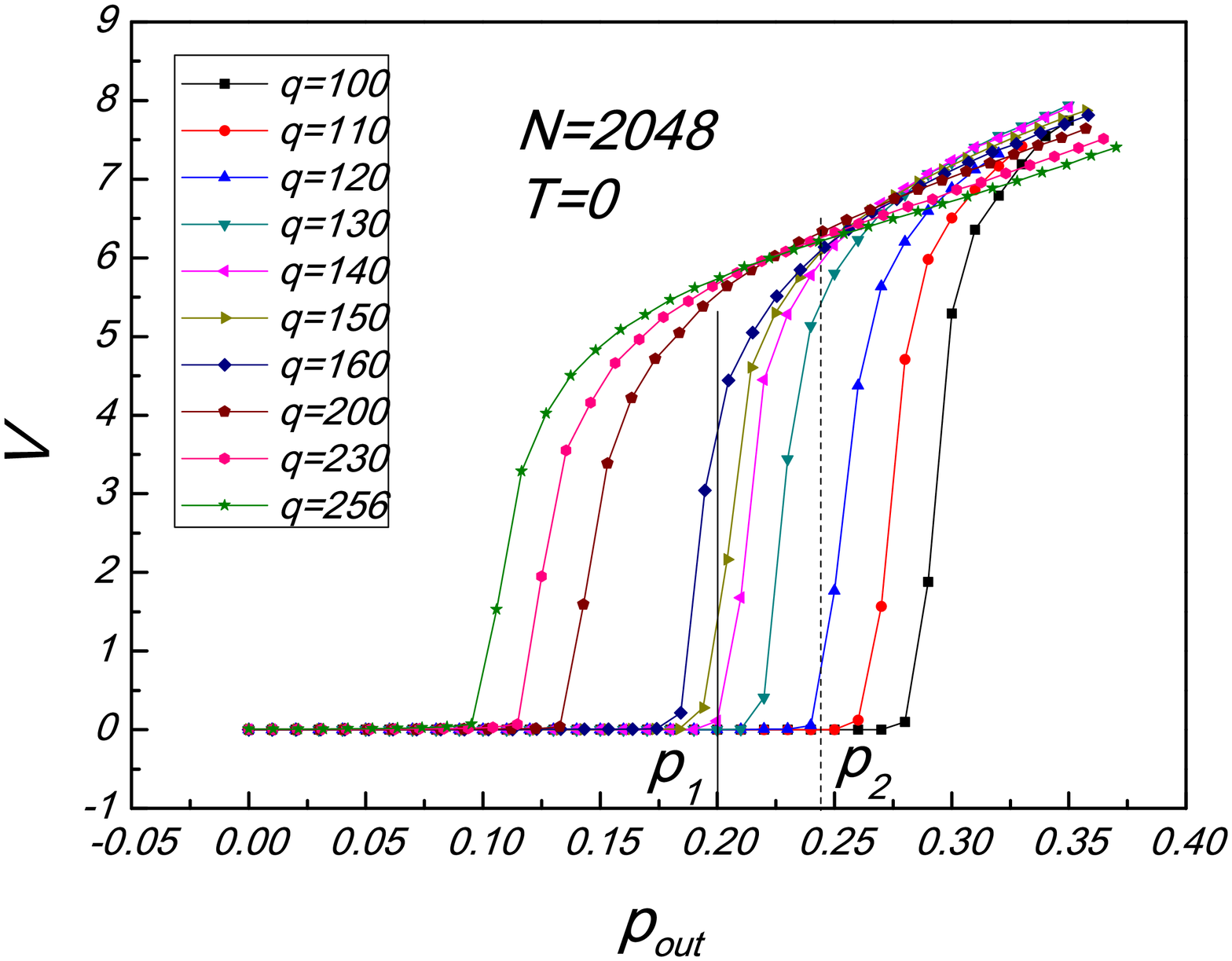}}
\subfigure[The Shannon Entropy $H$ versus $p_{out}$. $H$ starts to
increase at precisely $p_{out}=p_{1}$. Beyond$p_{out} =p_{2}$, the
entropy monotonically decreases and veers towards a universal curve
appearing for all values of $q$.]{\includegraphics[width=
2.45in]{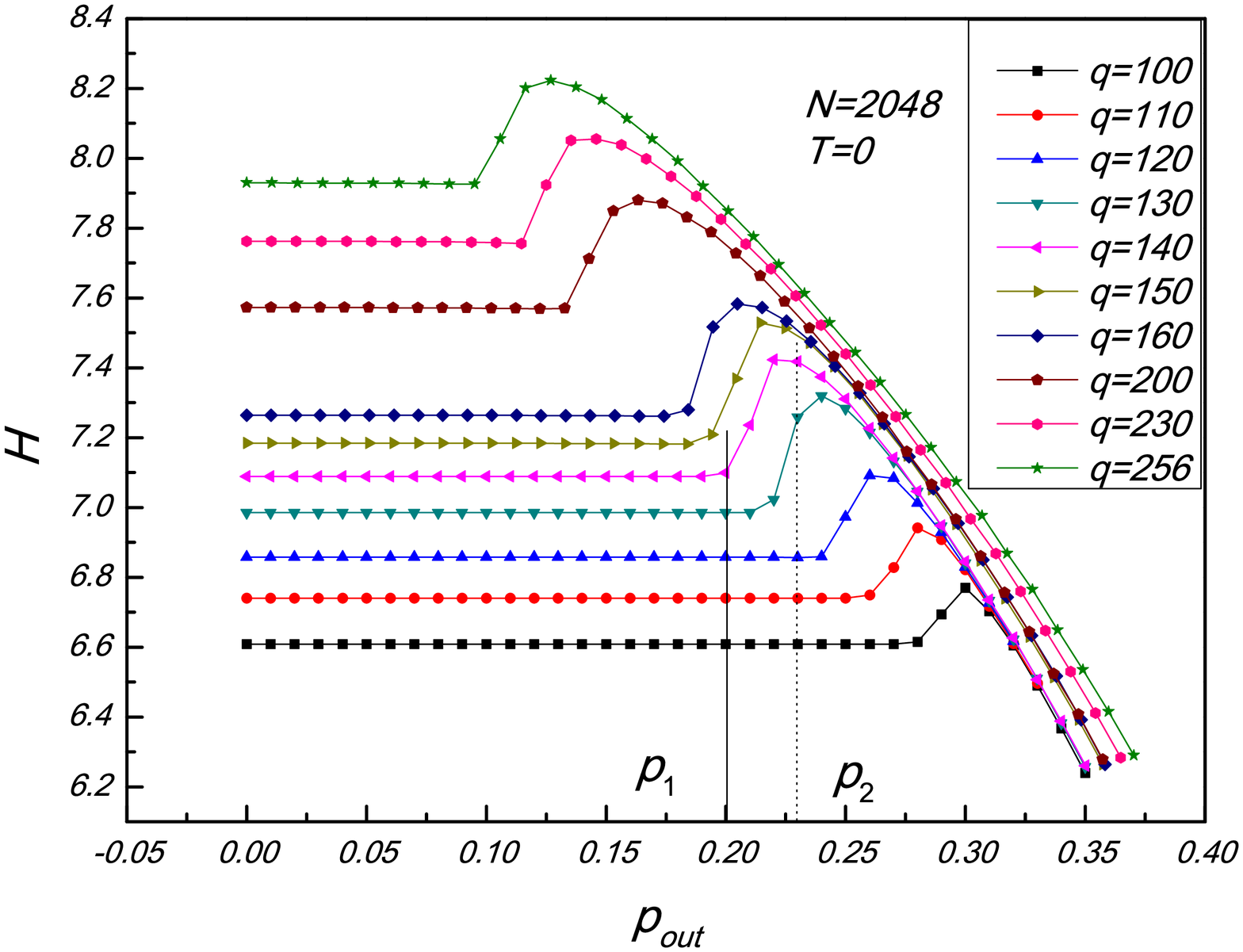}}
\end{center}
\end{figure}

\begin{figure}
\begin{center}
\subfigure[
The convergence time $\tau$ (following a greedy algorithm) to {\em a
local minimum} (as shown in panel (a) of
\figref{fig:schematicplot2}) exhibits a sharp maximum at the
transition between the easy and hard phases at precisely $p_{out}
=p_{1}$. The hard phase is marked not only by a large convergence
time to local minima $\tau$ but rather by a large complexity (a high
degree of metastable minima). This leads to a more difficult
convergence to {\em the global energy minimum} (requiring many
trials to achieve the desired accuracy (see text)). At $p_{out} =
p_{2}$, the convergence time collapses onto the universal curve
appearing for all $q$ (for high $p_{out}$).]{\includegraphics[width=
2.45in]{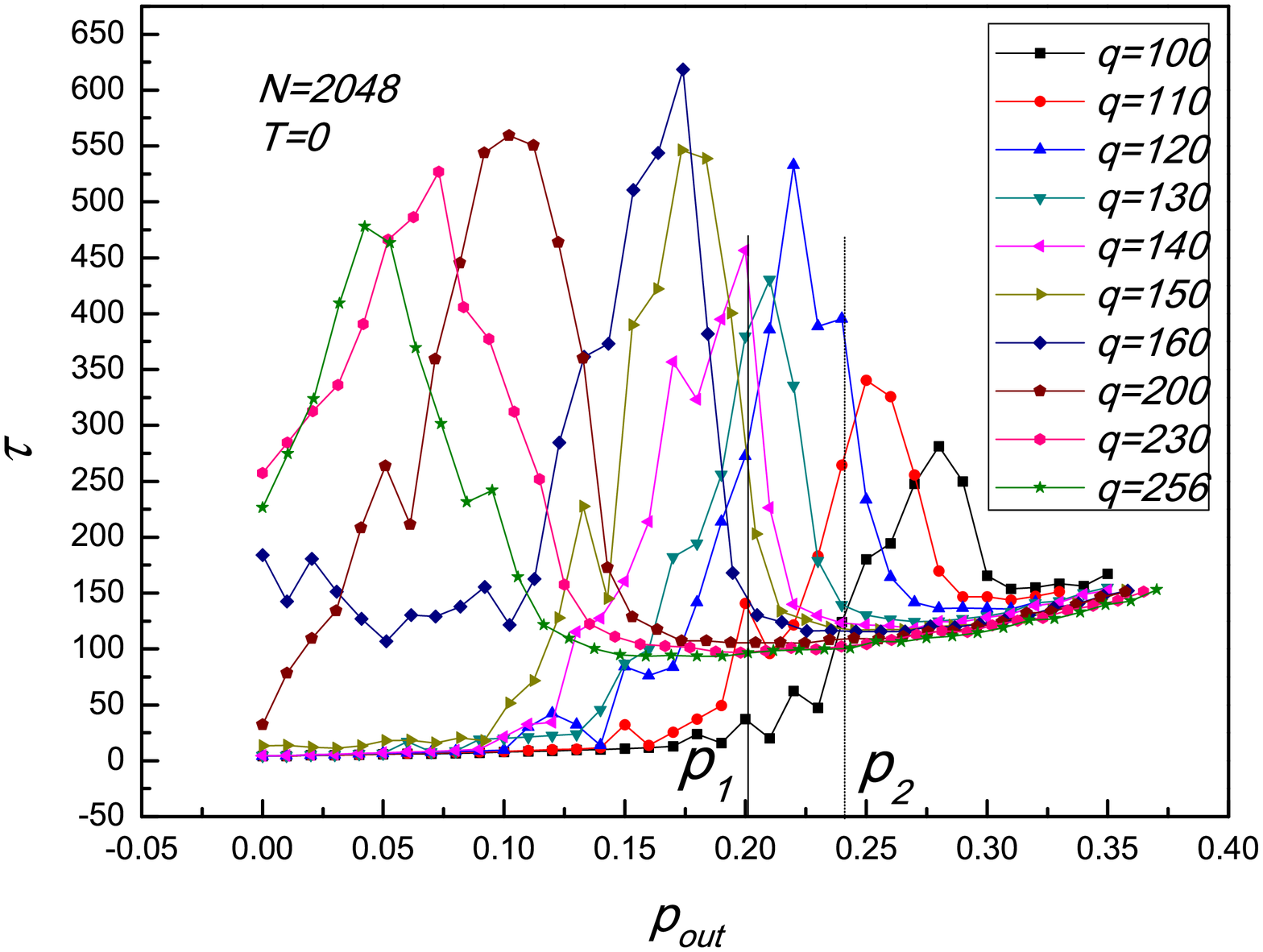}} \subfigure[The ``computational
susceptibility'' $\chi$ of Eq. (\ref{eq:susceptiblity}) (as shown in
panel (b) of \figref{fig:schematicplot2}) versus $p_{out}$ for
different trial numbers. This quantity monitors the complexity or
number of metastable local minima. Note that $\chi$ increases from
zero at precisely $p_{out} =p_{1}$. The computational susceptibility
markedly diminishes for $p_{out} =p_{2}$. As is evident here, a
higher number of trials (a higher number of starting points in the
high energy energy landscape) is required in order to achieve ever
more accurate solutions.]{\includegraphics[width=
2.45in]{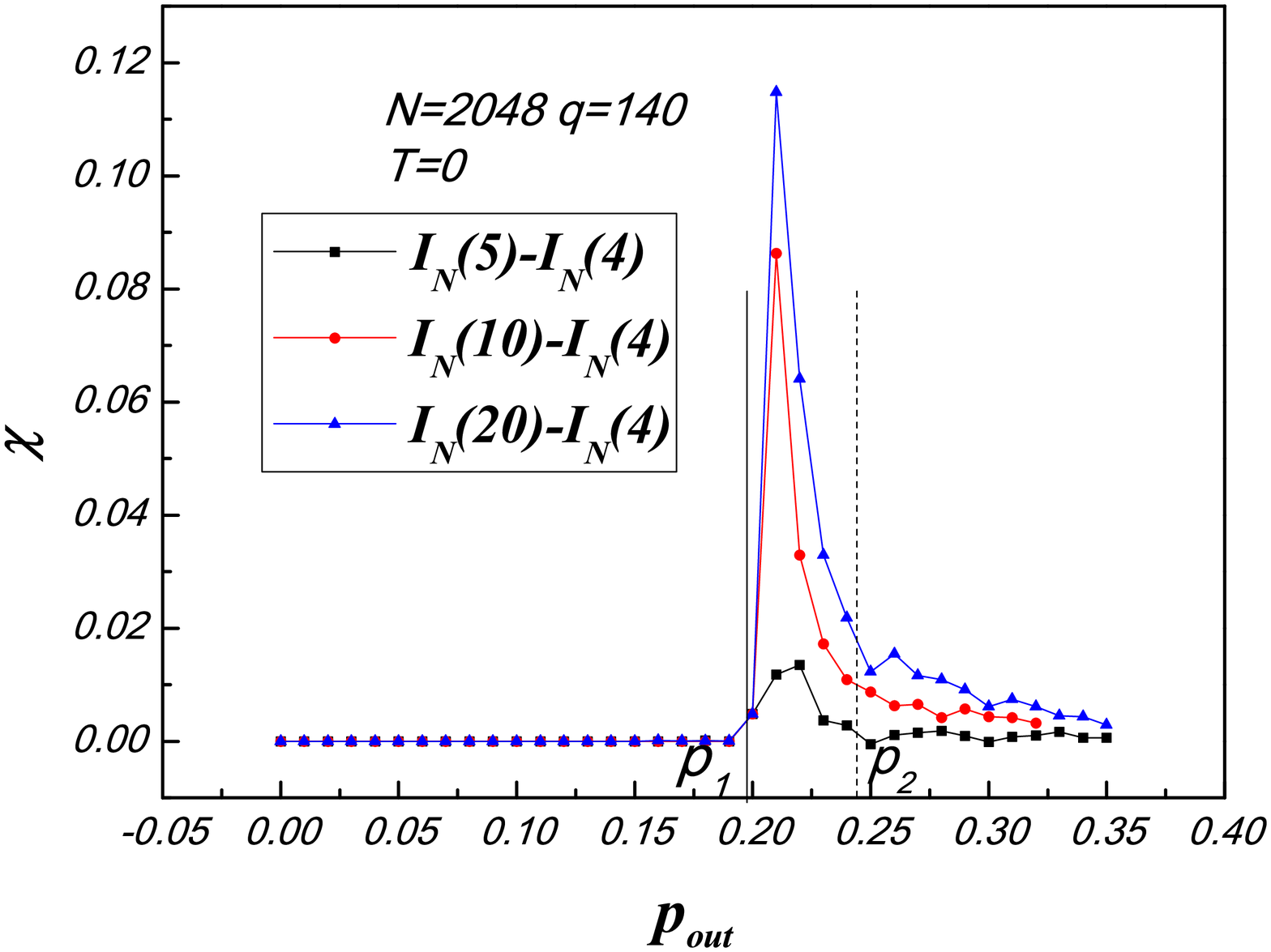}}\caption{Plots of various measures as a
function of the noise level $p_{out}$. $V$ is the variation of
information. $H$ is the shannon entropy (\cite{peter1}). $\tau$ is
the number of steps needed to reach local low energy state (see also
Fig. \ref{fig:schematicplot1}).  The ``computational
susceptibility'' $\chi$ is defined in \eqnref{eq:susceptiblity}. In
the examined system of $N=2048$ nodes with $q=140$ communities, all
of the plots show three phases as noise varies. (1) Below a noise
threshold value of $p_{1}=0.2$, the system can be ``easily'' solved.
(2) When $0.2<p_{out}<0.24$, the benefit of extra trials is most
significant (shown in (c)) and it is ``hard'' to solve the system.
 (3) Above noise levels about $p_{2}=0.24$, the system cannot
 be perfectly solved. As we will outline, the two
transitions at $p_{out} =p_{1},p_{2}$ are both of the spin-glass
type.}\label{fig:3phases}
\end{center}
\end{figure}

\section{Noise tests}
Similar to \cite{lfr_bench}, we will use a ``noise
test''  benchmark as a workhorse to study phase transitions in random graphs
\cite{peter2}.

We define the system ``noise'' in community detection problem as
edges that connect a given node to communities other than its
original community assignment
(``inter-community''
edges). In general \cite{peter2}, we cannot
initially distinguish between edges contributing to noise and those
constituting edges within communities of the best partition(s).

Specifically, for each constructed benchmark graph, we start with
$N$ nodes divided into $q$ communities with a power law size
distribution (with the exponent determining the community size
distribution \cite{lfr_bench} set equal to $(-1)$, i.e., the
community size $n$ scales as $n^{ {\overline{\beta}} }$ with
${\overline{\beta}}=-1$). We connect all ``intra-community'' edges
at a high average edge density $p_{in}= 0.95$.
This, when $p_{out}=0$ we have decoupled clusters
with no inter-community links. We then add random
``inter-community'' edges (``noise'') at a density of $p_{out}<0.5$.
Specifically, $p_{in}$ is defined as the ratio of the existing
intra-community edges over the maximal intra-community edges, and
$p_{out}$ is defined as the ratio of the existing
``inter-community'' edges over the maximal inter-community edges. If
we denote the {\em average external degree} for each node by
$Z_{out}$ (i.e., the average number of links between a given node to
nodes in communities other than its own) and the {\em average
internal degree} by $Z_{in}$ (i.e., the average number of links to
nodes in the same community- $Z_{in} + Z_{out} = Z$ with $Z$ the
average coordination number), then we define \cite{peter2}

\begin{eqnarray}
p_{in}=\frac{NZ_{in}}{\sum_{a=1}^q n_a(n_a-1)}, \label{eq:pin}
\end{eqnarray}

and

\begin{eqnarray}
p_{out}=\frac{NZ_{out}}{\sum_{a=1}^q\sum_{b \neq a}^q n_a n_b}.
\label{eq:pout}
\end{eqnarray}

In the above, as throughout, $n_a$ denotes the number of nodes in community $a$.

When the noise is low (i.e., when
$p_{out}$ is small), all the communities are well defined. As more
and more external links are progressively added to the system ($p_{out}$
increases), the communities become harder and harder to detect. In
some stage, when the external link density is efficiently high, the
system cannot be detected. As alluded to earlier, we investigate the phase transition from
the ``solvable'' to ``unsolvable'' at \emph{both} the low and high
temperature with the use of the heat bath algorithm (``HBA'' in Appendix B) in
the following section.

\section{Spin glass type transitions} \label{sec:transition}

\subsection{Results for information theory correlation and thermodynamic
quantities}

With all of the preliminaries now in place, we now
report our findings.  The upshot of the results to be presented
is evidence for the existence of \emph{two}
spin glass type transitions in general random graphs. Evidence for these transition is
afforded by changes in the accuracy of the solution obtained by the ``APM''
in \eqnref{eq:ourpotts} when noise is introduced. This is shown in
\figref{fig:compare}. The variation of information $V$ between the
test system result and the solution displays a phase transition as
the noise $p_{out}$ increases. A transition is also manifest in the
sudden jump of $V$. The variation of information $V$ remains zero (indicating,
essentially, perfect
solutions) up to a threshold value of the noise where a very sharp transition
is seen. We compared this transition to similar transitions that we detected
via more standard,
methods. These are labeled,
in \figref{fig:compare},
 by ``Q-opt SA'' (maximization of modularity (Q), set
by a comparison to a null model, \cite{mod} as solved by simulated
annealing (SA)) and ``RBPM'' (the Potts model of \cite{RB} wherein
the parameters in the Hamiltonian are also defined by a null model).
As seen, our ``APM" of \eqnref{eq:ourpotts} (which is free  \cite{peter1} of the so-called
``resolution limit'' \cite{fortunato,res_lim} that appears in
systems with null models) can be used to examine graphs
with high levels of noise \cite{peter2}. By
comparison to other models compared to null models, the APM exhibits
a sharper transition as the number of nodes $N$ is increased
\cite{peter1}.

\subsection{General features of the phase diagram as ascertained by numerical data}

As is evident in \figref{fig:3phases}, there are three different
phases. We denote these phases by the qualifiers of (i){\em ``easy''}, (ii){\em ``hard''}, and
(iii){\em ``unsolvable''}. These three phases are the analogs of the three
phases (i)``SAT'', (ii)``hard'', (iii)``unSAT'' in the k-SAT problem
\cite{mezard}. In later discussions, we elaborate
on their possible physical significance of these
phases in disparate arenas such as that
of supercooled liquids. In what follows,
we first present our results.  We first
discuss the zero temperature case ($T=0$)
and then explore the physics at $T>0$.

\begin{figure}[t]
\begin{center}
\subfigure[The computational susceptibility of Eq.
\ref{eq:susceptiblity} $\chi(T,p_{out})$ as a function of the heat
bath temperature $T$ and the level ``noise'' $p_{out}$ (density of
inter-community links) for the system with $N=2048$ nodes and
$q=140$ communities.]{\includegraphics[width=
2.1in]{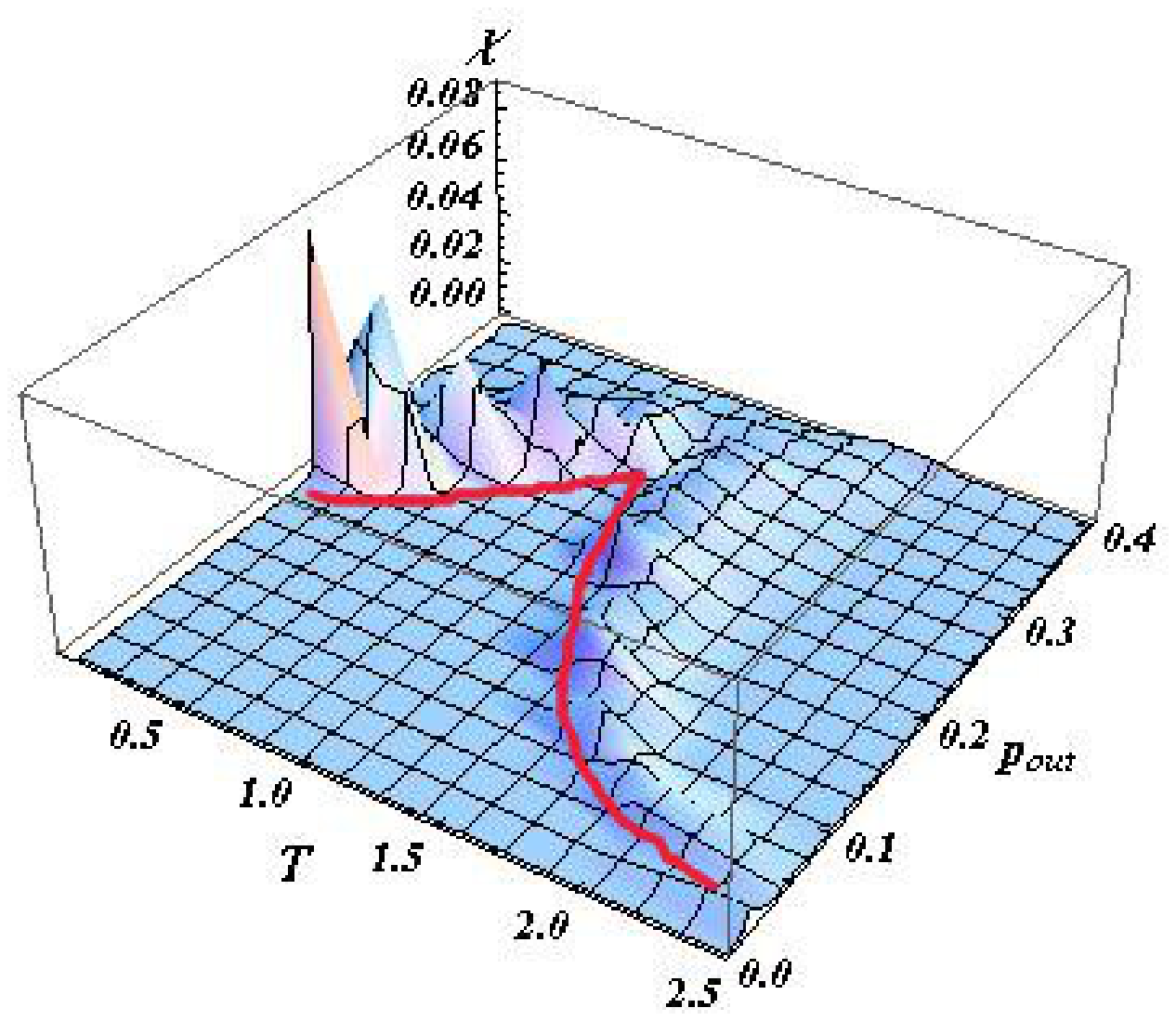}} \subfigure[The
normalized mutual information $I_N(T,p_{out})$ for the same system.
]{\includegraphics[width= 2.2in]{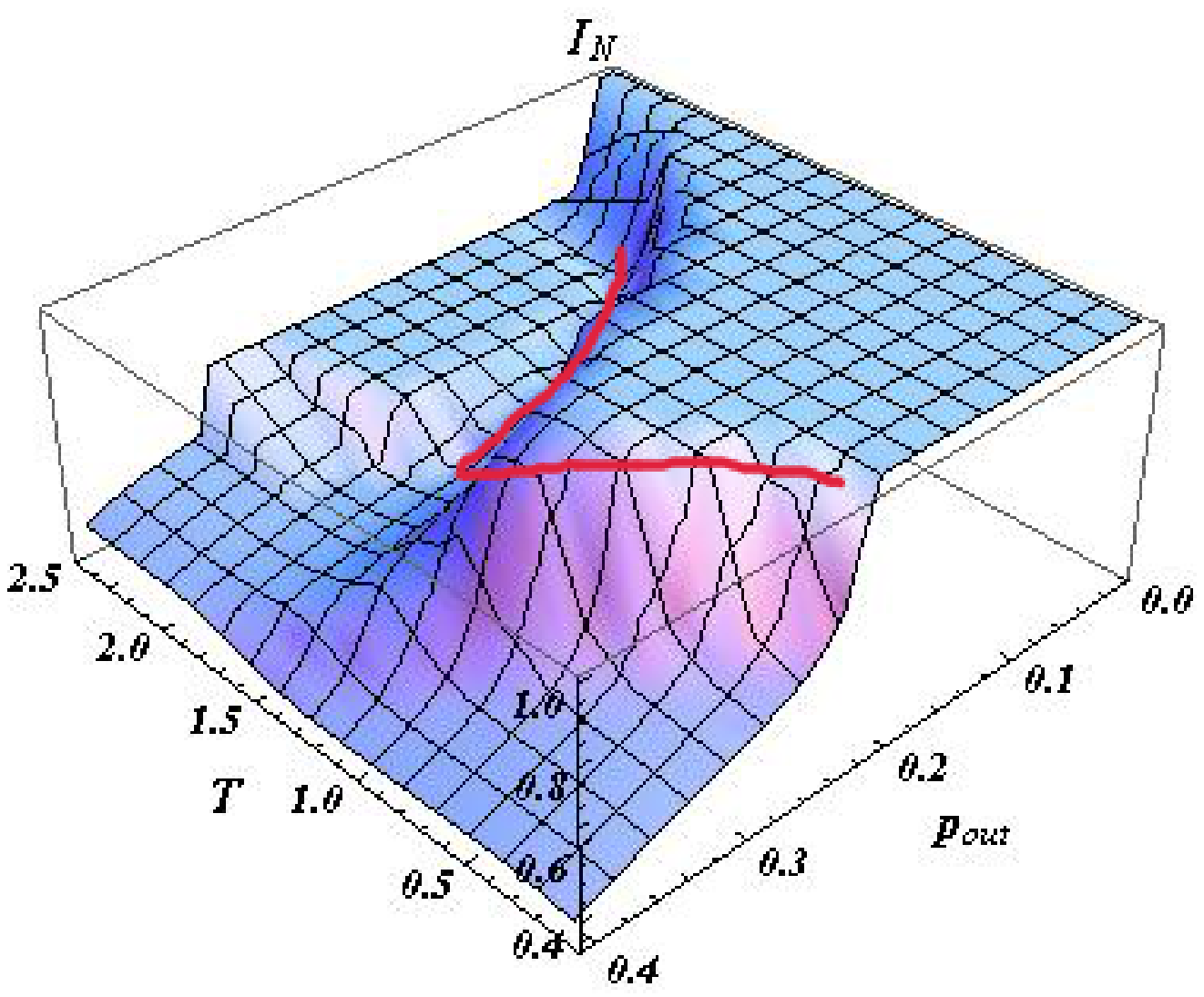}}
\end{center}
\end{figure}

\begin{figure}
\begin{center}
\subfigure[A plot of the energy $E(T,p_{out})$. The energy here is
an ensemble average energy over $100$ replicas at time
$t=1000$.]{\includegraphics[width= 2in]{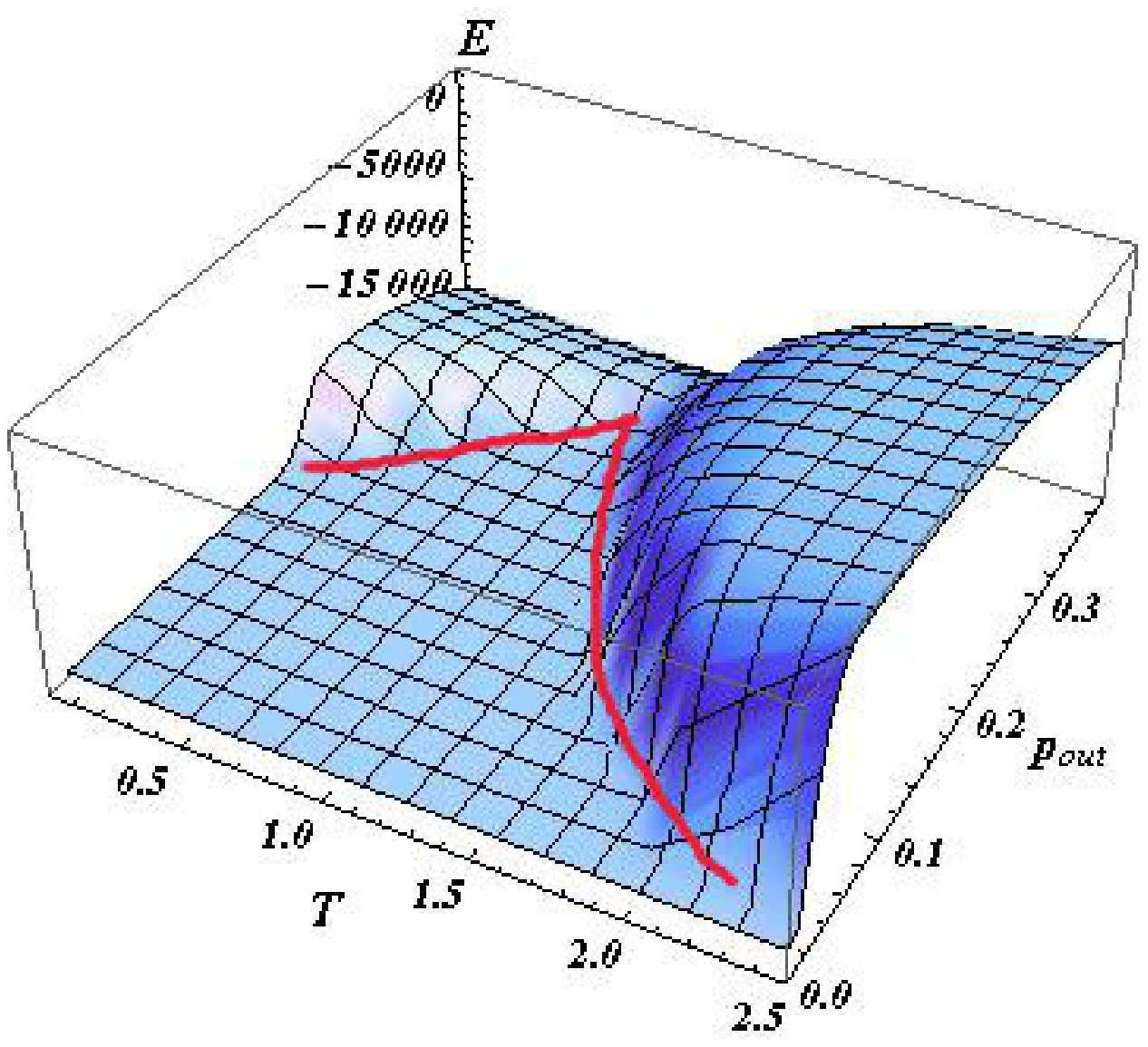}}
\subfigure[Plot of Shannon entropy $H(T,p_{out})$.
]{\includegraphics[width=
2in]{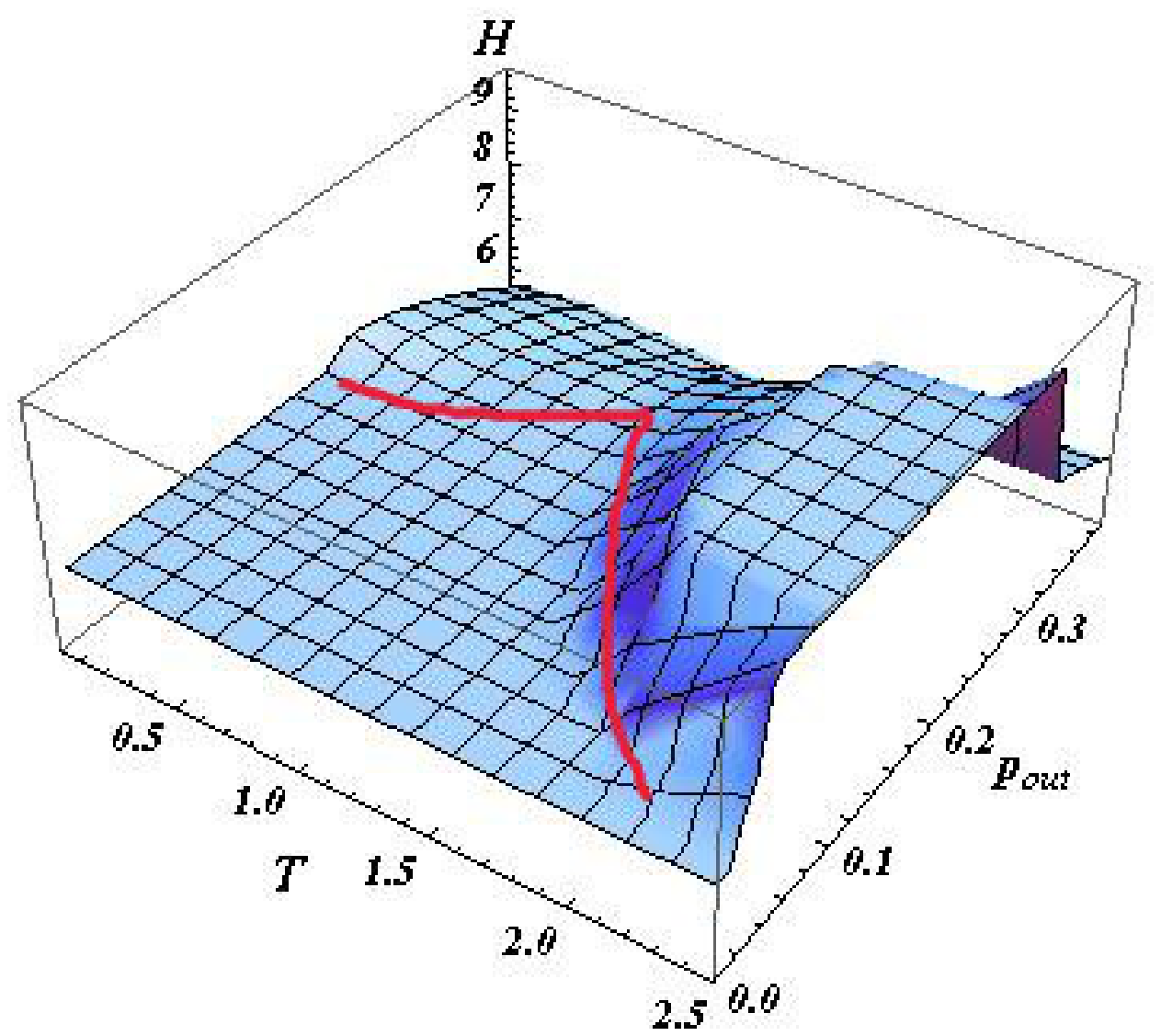}}\caption{The computational
susceptibility $\chi$, normalized mutual information $I_N$, Shannon
entropy $H$ and energy $E$ in terms of temperature $T$ and the
inter-community noise $p_{out}$ for systems with $N=2048$ nodes and
$q=140$ communities. All of the plots show three different phases
which correspond to the three panels ((a)-(c)) shown in
\figref{fig:schematicplot1}, denoted as ``hard-easy-hard''. The
first ``ridge'' in the low temperature in panel (a)-(d)
(computational susceptibility $\chi$/normalized mutual information
$I_N$/energy $E$/entropy $H$) corresponds to the ``hard'' phase
shown in panel (a) in \figref{fig:schematicplot1}. A higher
temperature hard phase is also present. A guide to the eye is drawn
to emphasize the manifestation of the hard phases in all measured
quantiities. The middle ``flat'' region in panels (a)-(d) is the
``easy'' phase.} \label{fig:3dplot}
\end{center}
\end{figure}

\subsubsection{$T=0$}

In \figref{fig:3phases}, the low noise region ($p_{out}<p_1$) is
seen to be in the ``easy'' phase. In this phase,  the accuracy
($V$), entropy ($H$), and the computational susceptibility ($\chi$)
are constant. Within this regime, the algorithm is able to correctly
distribute nodes into their correct communities. We test several
systems with different system size $N$ and number of communities
$q$, and in Appendix E, we plot the first transition point $p_1$ in
terms of $N$ and $q$.

As the noise $p_{out}$ is further increased beyond a threshold value
of $p_{1}$, the system enters the ``hard'' phase. The existence of
the ``hard'' phase is reflected by the rapid growth (decrease) in
the entropy and computational susceptibility (accuracy) curves. Even
though, we can increase the number of trials in order to improve the
accuracy of our solutions (as seen in panel (c) of
\figref{fig:3phases}), it is, nevertheless, still hard to obtain
exact solutions.

As the noise is yet further increased and exceeds a second threshold
value ($p_{out}>p_{2}$), the system undergoes another phase
transition from the ``hard'' phase to an ``unsolvable'' phase. This
``unsolvable'' region is reflected, amongst other things, by the
collapse of all of the curves in each panel of \figref{fig:3phases}.
In this regime, it is impossible to solve the system correctly
without infinite time in the third region.

\subsubsection{$T \ge 0$}
A more detailed, higher dimensional perspective, that includes the
effects of temperature is provided in \figref{fig:3dplot}. In this
figure, summarizing our results for the computational susceptibility
$\chi(T,p_{out})$, Shannon entropy $H(T,p_{out})$, normalized mutual
information $I_N(T,p_{out})$ and system energy $E(T,p_{out})$ at
general finite temperatures $T \ge 0$, we plot the loci of point
marking the boundaries between the different phases. The ``flat''
phase that lies in the middle of these panels is the ``easy'' phase.
[Within the ``easy'' phase, the system is easily solvable and the
planted communities are perfectly detected.] This ``easy'' phase is
separated by ``ridges'' of high computational susceptibility
(marking the ''hard'' regions) from the ``unsolvable'' phases. As
expected, the computational susceptibility/energy/entropy/$I_N$
exhibit a precipitous jump as the noise $p_{out}$ exceeds some
threshold value $p_{1}(T)$. A low temperature hard phase appear for
noise levels $p_{1}(T) \le p_{out} \le p_{2}(T)$. We can determine
the boundaries of the``hard'' phase, whenever it generally exists,
by seeing for which values of $p_{out}$ and $T$ there is a rapid
increase of $\chi$ and $E$. An additional high temperature bump in
the computational complexity $\chi$ and $E$  appears for noise
levels $p_{3}(T) \le p_{out} \le p_{4}(T)$. In this phase,  the
minimization of Eq. (\ref{eq:ourpotts}) is non-trivial. At yet
higher temperatures/noise levels, it is generally impossible to
solve the system. Thus, the two loci of ``ridges''  in the
computational complexity (i.e., $p_{1}(T) \le p_{out} \le p_{2}(T)$
or $p_{3}(T) \le p_{out} \le p_{4}(T)$) delineate the ``hard''
phases. To emphasize the appearance of this ridges and their
manifestation in all measured quantities, a guide to the eye is
drawn. Within the low temperature hard phase ($p_{1}(T) \le p_{out}
\le p_{2}(T)$), the system becomes trapped in the local energy
minima (panel (a) of Fig. (\ref{fig:schematicplot1})).  At low
temperatures, we find from the exact and extensive numerical
calculations (as shown in panel (d) in \figref{fig:3phases}), a very
dramatic increase in complexity just at the transition $p_{1}$
followed by a much more gradual decrease up to $p_{2}$. The
convergence time for a local greedy algorithm (such as ours shown in
(c) of \figref{fig:3phases}) does not correlate with the complexity
as the system. This is so as the system can easily converge to a
wrong local metastable minimum (while the number of such minima is
given by the complexity).

\myfig{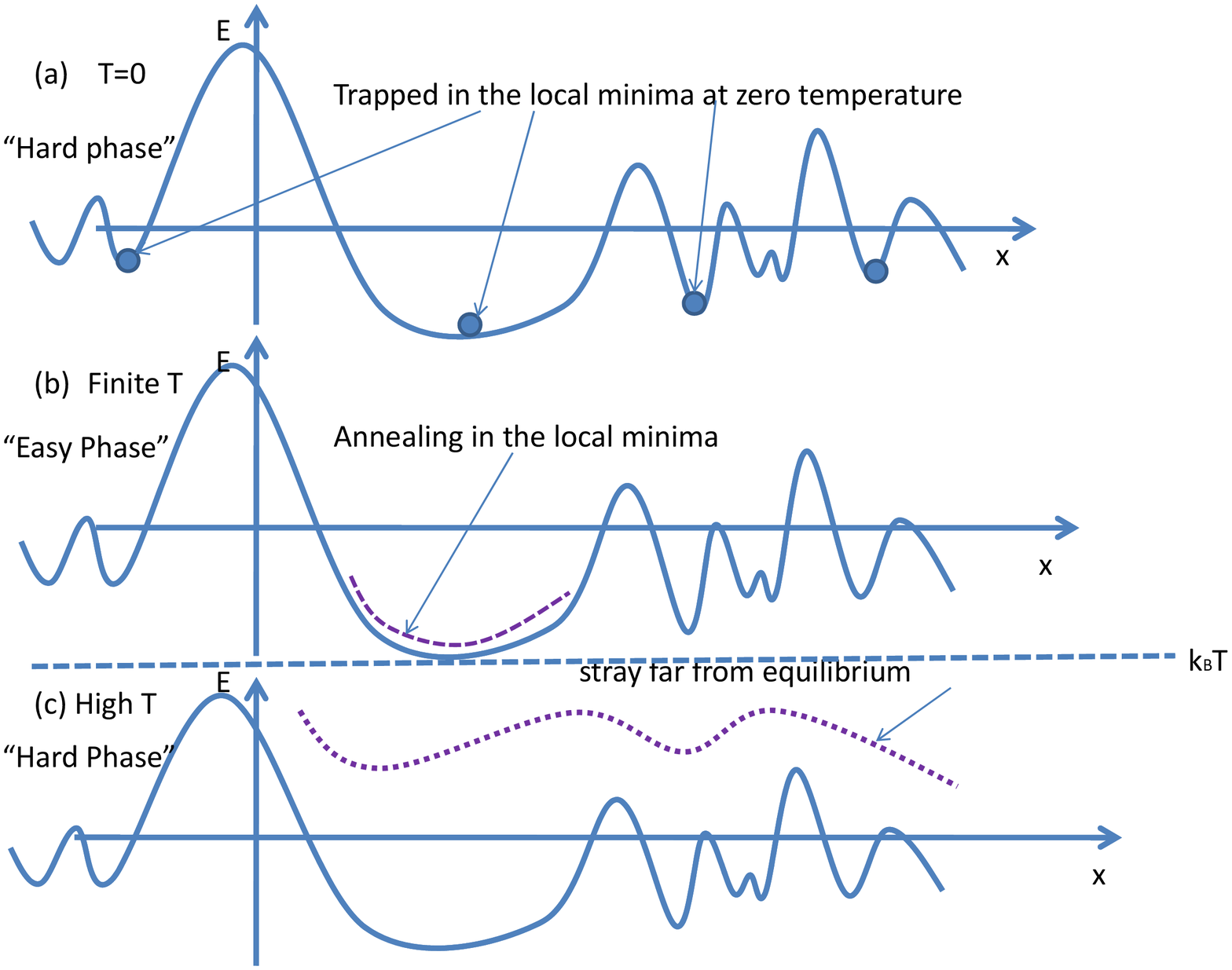}{A caricature of the accessible energy landscape
at different temperatures for a system, such as that examined in
Fig. (\ref{fig:3dplot}) with a fixed noise level $p_{out}$ which
slightly exceeds $p_{1}(T=0)$. In panel(a), at zero temperature, the
system is trapped in local minima. Panel(b) shows the system at
temperatures that are sufficiently high for the system to anneal
and better access regions in the vicinity of the lowest energy states. This
situation corresponds to the
intermediate region that lies between the two ``ridges'' in
\figref{fig:3dplot}. Panel(c) shows the system in a high temperature
phase where,
thermal fluctuations are exceedingly large and the system does not
veer towards low energy states.
 }{fig:schematicplot1}{1\linewidth}{}

\myfig{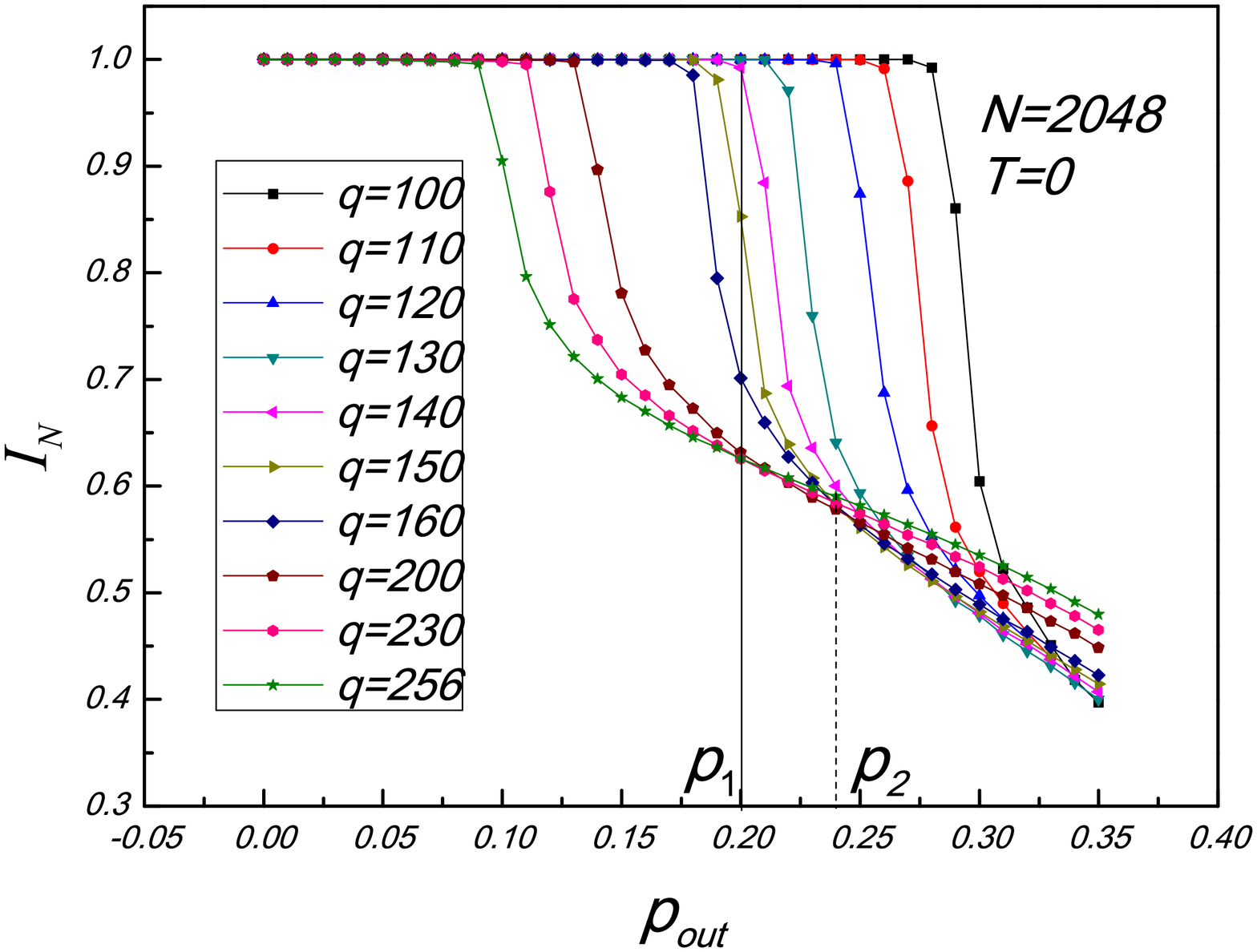}{The normalized mutual information $I_N$ as a
function of $p_{out}$ for system $N=2048$ at temperature $T=0$.
The noise levels
$p_{1}$ and $p_{2}$ are the first and the second transition points
for the particular displayed system of $N=2048$ and $q=140$. The
inferred values of $p_{1}=0.2$ and $p_{2}=0.24$ are consistent
with \figref{fig:3phases}.
The normalized mutual information $I_N$ records the overlap between the ``important partitions'' (the
optimal partition corresponding to the lowest energy state of
\eqnref{eq:ourpotts}) and the contending partitions found by the algorithm.
 }{fig:NMI2048}{0.8\linewidth}{t}

\myfig{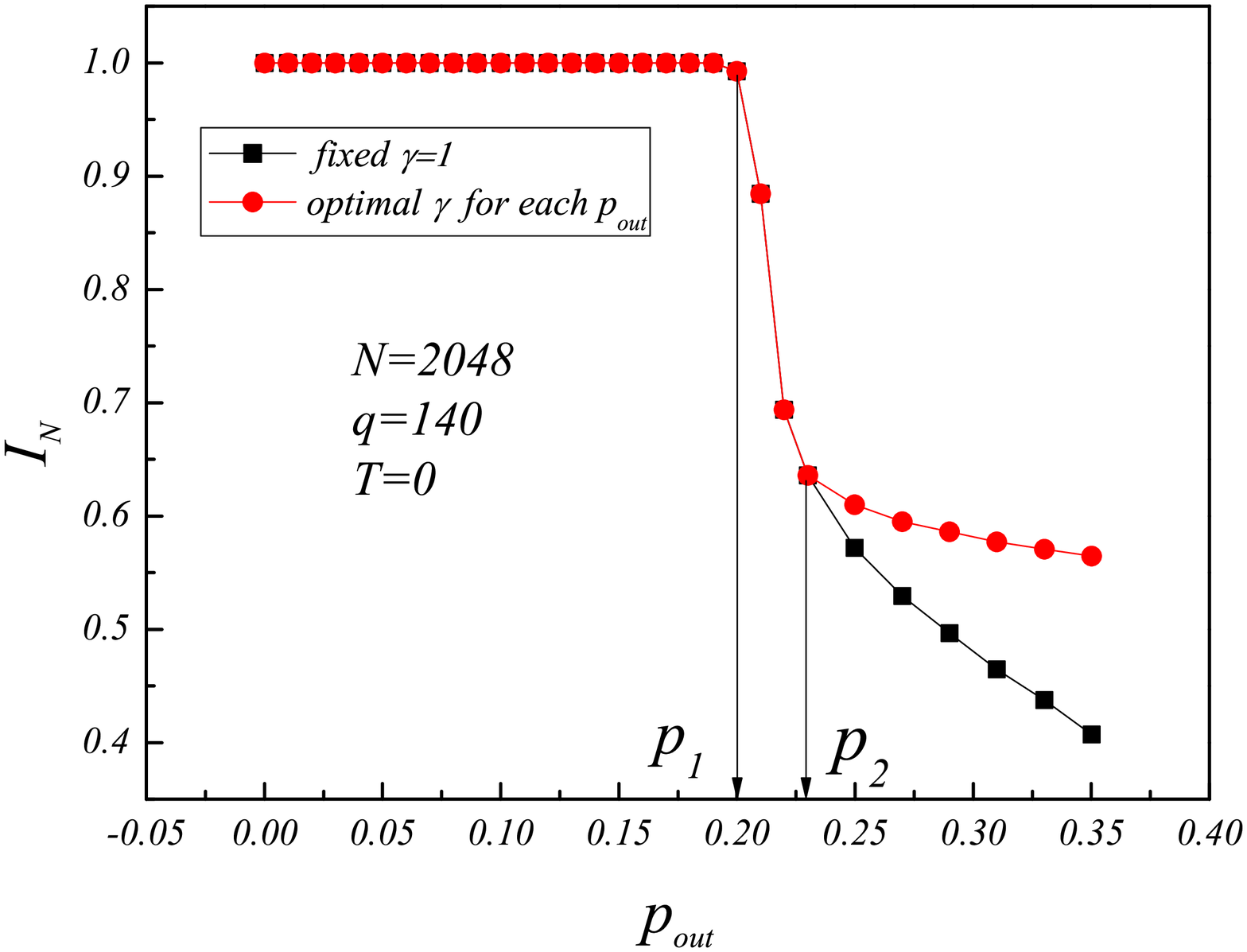}{A comparison of the normalized mutual
information $I_N$ as a function of noise $p_{out}$ between two cases:
(i) one with a fixed
resolution parameter $\gamma=1$ (see Eq. (\ref{eq:ourpotts}))
and (ii) a computation with the optimal $\gamma$ determined by the
maximal $I_N$/minimal $V$ (minimal variation of information).
No change in the transition points $p_{1}$ and $p_{2}$ occurs by optimizing $\gamma$
in this zero temperature system. Indeed, for this system $\gamma=1$ is the optimal
value of $\gamma$
for noise levels $p_{out} <p_2$. The two curves
start to separate for higher noise levels.
 }{fig:optimal}{0.8\linewidth}{t}

In \figref{fig:schematicplot1}, we provide caricatures of the
underlying physics in these phases and the low temperature/low noise
transitions. At low temperatures, for noise
$p_{out}$ slightly above $p_{1}$ (at zero temperature),
 the system becomes quenched in metastable local minima at low
 temperatures.
This is schematically illustrated in panel (a). As the temperature
is increased, the system may, as depicted in panel (b) of \figref{fig:schematicplot1}, veer towards its
global minimum by annealing. Physically, a similar mechanism
is at work in many frustrated physical system where it goes under
the name of  ``order by disorder'' . In such cases, by virtue of
entropic fluctuations, quenching is thwarted and the system may
probe low lying states and indeed order
\cite{villain,shender,henley,biskup}.
Thus, the energy and computational susceptibility may remain constant (there is only one global
energy minimum, i.e., one state or a finite set of such states). However, it does take
progressively more time to locate the global minimum state ((c) in
\figref{fig:3phases}). As the noise is further increased, the system is still ergodic.
However, it takes
a very long time to find the lowest energy state. On finite time
scales,
the system stays in the vicinity of local minima thus yielding a
higher observed energy. Only on sufficiently long time scales
does the system veer towards its global minimum (or minima).
Within this ``hard'' region, there are many
metastable states. This leads to a significant increase in the complexity as
is made evident by the rapid
growth of the computational susceptibility $\chi$ of  \eqnref{eq:susceptiblity}.
The large computational complexity marks the initial rapid climb of the complexity.

We now return to the results of \figref{fig:3dplot} at yet
higher temperatures and values of the noise $p_{out}$.
The high temperature ``ridge'' in \figref{fig:3dplot} ($p_{3}(T) \le p_{out} \le p_{4}(T)$)
corresponds to the system being far away from the minimum energy
state. As we remarked earlier, this delineates yet another ``hard'' phase.
According to the above explanation and the corresponding caricature of
\figref{fig:schematicplot1}, increasing the running time and/or
number of trials should help increase the accuracy of the solution
in this region (the peak area of the computational susceptibility). Beyond this region, at
higher temperatures, the system is unsolvable.
This corresponds to panel (c) in Fig. (\ref{fig:schematicplot1}).

At low temperatures and high noise,
due to the proliferation of metastable states, (i) the convergence
time $\tau$ (as seen in panel (c) of Fig. (\ref{fig:3phases}))
can be low while (ii) the increase in accuracy by performing
more and more trials is, essentially, nil [as seen by the low
value of $\chi$ in panel (d) of \figref{fig:3dplot}].
Similar conclusions can be arrived at finite
temperatures by examining constant $T$
slices of $\chi(T,p_{out})$.


\myfig{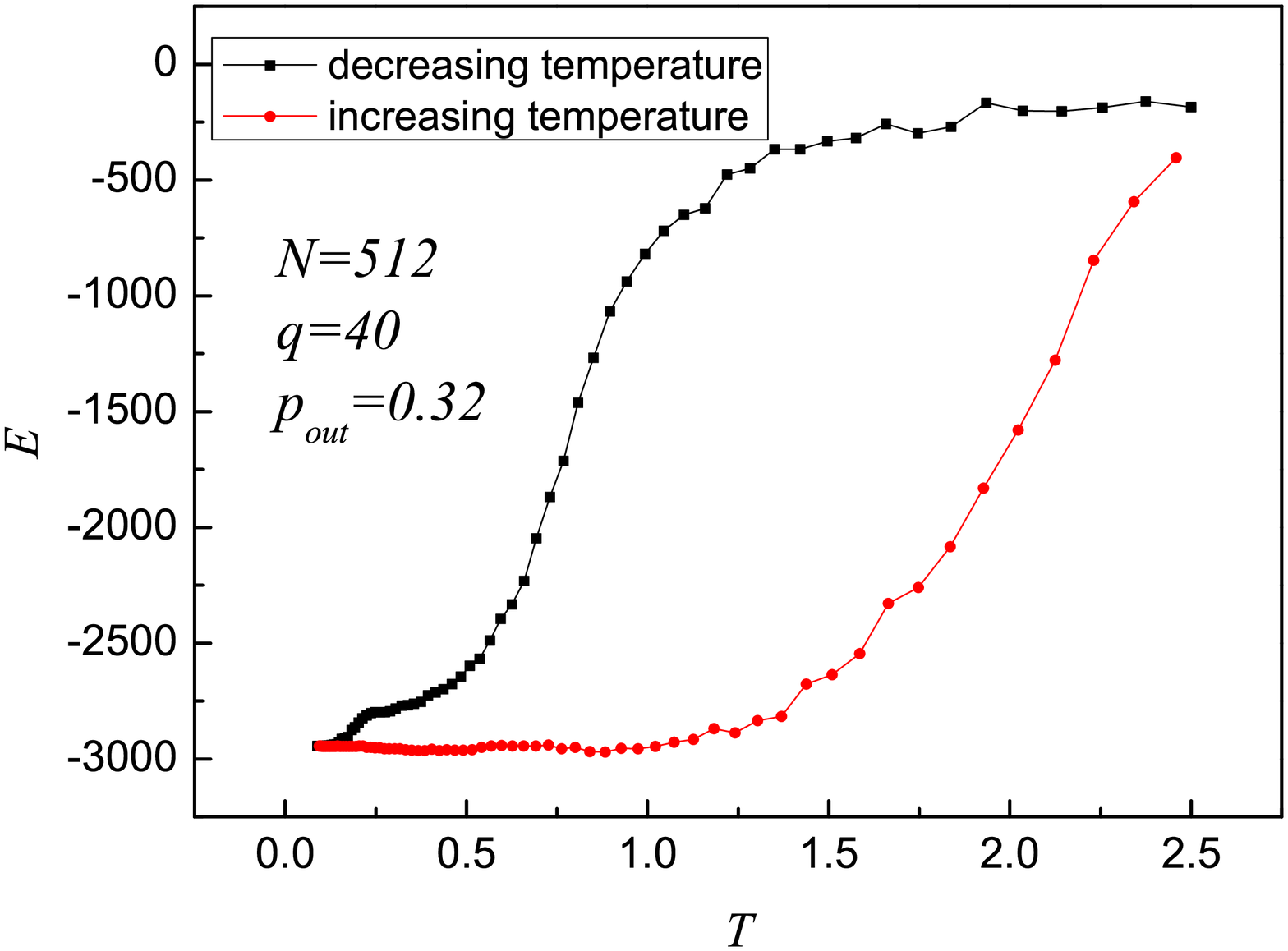}{The energy $E$ as a function of the temperature $T$ for a
system with $N=512$ nodes, $q=40$ communities, and noise $p_{out}=0.32$.
(This noise level exceeds the zero temperature $p_{1}=0.29$ for this system.)
We perform a computational experiment at $T=2.5$ and lower the temperature according to $T_{k+1}=0.95T_{k}$
in consecutive time steps $k$. After a steady-state is obtained, the process
is reversed.  A clear hysteresis-like effect
is evident.
}{fig:EvsTemp}{0.8\linewidth}{t}

\myfig{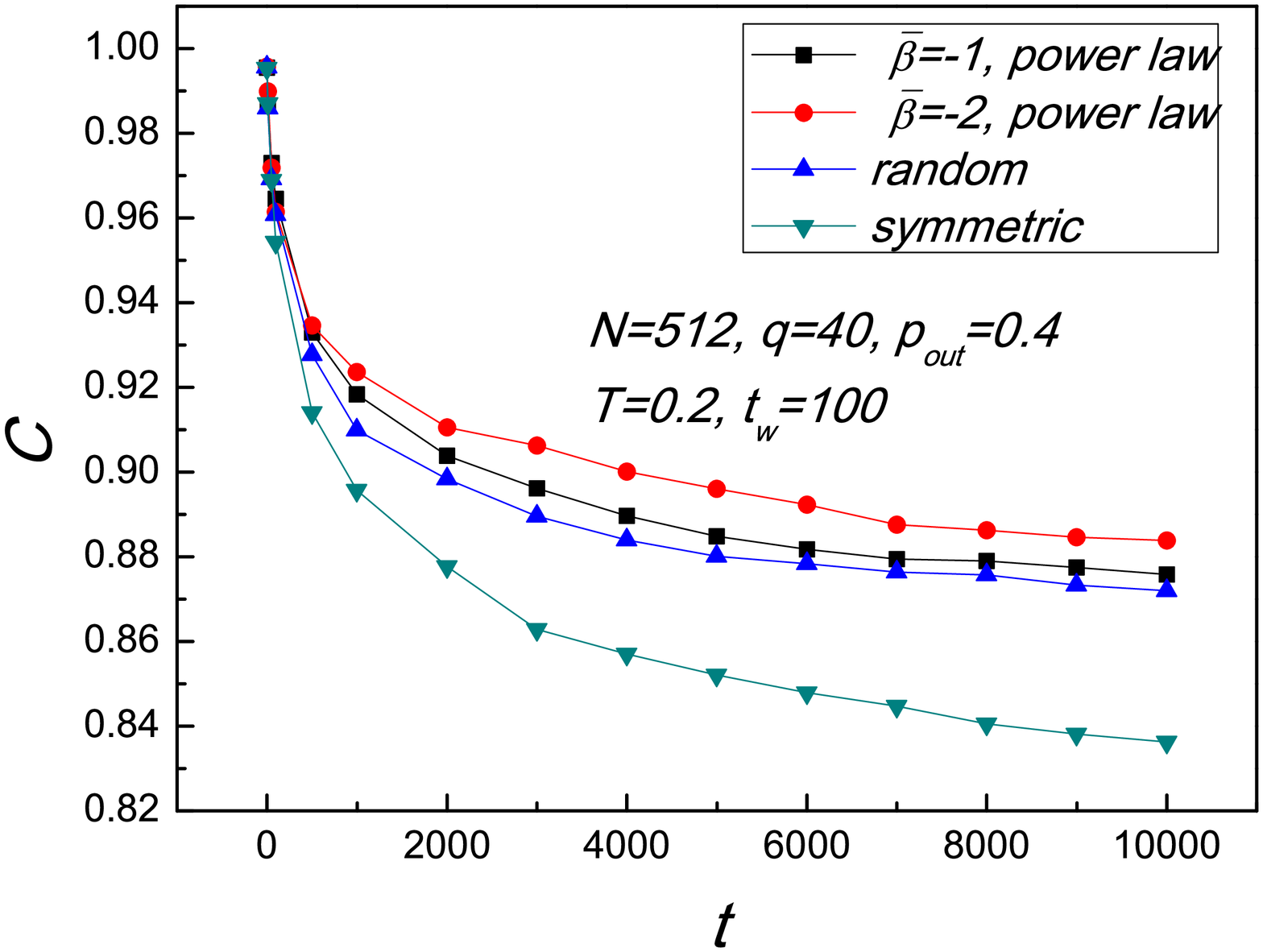}{The autocorrelation function
(\eqnref{eq:correlation}) as a function of time for system of
$N=512$ nodes consisting of $q=40$ communities with a noise level of
$p_{out}=0.4$. The waiting time $t_w=100$ and the temperature
$T=0.2$. The four displayed curves represent four different
initializations for the studied system. ``Symmetric'' initialization
means that each node forms its own community, so there are $N$
communities as a starting point for the algorithm. ``Random'' means
randomly filling $q_0$ communities with nodes, where $q_0$ is a
random number generated between $2$ and $\frac{N}{2}$. ``Power law
distribution'' means separating $N$ nodes into different
communities, whose size satisfy power law distribution with a
negative exponent, set to be ${\overline{\beta}}=-1$, $-2$. also,
the maximal community size is set to be $50$, the minimal community
size is $8$ in the above simulation results. At low temperature
($T=0.2$), all of the curves with different initialization separate
from each other even up to times of $t=10000$ steps. As this figure
makes clear, different sorts of randomness lead to different
behaviors.}{fig:CI1}{0.8\linewidth}{t}

We now examine, in further detail, several aspects of these transitions at $T=0$.
The (zero-temperature) normalized mutual information is displayed in \figref{fig:NMI2048}.
As evident from the figure, $I_N$ starts to drop below its maximal
value of $I_N=1$ (which indicates perfect agreement with the optimal
solution) when $p_{out}=p_{1}$ (i.e., at the very same value of the
noise $p_{out}=p_1$ where the relaxation time is maximal and the
complexity increases) and $I_N$ levels off at a higher value of the noise
$p_{out}=p_{2}$ (coincident with the transition value as ascertained from the
energy, entropy and complexity in \figref{fig:3phases}). Amongst other
collapses that we observed, systems with differing number of communities q all collapse onto
at $p_{out}=p_2$.


Before we turn to a more detailed analysis of spin-glass character of the transitions,
we make one remark. A possible concern is that we did not examine transitions the
optimal value of $\gamma$. Indeed, the central thesis of
\cite{peter2} was that there are optimal values of $\gamma$ that
signify the natural scales in the system. In general, transitions as
a function of $\gamma$ correspond to transitions in structure that
appear as the system is examined on larger and larger scales
as we have examined in detail in earlier works
\cite{peter2,peter3,peter4,image}. To ascertain the changes that
occur in the random systems that we investigated in this article
for a broad spectrum of different values of $\gamma$ (i.e., containing general
$\gamma \neq 1$), we re-investigated these systems
with $\gamma$ values within the range $10^{-2}\leq\gamma\leq 100$. The
``best '' values of $\gamma$ are ascertained by maxima of the
normalized mutual information $I_N$ \cite{peter1}. In
\figref{fig:optimal}, we display $I_N$ as the function of the noise
$p_{out}$ for both the fixed $\gamma=1$ and the optimal $\gamma$
determined by the multiresolution algorithm. The first transition
point $p_1$ is the same in both cases, and the two curves start to
separate around the second transition point $p_2$. This indicates
that, as it so happens to be in this case, $\gamma=1$ is the best
value of resolution parameter for noise levels below $p_2$ in this
example system ($N=2048, q=140$) at zero temperature.

\begin{figure}[]
\begin{center}
\subfigure[$p_{out}=0.22$ is within the low temperature ``hard''
region, where the collapse is perfect. ]{\includegraphics[width=
2.5in]{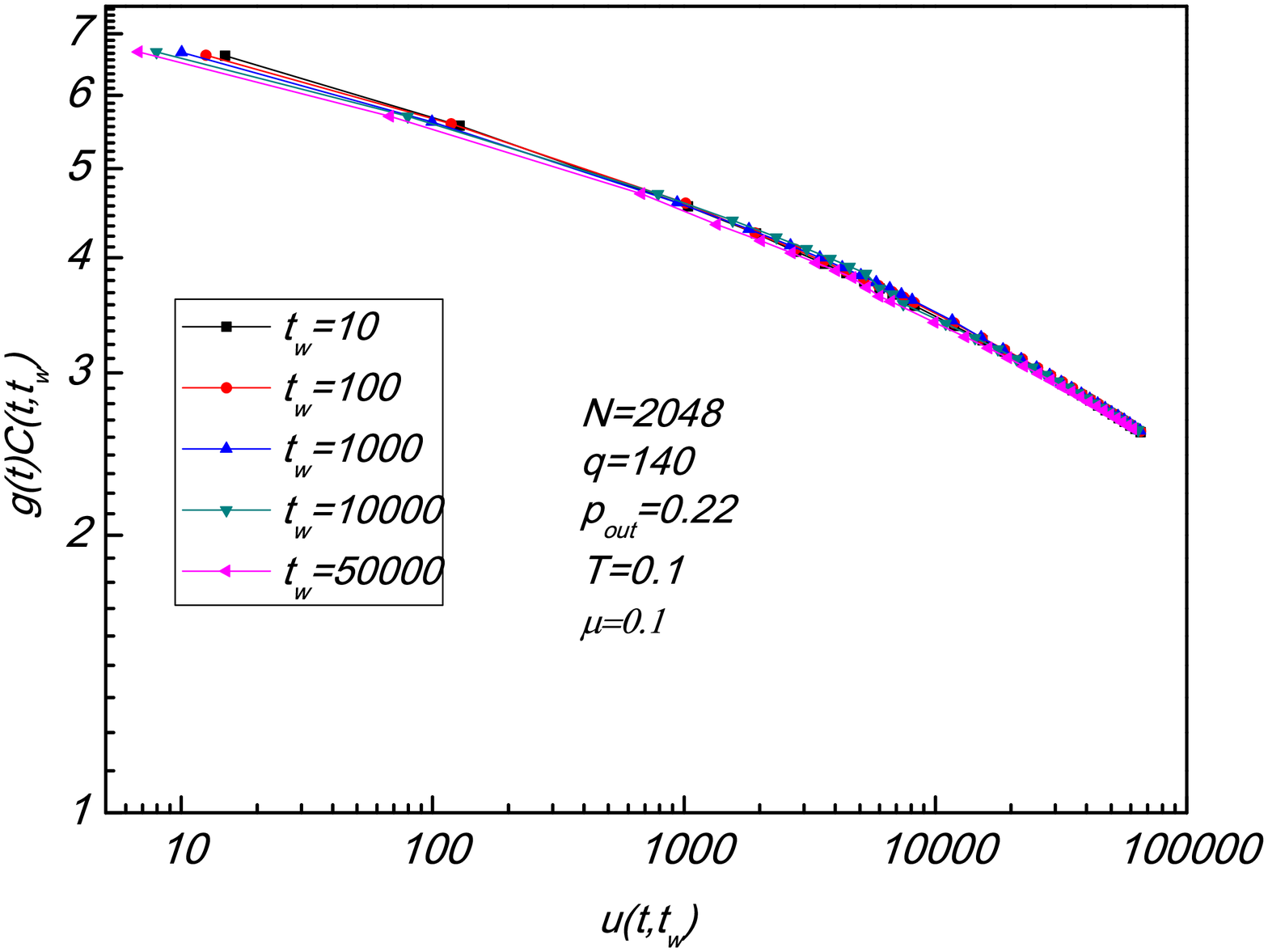}} \subfigure[$p_{out}=0.24$ is around the
second transition point, where the collapse starts to wane.
]{\includegraphics[width= 2.5in]{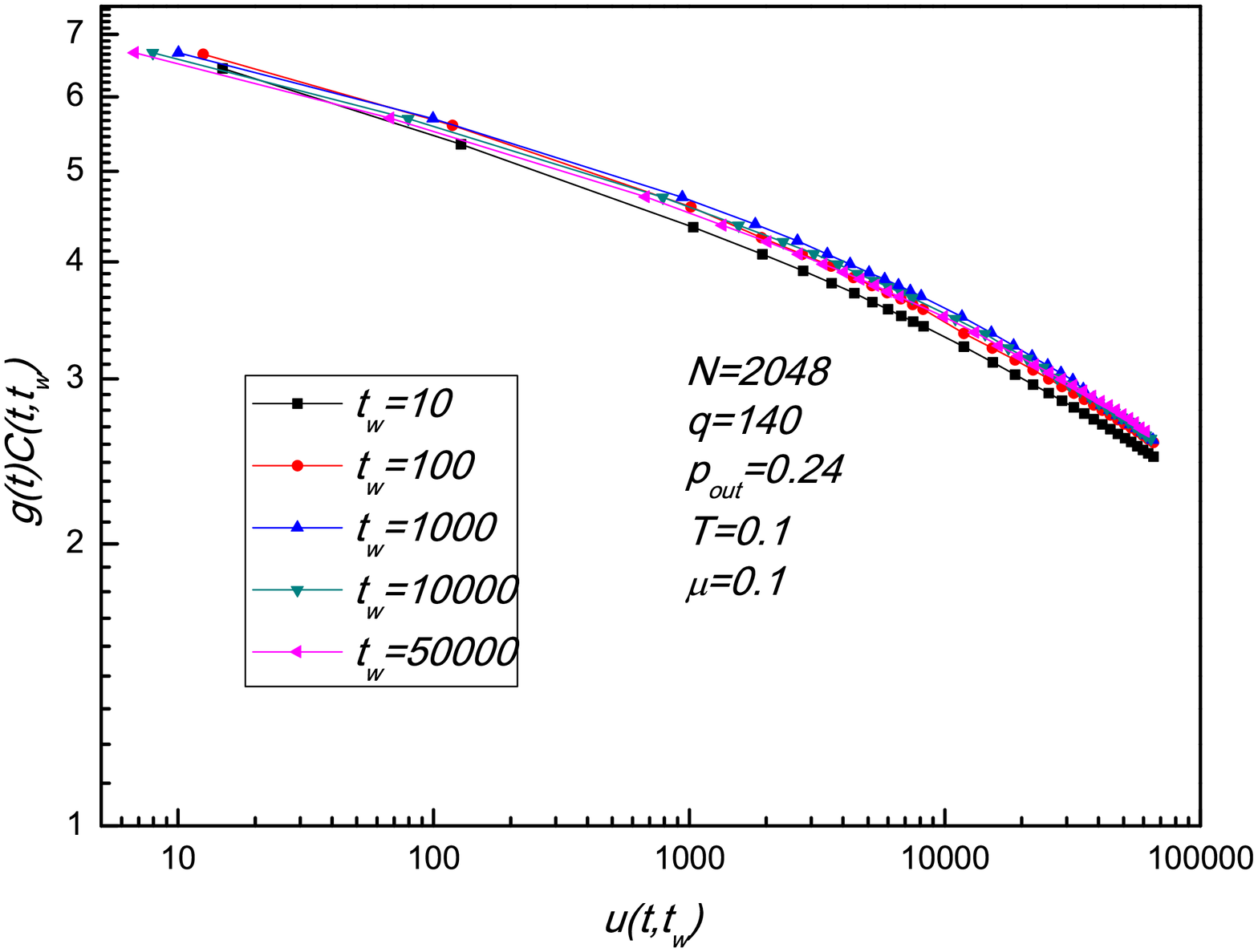}}
\end{center}
\end{figure}

\begin{figure}
\begin{center}
\subfigure[$p_{out}=0.25$ is around the second transition point,
where the collapse becomes fainter.]{\includegraphics[width=
2.5in]{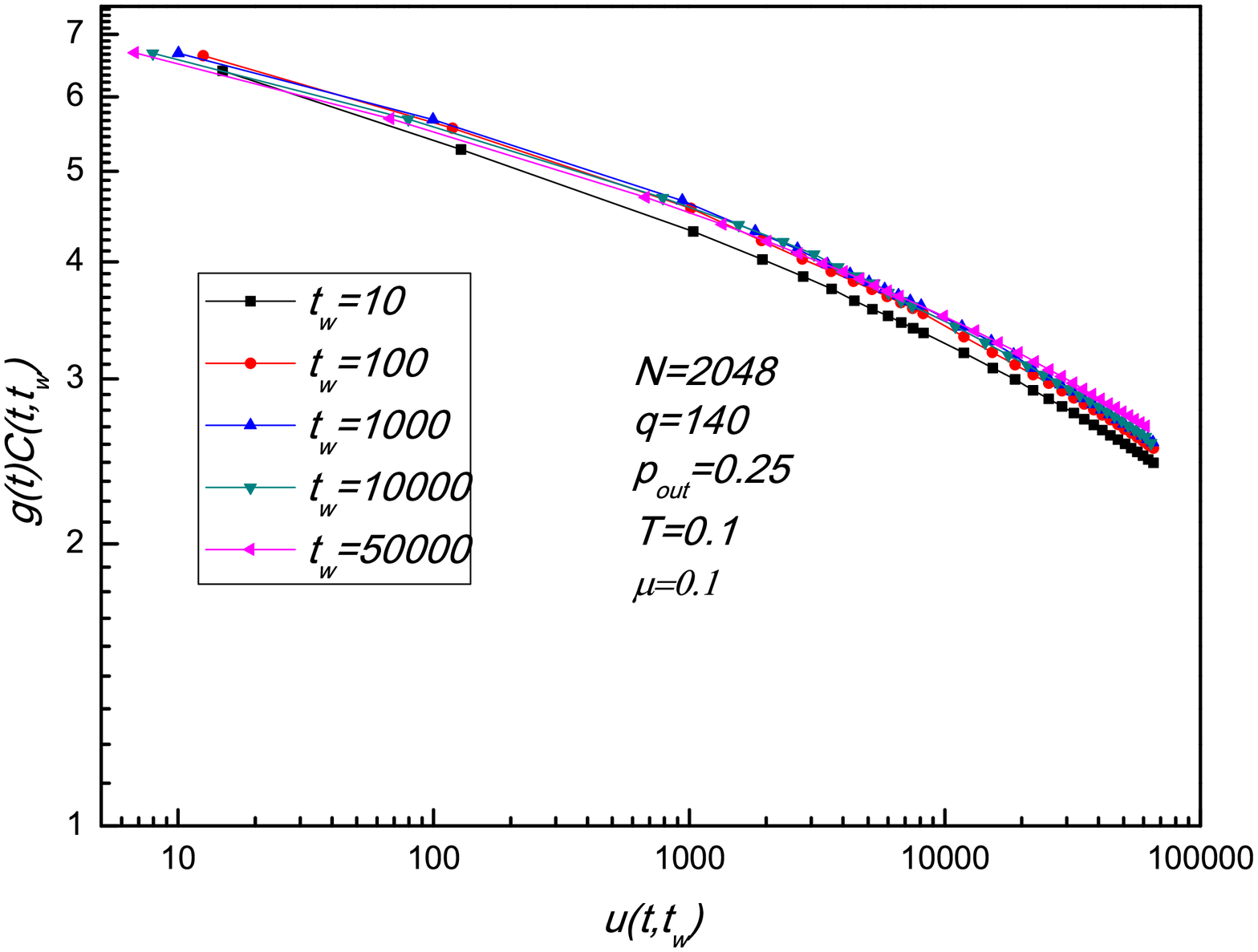}} \subfigure[$p_{out}=0.28$ is within the
``unsolvable'' region, where the collapse is poor.
]{\includegraphics[width= 2.5in]{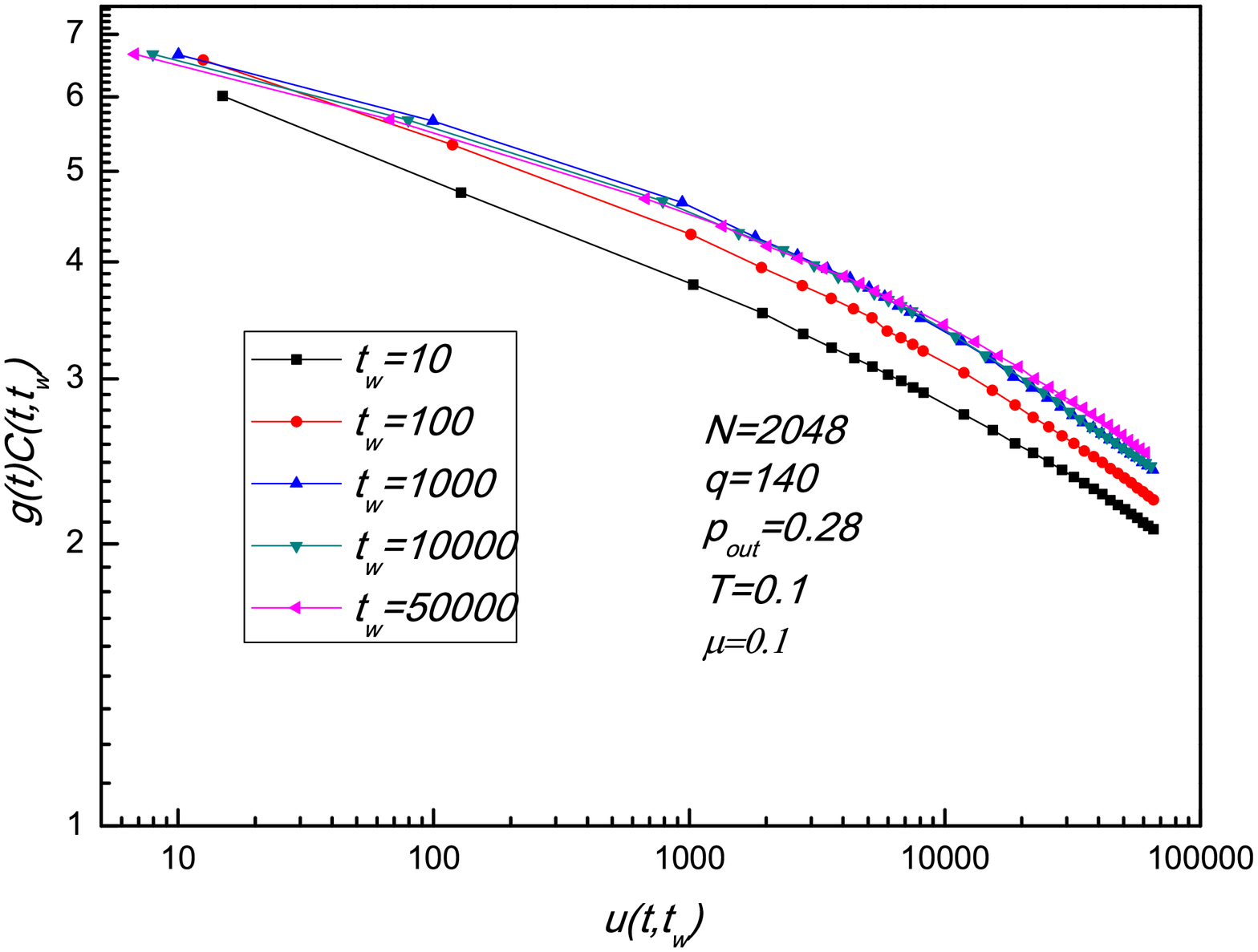}} \caption{A
validation of the spin glass character of the low temperature hard
phase. We show a collapse of the autocorrelation curves for the for
different waiting times $t_w$ for a system with $N=2048$ nodes,
$q=140$ communities, and $p_{out}$ varies from $0.22$ to $0.28$. The
first and second transition points for this system are $p_1=0.2$ and
$p_2=0.24$. The heat bath temperature is $T=0.1$ in all these
panels. The vertical axis is $g(t)C(t_w,t)$ where
$g(t)=8-\log_{10}(t)$. The horizontal axis is
$u(t_w,t)=\frac{1}{1-\mu}[(t+t_w)^{1-\mu}-t_w^{1-\mu}]$ where
$\mu=0.1$. (See text.) The noise $p_{out}=0.22$ in panel (a) lies
within the ``hard'' region where the collapse of correlation
function is perfect. The noise values of  $p_{out}=0.24$ and
$p_{out}=0.25$ in panels (b) and (c) respectively are around the
second transition point, where the collapse becomes fainter. The
noise of $p_{out}=0.28$ in panel(d) is above the second transition
point $p_2$-i.e.-in the ``unsolvable'' region, where the collapse
becomes very poor. That the collapse of the correlation function
starts to degrade right after the second transition point $p_2$ at
\emph{low temperature} indicates that this transition is of the
spin-glass type. }\label{fig:fit}
\end{center}
\end{figure}

\subsection{Numerical validation of the spin glass character of the two transitions}

The proliferation of metastable states thwarts equilibration.
A specific facet of this is detailed in Appendix F wherein,
by energy measurements, the lack of equilibration at short
times is evident. As is well
appreciated, this absence of equilibration due to
multiple metastable states may lead to spin-glass-like
(as well as structural glass like) properties. Amongst
other traits, these include memory effects previously studied for other
systems \cite{jonason,jonason2}. When a spin glass is cooled down, a
memory of the cooling process is imprinted in the spin structure,
and this process will be reproduced if one heats the system up.

We conduct a similar computational ``experiment''.  We immerse our
system (with a fixed value of the noise $p_{out}$)
in a heat bath. We then lower the heat bath temperature $T$ by
small increments at consecutive time steps $k$.
(Each time step corresponds to a single iteration through all nodes
according to the minimization algorithm of \cite{peter2,peter1}.)
In this case, we set $T_{k+1}=0.95T_k$. After attaining a steady-state solution, we
then reverse the process and increase $T$ after each step via
$T_{k+1}=1.05T_k$. In \figref{fig:EvsTemp}, we plot the long time system
energy $E$ as a function of $T$ during this process. The energy
curve as $T$ decreases follows a different path than when $T$
increases which strongly implies a hysteresis-like effect. This
memory effect as the temperature is cycled between high and
low $T$ reinforces the similarity between the community detection and
a spin glass system.

The behavior of the energy displayed in \figref{fig:EvsTemp} suggests
the same three regions that we ascertained earlier: (i) When the two
curves overlap at low temperatures (i.e., $T<0.1$), the system is in
its  ``frozen phase''. (ii) When the two curves separate in a medium
temperature range (i.e., $0.1<T<2.5$), the system is in
a``spin-glass'' phase. (iii) At yet higher temperature ($T>2.5$),
the two curves overlap once again. This marks the onset of the
``disordered'' high temperature regime.

\begin{figure}[]
\begin{center}
\subfigure[The systems has a noise value of $p_{out}=0.21$ and is at a temperature $T=1.3$.
 With these parameters, the system is in the hard phase
 (or the region of soaring computational susceptibility in
\figref{fig:3dplot}). Here, the collapse is nearly perfect.
]{\includegraphics[width= 2.5in]{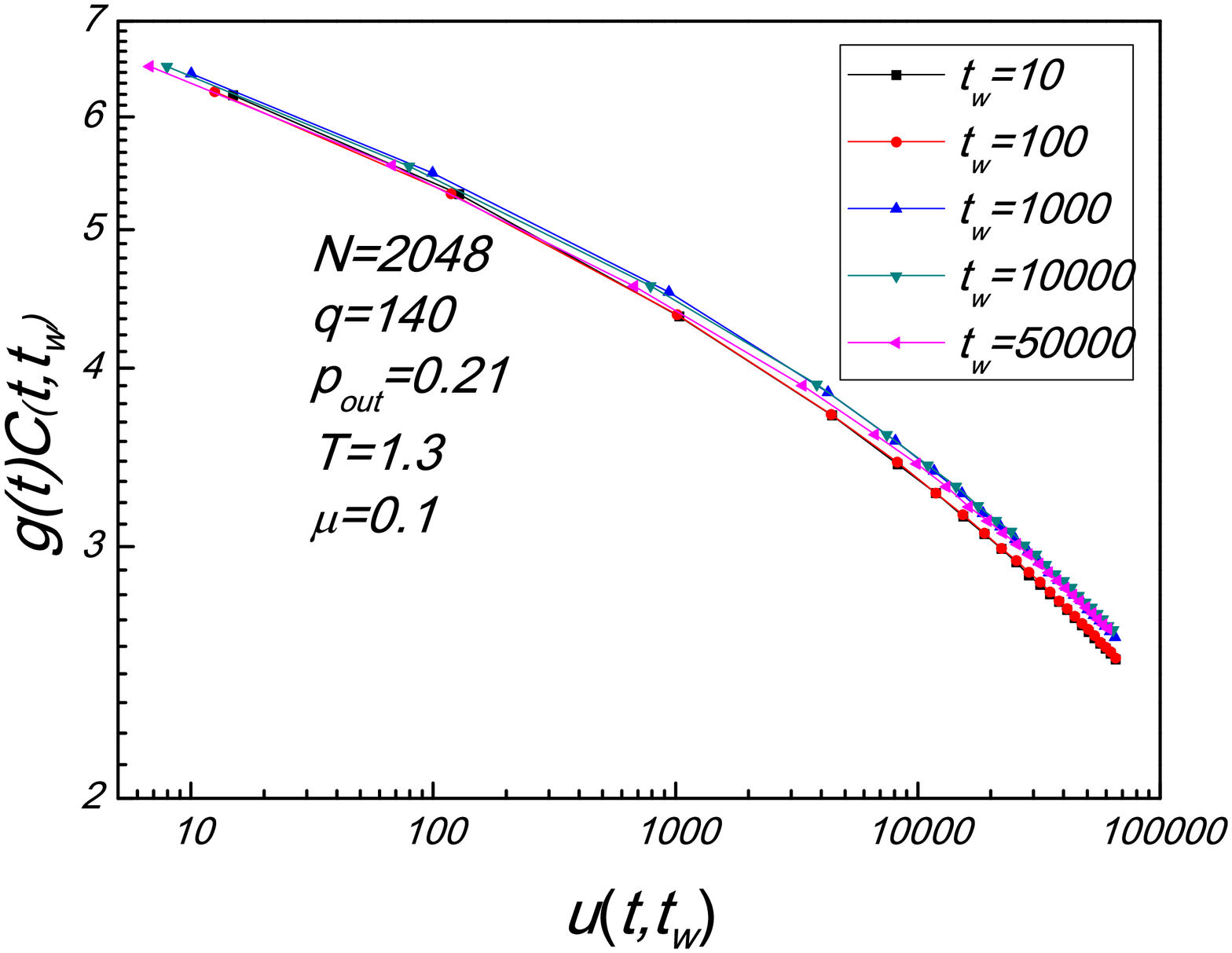}}
\subfigure[$p_{out}=0.3$ at temperature $T=1.3$ is around the
boundary of the hard phase. The collapse starts to lose its
perfection. ]{\includegraphics[width= 2.5in]{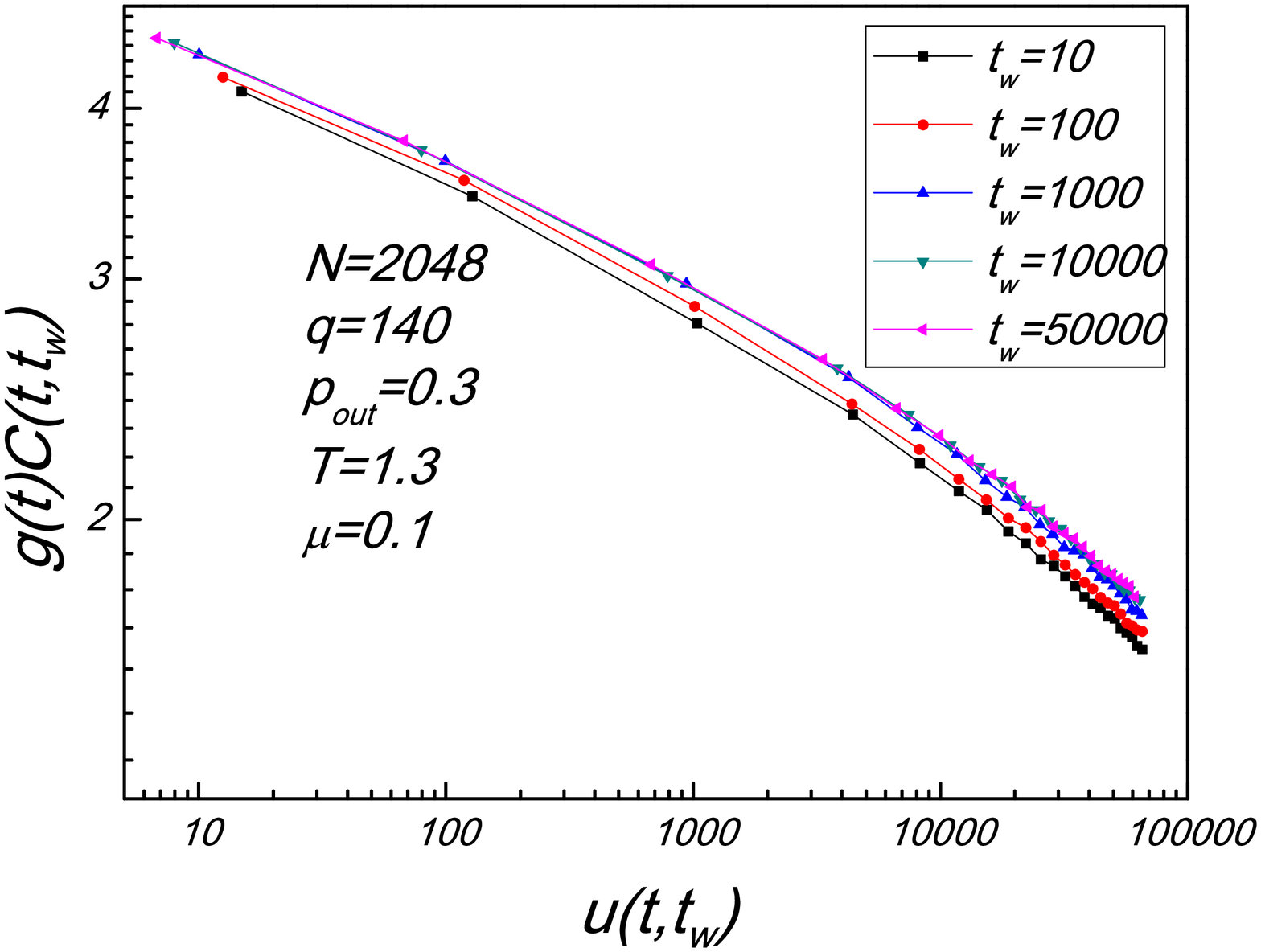}}
\end{center}
\end{figure}

\begin{figure}
\begin{center}
\subfigure[$p_{out}=0.35$ at temperature $T=1.3$. Here, the system is outside the hard
phase. In this case, the
collapse is poor.]{\includegraphics[width=
2.5in]{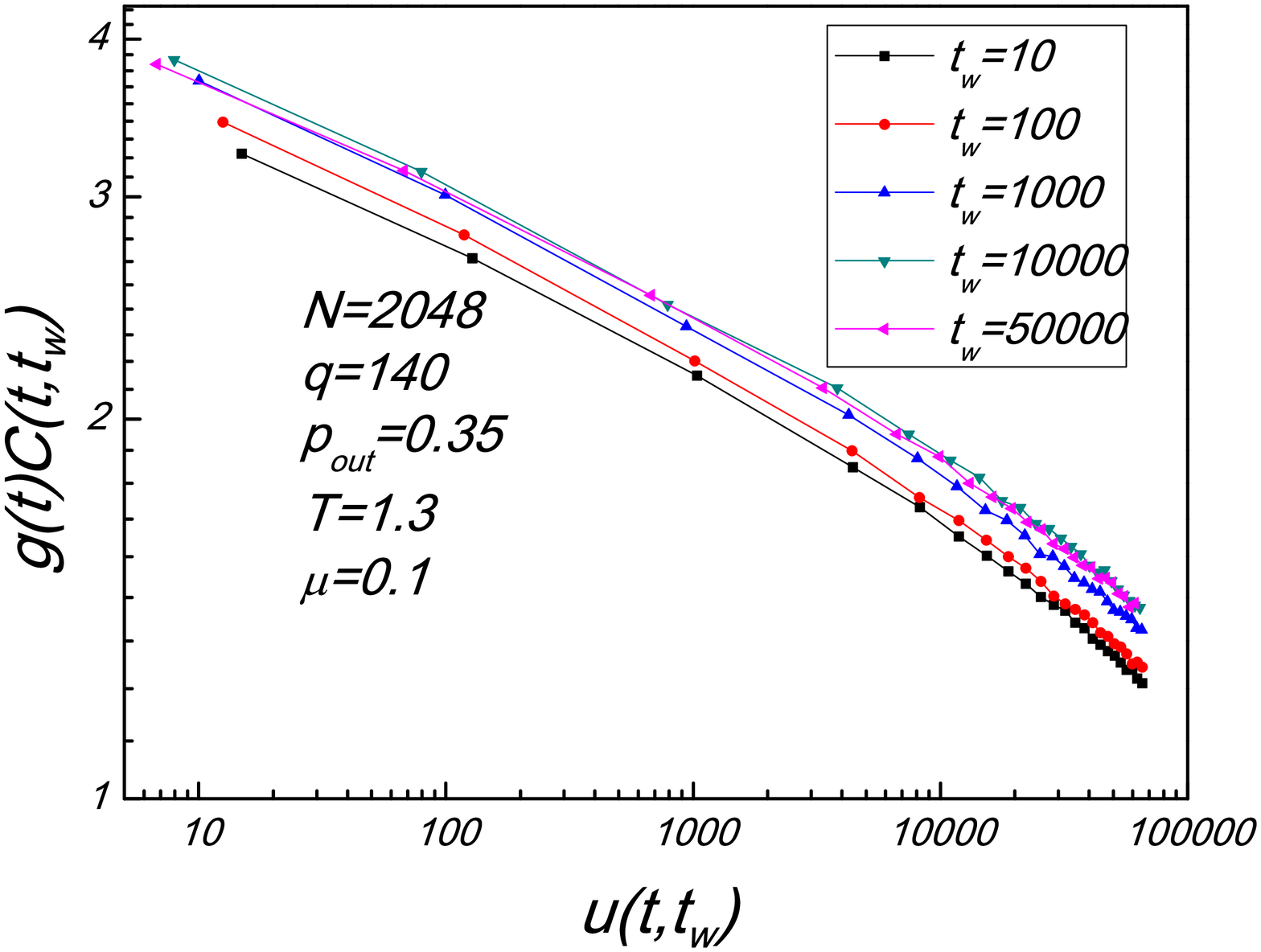}} \subfigure[$p_{out}=0.4$ at temperature
$T=1.3$-- far away from the hard phase. The collapse is non-existent
]{\includegraphics[width= 2.5in]{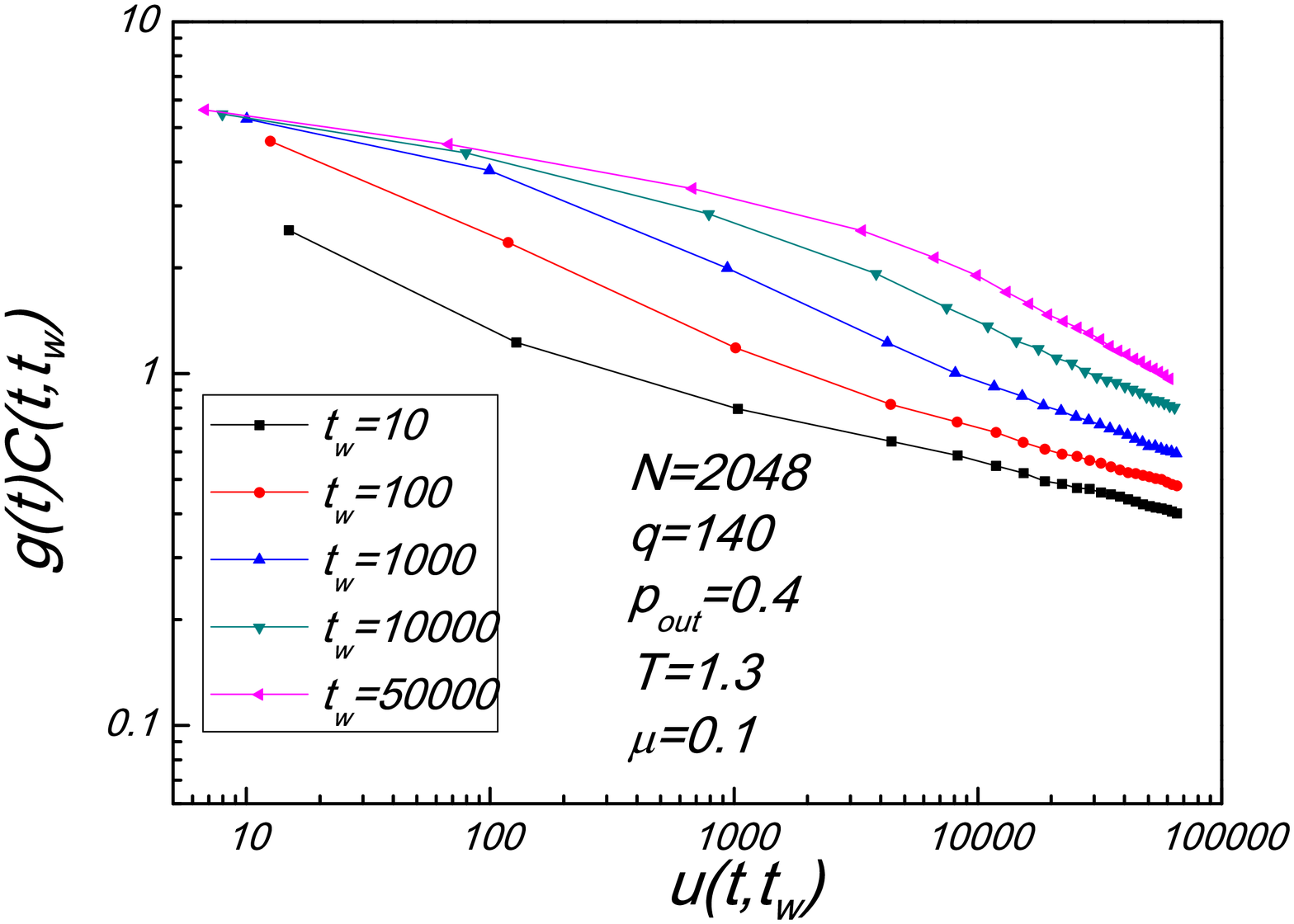}} \caption{An
illustration of the spin glass character of the high temperature
hard phase. Shown is a collapse of the autocorrelation curves for
different waiting times $t_w$ for the system of $N=2048$ nodes and
$q=140$ communities. The heat bath temperature is $T=1.3$ in all
panels. In this collapse (see text), the vertical-axis is
$g(t)C(t_w,t)$ where $g(t)=8-\log_{10}(t)$ and the horizontal-axis
is $u(t_w,t)=\frac{1}{1-\mu}[(t+t_w)^{1-\mu}-t_w^{1-\mu}]$ where
$\mu=0.1$. Panel(a) of $p_{out}=0.21$ is within the high temperature
hard phase (evident as the higher temperature ``bump'' in the 3D
plot of computational susceptibility $\chi(p_{out},T)$
(\figref{fig:3dplot})). Within the hard phase, the collapse is
perfect. Panel(b) of $p_{out}=0.3$ is around the boundary of the
hard phase. Correspondingly, the collapse starts to lose its
precision. Panel(c) corresponds to $p_{out}=0.35$-- outside the
hard phase. A poor collapse is seen. Panel(d) corresponds to
$p_{out}=0.4$
is far from the hard phase. No collapse is seen. 
The collapse of the auto-correlation function loses its perfection right
after the second transition point $p_4$ at \emph{high temperature}
indicates that this transition is also of the spin-glass
type.}\label{fig:fitT}
\end{center}
\end{figure}

As illustrated in
Figs.(\ref{fig:3phases},\ref{fig:3dplot}) (and as will be further
discussed in Figs. (\ref{fig:fit},\ref{fig:fitT})), the hard phases at both
low and high temperatures
do not extend over all temperatures. Rather, as we have emphasized
above, the hard phases only appear in the
``complexity'' ridges as shown in panel (a) of \figref{fig:3dplot}.
However, in \figref{fig:EvsTemp}, the hysteresis occurs in the temperature range
$0.1<T<2.5$. This range is considerably larger than that
of the hard phases. To
understand this, we remark on the ``experimental'' differences
between the results displayed in \figref{fig:3dplot} and those in \figref{fig:EvsTemp}. In
constructing the 3D plot of the ``complexity'' (panel (a)) in
\figref{fig:3dplot}, we apply the ``HBA'' at each temperature.
The systems at different temperatures are independent of one another.
That is, each
system is solved afresh from the symmetric initial state. In the hysteresis
loop in \figref{fig:EvsTemp} on, e.g., the decreasing temperature curve,
a system at higher temperature provides the initial state
for a lower temperature system. Thus, in this case, the systems at different
temperatures are not independent but rather serve as ``seed'' states for one another.

Aspects of the memory effect are evidently not limited to those of, e.g.,
\figref{fig:EvsTemp}. For instance, if we incorporate the effects of increasing and
decreasing noise to the same
system  \cite{david} instead of temperature, the accuracy
of the solution also forms a hysteresis loop at low temperature (see
Appendix C). Similar to a real spin glass system, the magnitude of
this effect also decreases as the temperature increases and
finally disappears beyond a threshold temperature.

A general quantitative measure of the memory,
the two-time autocorrelation function
between the system at times $t_w$ and time $t+t_w$,
\begin{equation}
C(t_w,t) = \frac{1}{N}\sum_{i=1}^N\delta_{\sigma_i(t_{w}),
           \sigma_i(t_{w}+t)}, \label{eq:correlation}
\end{equation}
can be used to explore the spin-glass-like behavior.
The upshot of the below discussion is that the autocorrelation
function data only within the hard phases
(both at low and at high temperatures- coincident,
as emphasized earlier, with the ``ridges'' in Fig. (\ref{fig:3dplot}))
adheres to a spin-glass type collapse.
This affirms, once again, the spin glass character
of the transitions.

If we apply the HBA starting from {\em different initial
configurations} at low temperature (as elaborated on in Appendix D), all of
the auto-correlation
curves with different initializations separate from each other even
up to times $t$
[in units of the iteration through all nodes according to the algorithm of \cite{peter2,peter1}]
 as large as $t=10000$ (\figref{fig:CI1}). This indicates
that disparate sorts of randomness can, generally, lead to different
results. As the temperature $T$ is increased, all of the curves
ultimately {\em collapse} onto one another. The temperature at which
the different initial configurations overlap indicates when the
respective systems start losing memory of their initial
configurations directly relates to the transition temperature in the
hysteresis loop for the same system. This further establishes the
existence of spin glass transition in the community detection
problem.

We use the HBA starting from a symmetric initial state and calculate
the autocorrelation in \eqnref{eq:correlation} for different waiting
times $t_w$ and temperatures $T$. We further found that
each auto-correlation curve $C(t,t_{w})$ corresponding to longer waiting time $t_{w}$ lies above
those with shorter waiting times, and all the curves (with different
waiting times) are non-zero for a long period of simulation time
indicating a memory effect. Moreover, we can predict the long time
behavior of $C(t_w,t)$ by fitting the curves using a commonly-used
equation in Fig. (\ref{fig:fit}, \ref{fig:fitT}), for more details,
see \cite{stariolo,young}.

Towards this end, we set
\begin{eqnarray}
 g(t)=a-b\log_{10}(t) \label{eq:g(t)},
 \label{eq:g(t)}
\end{eqnarray}
and
\begin{eqnarray}
u(t_w,t)=\frac{1}{1-\mu}[(t+t_w)^{1-\mu}-t_w^{1-\mu}].
\label{eq:u(t)}
\end{eqnarray}
In the above equations, $a$,$b$ and $\mu$ are parameters that need
to be optimized in order to ascertain whether a generic
spin glass type collapse occurs \cite{stariolo,young}.
In searching for a collapse of the data points at different waiting times $t_{w}$, we use $g(t)C(t_w,t)$ as
a vertical-axis and $u(t_w,t)$ as a horizontal-axis. As seen in
Figs. (\ref{fig:fit}, \ref{fig:fitT}), a collapse indeed occurs over
4 decades in values of $u(t_w,t)$ in
{\em both the high and low temperature hard phases}.

We discuss several features of this collapse and its coincidence
with the hard phase below. \figref{fig:fit} corresponds to the low
temperature hard phase and \figref{fig:fitT} corresponds to the high
temperature hard phase [see, e.g., the 3D computational
susceptibility plot $\chi(p_{out},T)$ in \figref{fig:3dplot}]. As
seen in Figs.(\ref{fig:fit}, \ref{fig:fitT}),  {\it both the high
and low temperature cases}, the autocorrelation functions with
different waiting times $t_{w}$ exhibit spin-glass collapse when the
value of $p_{out}$ lies within the ``ridge'' area of the hard
phases. This collapse wanes when $p_{out}$ veers towards the
``foot'' of the complexity ridge just at the onset of the hard
phase. The collapse ultimately becomes non-existent when $p_{out}$
is further away from the ``ridge'' area. The regime where the
correlation functions satisfy the spin-glass collapse is consistent
with the parameters corresponding to the hard phase (or ``ridge'' in
the 3D computational susceptibility plot of Fig.
(\ref{fig:3dplot})). Putting all of the pieces together, we see from
our scaling and collapse in Figs. (\ref{fig:fit}, \ref{fig:fitT}),
that both high and low temperature transitions of the
\emph{spin-glass type}.

In the random graphs, we reported on \textit{spin-glass} type
transitions. Although trivial, for completeness, we should however
note that a graph can, obviously, also be very regular. A
prototypical example is that of the two-dimensional square lattice
\cite{square_lattice}. For such regular unfrustrated lattice
systems,  the Potts model of Eq. (\ref{eq:ourpotts}) becomes the
``standard'' Potts model of lattice systems. In these instances, we
generally have single \textit{first or second order} transitions
instead of spin-glass type transitions. We briefly elaborate on this
point. Simple regular lattices are a particular realization of a
graph (one with the fixed coordination and translational symmetry).
As is well known, on,
e.g., the square lattice, 
the Potts model, which we use throughout, exhibits as a function of
the temperature $T$, two phases with an intervening {\em critical
point} for small $q$ ($q \le 4$); for larger $q$ ($q>4$), {\em a
first order transition} appears. Thus, particular realizations of
our hamiltonian for these graphs display (usual) critical points and
first order phase transitions. For more generic random graphs with
high coordination, the system displays (as we showed above
and will further elaborate on), spin-glass type transitions
appear along with intervening hard phases.

We further reiterate an earlier remark and note
that in systems with well defined structures on multiple scales,
{\em additional transitions} may appear as the resolution parameter $\gamma$
of Eq. (\ref{eq:ourpotts}) is varied. In earlier works, we reported on these
transitions and further employed these in the analysis of disparate systems
\cite{peter2,peter3,peter4,image}.

\subsection{General discussion}

In this subsection, we detail general considerations directly related to the spin-glass Potts analysis
thus far. In the next section, we will further discuss dynamics which
further relates to aspects that we detail herein. This subsection
is different from others in that here (and only in this subsection),
we present a general discussion and some speculations
and not present data.

\subsubsection{Theoretical Expectations from NP-completeness}

In \cite{brandes}, it was shown that maximizing modularity
(an earlier alluded to prominent approach for the community detection problem
\cite{fortunato,mod,goodMC})
is NP-complete. Thus, as all NP complete problems may (by their very definition) be
mapped onto one another, maximizing
modularity on the most general graphs must span the three phases
(solvable and unsolvable with the further division of the solvable
problems into the ``easily solvable phase'' and the ``hard phase'')
that appear, e.g., in k-SAT problem \cite{mezard} which is known to
be NP-complete \cite{cook}.
Similarly, if other approaches to community detection are,
ultimately, equally hard as maximizing modularity, then all of these
approaches may in general display three phases. It may be, as in the
k-SAT problem, that for simple problems, we have only an ``easy
phase'' and an ``unsolvable'' (or ``unSAT'') phase. This does indeed
occur for some graphs. In general, though, we find the three different phases
(as expected) that we reported on in this work.

\subsubsection{Physical content of the transition in many body systems}

$\bullet$ {\em Approximate decoupling} \newline

We briefly speculate, in this subsection alone, on potential physical consequences of the phase
transition that we find in the community detection problem. As elaborated on in
\cite{peter3,peter4} a general
many body system with two particle interactions may be regarded
as a network with edge weights determined by the interactions.
In the easy phase in the extreme limit of $p_{out}=0$,
the system is essentially that of disjoint non-interacting
clusters. This point is analytically connected to any other point
in the easy phase. More generally,
The Potts model Hamiltonian \eqnref{eq:ourpotts} can be written as:
$H(\{\sigma\}) =\sum_{k=1}^{q}H_k$. Thus, the partition function
becomes:

\begin{eqnarray}
Z(\beta)=\sum_{\{\sigma\}}e^{-\beta
H}=\sum_{\{\sigma\}}e^{-\beta\sum_{k=1}^{q}
H_k(\{\sigma_i\}\in k)}
\nonumber\\ =\sum_{\{\Lambda\}}\prod_{k=1}^{q} (\sum_{\{\sigma_i\}\in
k}e^{-\beta H_k})=\sum_{\{\Lambda\}}(\prod_{k=1}^{q}
Z_k).\label{eq:pf}
\end{eqnarray}
In Eq. (\ref{eq:pf}), $Z_k$ is the partition function as computed with the Hamiltonian of
the entire system for the particles in community k, and
$\{\Lambda\}$ denote partitions of the system.

A similar form
was proposed for many body systems in \cite{peter3}
when partitioning a general interacting system into
decoupled clusters.
Even though, we sum over all partitions, we may have an
important subset of partitions, denoted as $\{\Lambda'\}$
(each with a corresponding number of clusters equal to
$q_{\Lambda'}$), which will have
in general instances, high Boltzmann
weights and/or frequencies and will dominate
the sum. These partitions will, correspondingly,
have a significant lower free energy relative to other partitions.
In such cases, the partition function can be
approximated as
\begin{eqnarray}
Z\simeq \sum_{\{\Lambda'\}}\prod_{c=1}^{q_{\Lambda'}}
Z_c\label{eq:gpfi}
\end{eqnarray}
\eqnref{eq:gpfi} is exact in the limit of $T=0$ where $\Lambda$
denotes the ground-state(s) of our Potts type Hamiltonian. If
$p_{out}$ is small (in particular if $p_{out}=0$), then there will generally exist a
small number of sharply defined ground-states $\{\Lambda'\}$
pertaining to partitions into completely
disconnected communities. This general trend of dominant subsets
may persist within the easily solvable phase.

The possible upshot of this discussion is that we might, in easy phases,
approximate many body interacting systems (such as supercooled liquids
that we will briefly discuss next) as effectively composed of disjoint
non-interacting clusters. This picture may badly break down
once transition lines between the easy phase and the
hard or unsolvable phases are traversed. \newline

$\bullet$ {\em Possible relation to structural glasses and other complex physical systems}  \newline

Glasses
(according to the theories such as the random first order transition
theory of glass (RFOT) in \cite{wolynes}) may have three phases as
a function of temperature. In the intermediate phase, the system
displays a large complexity (as manifest in the configurational
entropy being extensive). If we replace the interacting particles in
a supercooled liquid (that form a glass at low temperatures) by
decoupled communities \cite{peter3,peter4}, then the three phases found in the
computational community detection problem may be manifest as three
disparate phases of supercooled liquids as a function of
temperature. Within RFOT, at temperatures in an
intermediate region $(T_{0}<T<T_{A}$), the system physically
displays an extensive configurational entropy (which is tantamount
to an extremely large complexity in the current context). This configurational entropy
precipitously onsets at $T=T_{A}$ and gradually diminishes until it
no longer becomes extensive a lower temperature ($T=T_{0}$) whence
the system freezes into an ``ideal glass'' that is permanently stuck
in a metastable state.

We will discuss, in Section \ref{dee} dynamical aspects that directly relate
to the Potts model Hamiltonian. Insofar
as additional general related aspects of the results of our community detection
analysis the implication of phase boundaries, we make a brief comment.
When, as discussed in \cite{peter3,peter4}, a weighted version of  Eq.(\ref{eq:ourpotts})
is used with edge weights that are set by forces then in overdamped viscuous
systems (where the total force on a particle is proportional to its
velocity, $\vec{f}_{i} = c \vec{v}_{i}$), particles that experience a similar total force,
will tend to move in unison. Thus, in the easy phase motion of decoupled
cohesively moving particles will occur. In the unsolvable phase,
the particle motion will be more complicated. In earlier work, forces were used
to study community detection with, overall, similar results
to the one afforded by our spin glass approach in this work \cite{gudkov}.
When other weights are used (such as potentials, two-body correlations or other metrics), similar
decoupling within the solvable phase signifies a tendency of the clusters not to be related
insofar as the metric being used.

\subsubsection{Image segmentation}

Recently, we investigated and invoked the features of the phase diagram in order to
address the computer vision problem of detecting objects in general
images \cite{image} (including notably challenging ones).  As
in in this work for random graphs,
by varying parameters
such the temperature,the (graph) resolution parameter $\gamma$ and physical
length scales, we explored the community detection phase diagrams for image
segmentation. Within the
easy phase, disparate objects were clearly seen. As the system moved
into the hard phase, the sharpness of the objects became more fragmented.
These ultimately became very noisy in the unsolvable phase.

In summary, whenever a decomposition of an interacting
many body system into nearly decoupled communities is possible
(indeed, as alluded to above, such a decomposition is exact for Potts model systems
wherein the exchange energy between spins in different domains is
zero) then the phase transitions that we report on here for the
community detection problem may carry direct physical consequences.
This may afford a direct link between the phase diagrams of hard
computational problems (as ascertained by physically inspired
approaches) and the phase diagrams of physical systems that may be
investigated via solutions to these related computational problems.
It is important to note that the direct relation between complexity
and glassiness is not simple as some problems that may be
investigated by sub-optimal algorithms (such as physical stochastic
systems) may appear to have a ``hard phase'' while if investigated
by a more efficient algorithm do not have a ``hard phase''
\cite{science}. Nevertheless, it may well be possible that the
decomposition of physical systems into simple elements will no
longer be simple at the onset into complex states such as those of
supercooled liquids. Indeed, in recent work, we applied the
community detection ideas to general many body systems (including
glasses) in order to flesh out prospective important structures on
all scales \cite{peter3,peter4,image}.

\section{Dynamical Aspects}
\label{dee}

In the following, we also study the related dynamical
transition. Dynamic approaches to community detection have been
suggested earlier \cite{gudkov,arenas2}. To describe the dynamical
process, we need to calculate the trajectory (of community
memberships) for each node as a function of time. Specifically, we
use the correspondence between the $q$-state Potts model and a
clock-type model in $(q-1)$ dimensions. We replace the Kronecker
delta $\delta(\sigma_i,\sigma_j)$ in \eqnref{eq:ourpotts} by a
product $\vec{n}_i\cdot\vec{n}_j$ where $\vec{n}_i$ and $\vec{n}_j$
are the vertices of a regular $(q-1)$-dimensional simplex. On such
simplifies (e.g., an equilateral triangle ($q=3$), tetrahedron
($q=4$), ...),
$\vec{n}_i\cdot\vec{n}_j=[1+1/(q-1)]\delta_{ij}-1/(q-1)$. Thus, as
is well known, we can cast the Hamiltonian of \eqnref{eq:ourpotts}
into the form $H=-\sum_{ij}A'_{ij}\vec{n}_i\cdot\vec{n}_j$ where
$A'_{ij}=(1+\gamma)A_{ij}-\gamma$ is the interaction weight. If we
insert an external field $\vec{h}_i$ into this simplified
Hamiltonian, then it becomes
\begin{eqnarray}
H=-\sum_{ij}A'_{ij}\vec{n}_i\cdot\vec{n}_j-\sum_i\vec{h}_i\cdot\vec{n}_i.
\label{HAij}
\end{eqnarray}

In what follows, we will first outline a very simple new general method for
relating a general statistical mechanics system (such as
the particular Potts model under consideration) and
a dynamical system from classical mechanics. Although,
this method will be specifically invoked to the Potts
model, all of its steps can be replicated
for other systems as well. We will then proceed to show the
results of our numerical analysis.
The final result of our analysis is that
{\em the spin glass transitions relate to
transitions to chaos in the
dynamics of the continuous
mechanical system}.

\subsection{Relating discrete Hamiltonians
to continuous dynamics}

We will in this subsection illustrate how it is possible to
relate the discrete Potts model Hamiltonian of Eq. (\ref{eq:ourpotts})
[and its clock model variant of Eq. (\ref{HAij})] to mechanical system with continuous dynamics. Many possible
similar variants of the method outlined below are possible.
Although our present aim is to investigate the Potts model Hamiltonians,
as noted above, our method can be applied mutatis mutandis to general discrete Hamiltonians.
A benefit of this mapping is that it bridges {\em chaos} in the more
standard mechanical sense to that reported in spin glass systems.

Starting with Eq.(\ref{HAij}),  we perform a Hubbard-Stratonovich transformation via
non-compact auxiliary fields $\vec{\eta}$ to arrive at the effective Hamiltonian (or, more precisely,
free energy)

\begin{eqnarray}
  \beta H_{eff}&=&-\ln Z\nonumber\\
  &=&\sum_{i\neq j}\vec{\eta}_i(\beta
    A')^{-1}_{ij}\vec{\eta}_j- \ln\left( Tr_{\vec{n}_i}
    e^{\vec{n}_i(\vec{\eta}_i+\beta\vec{h}_i)}\right) \label{eq:h1}
\end{eqnarray}
where $Z$ is the partition function.

The dynamical equation for a node moving under the effective field is, for a damped system, given by
\begin{eqnarray}
\frac{d\vec{\eta}_i}{dt} &=&
         \frac{\delta H_{eff}}{\delta \vec{\eta}_i}\bigg|_{\vec{h}_i=0}\nonumber\\
&=&\beta^{-1}\sum_j (\beta A')_{ij}^{-1} \vec{\eta}_j
         -\beta^{-1}\frac{\sum_{\vec{n}_i} \vec{n}_i e^{\vec{n}_i\vec{\eta}_i}
              }{\sum_{\vec{n}_i}
             e^{\vec{n}_i\vec{\eta}_i}}. \label{eq:dynamicalequation}
\end{eqnarray}
We initialize the auxiliary field $\vec{\eta}_i$ to be some constant
vector close to $0$. We can solve this dynamical relation to obtain
the non-compact auxiliary field $\vec{\eta}$ as a function of time
\cite{damp}.

We can obtain the expression for nodes trajectories
$\langle\vec{n}_i\rangle$ in terms of time by taking the derivative
of the partition function $Z$'s [\eqnref{eq:h1}] with respect to the
source $\vec{h}_i$, i.e.,
\begin{eqnarray}
\langle\vec{n}_i\rangle|_{\vec{h}_i=0}&=&\frac{\delta\ln
Z\{\vec{h}_i\}}{\delta
\beta\vec{h}_i}\bigg|_{\vec{h}_i=0}\nonumber\\
&= &\frac{\sum_{\vec{n}_i}\vec{n}_i
e^{\vec{n}_i\vec{\eta}_i}}{\sum_{\vec{n}_i}
e^{\vec{n}_i\vec{\eta}_i}}.\label{eq:derivative}
\end{eqnarray}


Substituting $\vec{\eta}$ in \eqnref{eq:dynamicalequation} into
\eqnref{eq:derivative}, we can determine the trajectory of the nodes.

\begin{figure}[t]
\begin{center}
\subfigure[$p_{out}=0.2$]{\includegraphics[width=
1.8in]{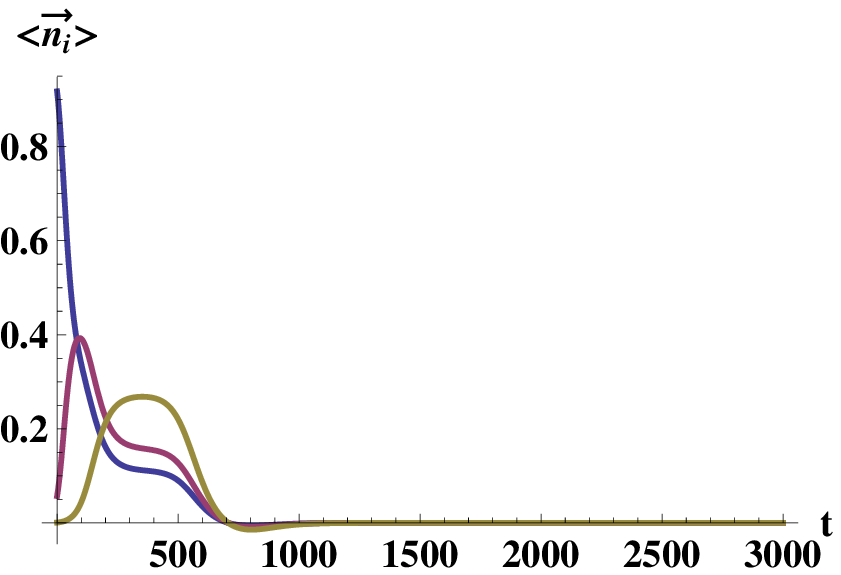}}
\subfigure[$p_{out}=0.3$]{\includegraphics[width=
1.8in]{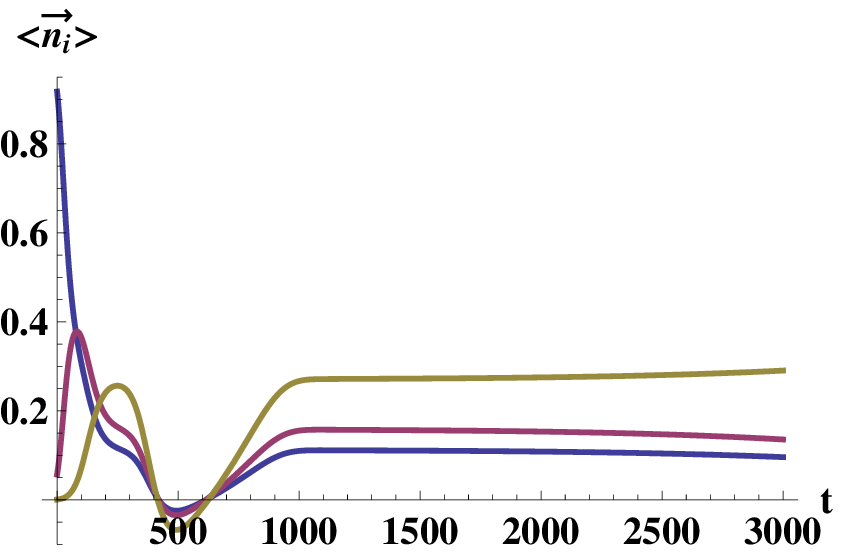}} \caption{Plots of node trajectories
$\langle\vec{n}_i\rangle$ as a function of time $t$ (number of
algorithm steps). The tested system has $N=24$ nodes, $q=4$
communities, and is solved at a temperature of $T=0.05$. According
to the description in the text, $\vec{n}_i$ is a $q-1=3$ dimensional
vector. In each plot, the three different Cartesian components of
$\langle\vec{n}_i\rangle$ marked by different colors (shades).
 Node $i$ is picked randomly
from the $24$ nodes.
In panel (a), the noise $p_{out}=0.2$ is below the transition point
$p_1=0.28$. In panel (b),
$p_{out}=0.3$ is above $p_1$.
Note that panel (a) shows a convergent solution for node $i$ where
panel (b) indicates the absence of a collapse.
}\label{fig:nt}
\end{center}
\end{figure}

\subsection{Numerical results for the continuous dynamical analog}

Eq. (\ref{eq:dynamicalequation}) describes overdamped (or
Aristotelian) dynamics.  It is, of course, possible to also define
the system in such a way that it evolves according to Newton's
equation. In overdamped systems, the energy of the system goes down
with time and thus the system veers towards a local (or global)
energy minimum. The system exhibits no dynamics once it gets stuck
in a local (global) minimum of the energy. For the shown system in
\figref{fig:nt} , in the absence of perturbing fields, at low noise
within the solvable region the system, the node coordinates
$\langle\vec{n}_i\rangle$ quickly collapse to the origin.
Conversely, at high values of the noise (i.e., large $p_{out}$) the
node coordinates do not converge (and indeed, as we elaborate on
below, the system is not solvable). As detailed in Appendix G, we
further applied weak perturbing fields $\{\vec{h}_{i}\}$ and found
that they can indeed veer the system, at low noise, towards the
correct solutions.

The system shown in \figref{fig:nt} contains only $N=24$ nodes with
$q=4$ communities at a temperature $T=0.05$. Prior to investigating
this system using the dynamical approach outlined above, we first
examined this system also using the entropy/energy/computational
susceptibility measures discussed in this article and found that, in
this system, there is no hard phase. Rather, there is a direct
transition
 (or, more precisely, crossover in this small $N$ system) from an
 easy solvable phase for $p_{out}<p_{1}=0.28$ to a disordered
unsolvable system for $p>p_{1}$. The result of our dynamic analysis
following Eq.(\ref{eq:derivative}), demonstrates the existence of a phase transition
(or crossover for this finite $N$ system) {\em at precisely the same values of $p_{out}$ found by the analysis of
the thermodynamic quantities associated with the Potts model for
this small system}. In this case, the dynamics of the nodes
illustrate that when $p_{out}$ exceeds $p_{1}$, the system exhibits
a transition from a stable system ($p<p_{1}$) to one which is
chaotic ($p>p_{1}$). Our dynamic approach to the community detection
transition may generally bridge such transitions in system dynamics
to thermodynamic phase transitions.

We illustrated via our dynamic approach,  how ergodic behavior can arise
depending on $p_{out}$ (and, similarly, also on temperature). This
relates to {\em ``chaotic''} behavior reflecting the sensitivity to
the temperature and in our case other parameters (such as $p_{out}$)
that define the computational problem in spin-glasses
\cite{fisher,bray,middleton} to real chaotic behavior of a
dynamical system. Further, in our spin-glass approach, \figref{fig:CI}
illustrates that auto-correlation functions corresponding to different initial conditions
(or randomness) remain different up to long times. This sensitive dependence
on the initial conditions is the hallmark of chaotic systems.
 Although, we have not observed such an
intermediate hard phase for the small $N$ system that we
investigated using this dynamic approach, we speculate the above
dynamic transition from more stable orbits to ``chaos'' may, for
larger systems, exhibit also indeed an intermediate region
corresponding ($p_{1}(T) \le p_{out} \le p_{2}(T)$ for low $T$
or also $p_{3}(T) \le p_{out} \le p_{4}(T)$ for higher $T$) where more and more
branching points may appear (or period doubling, etc.) as the system
transitions into chaos.  Ideas from KAM analysis may, hopefully,
be invoked in more sophisticated treatments.

\section{Conclusions}

We reported on disparate high and low temperature spin glass type
phase transitions in the community
detection problem and, by extension, rather general disordered Potts
spin systems. Our investigation involved several complementary approaches
and was not confined to systems with a small number of
Potts spin flavors or communities.
In the community detection setting, similar to other computational problems,
phase transitions occur between a solvable and unsolvable region. The solvable region
may further split into an ``easy'' and a ``hard'' region. We illustrated how thermal ``order out
of disorder'' may come into play in these systems and provided ample evidence
of the spin-glass character of the transitions that occur.  Amongst other results,
we found that different sorts of randomness can lead to different behaviors, e.g., ``chaos''.
We introduce a general correspondence between discrete spin systems and
mechanical systems with continuous dynamics. With the aid of this mapping,
we illustrated that spin glass type transitions in the disordered
system correspond to transitions to chaos in the mechanical system.
The mapping that we use to relate the
thermodynamics to the dynamics suggests how chaotic-type behavior in
thermodynamical system can indeed naturally arise in
hard-computational problem and spin-glasses. We further briefly
speculate on possible physical consequences (such as supercooled
liquids and glasses) of the transitions that we find here.
Recently, we indeed employed the transitions that we found here
in the analysis of such complex physical systems
\cite{peter3,peter4} as well as image segmentation
\cite{image}.

{\bf Acknowledgments.} This work was supported by NSF grant DMR-1106293 (ZN).
We also wish to thank S. Chakrabarty, R. Darst, P. Johnson,
B. Leonard, A. Middleton, M. E. J. Newman, D. Reichman, V. Tran,
and L. Zdeborova for discussions and ongoing work.

\textbf{Note added in Proof}: Some time after the initial appearance
of the current work \cite{us} and earlier reports of particular
aspects of a phase transition (Appendix E of \cite{peter2} and
Appendix B of \cite{peter1}), the authors of Ref. \cite{zdeborova}
investigated phase transitions in the community detection problem on
sparse graphs of small $q$ and reached similar
conclusions as we have for general graphs
with larger $q$ values.

\appendix

\section*{Appendix A: Theory analysis of the community detection problem}
\label{sec:cavity}

In this appendix, we follow the description of cavity method in \cite{jorgbook}
and merely generalize it to all graphs (general $q$ and unequal size communities).
The uninitiated reader is encouraged to peruse
\cite{jorgbook} in order to familiarize him/herself
with basic the cavity method (and the notations)
used that we expand on below. The brief introduction
below is not self-contained.

Within the cavity approach, each node passes  {\em a message} along
edges. A message from node $i$ to $j$ is a $q$-dimensional vector of
zeros and ones. Node $i$ takes the messages from all the other nodes
$k\neq j$ connected to $i$ and sums them. Then the cavity field
defined as $\mathbf{h}_{i\rightarrow j}=\sum_{k\neq
j}J_{ki}\mathbf{u}_{k\rightarrow i}$ is obtained through the above
process. Finally, node $i$ converts this cavity field into a message
to $j$ by picking and setting the maximal components in $\mathbf{h}$
to one and the rest to zero. The probability distribution of
messages being sent in the system is denoted as
$\mathcal{Q}^s(\mathbf{u})$. The superscript $s$ denotes a possible
dependence of this distribution on the index of the pre-defined
cluster to which the sending node belongs.

To be consistent with the notations in \cite{jorgbook},
in what follows in this appendix (and only in this appendix),
we will employ the same definition for $p_{in}$ and $p_{out}$ as that
of \cite{jorgbook}. For a fixed
cluster $A$, $p_{in}^A=p(A|A)$ is the conditional
probability that a link starting with a node in $A$ also ends in
$A$. Given two (different) clusters $A$ and $B$,
$p_{out}^{B|A}=p(B|A)$ denotes the conditional probability that a
link starting with a node in $A$ would end in $B$. It follows
directly from these definitions that,
\begin{eqnarray}
p_{in}^A+\sum_{B\neq A}p_{out}^{B|A}=1.
\end{eqnarray}

In particular, when $q=2$, there are only two clusters/states which $A,B$,
and we have $p_{in}^A+p_{out}^{B|A}=1.$

Following the same calculation process in \cite{jorgbook}, we also
test the phase transition of community detection in a random Bethe
lattice with exact degree $k=3$. But there is an essential
difference that our Hamiltonian {\em does not have the constraint of
equal-size clusters}, which means we do not have the symmetric
condition for the order parameter
$\mathcal{Q}^s(\mathbf{u})=\eta_{c\omega}$, where, $c=1$ denotes the
``correct'' component, and $\omega\in\{1-c,...,q-1\}$ denotes the
number label of a``wrong'' component. In our case, $\omega$ now can
not be necessarily written as $\omega=||\mathbf{u}||-c$; we now
write this as $\eta(\mathbf{u})^{state}$.

We first discuss the case of $q=2$ and then proceed to its generalization.

In systems with two clusters ($q=2$), the are two (Potts) spin states.
We will denote these herein as $A$
and $B$ (once again, we do so to be consistent with the notations in \cite{jorgbook}, in particular
Eqs. (6.60)-(6.63) therein). In this case, there are
$6$ different ``order parameters''. We will
denote these as  $\eta_{01}^A$, $\eta_{10}^A$, $\eta_{11}^A$,
$\eta_{01}^B$, $\eta_{10}^B$ and $\eta_{11}^B$.

In the following, we present the expressions for $\eta_{01}^A$, $\eta_{10}^A$ and
$\eta_{11}^A$. The expressions for
$\eta_{01}^B$, $\eta_{10}^B$ and $\eta_{11}^B$ have an identical form
with a permutation of the superscripts $A \leftrightarrow B$.
\begin{eqnarray}
\eta_{11}^A=&&(p_{in}^A\eta_{11}^A+p_{out}^{A|B}\eta_{11}^B)^2\nonumber\\&&+2(p_{in}^A\eta_{10}^A+p_{out}^{A|B}\eta_{10}^B)(p_{in}^A\eta_{01}^A+p_{out}^{A|B}\eta_{01}^B),
\label{eta_1}
\end{eqnarray}
\begin{eqnarray}
\eta_{10}^A=&&(p_{in}^A\eta_{10}^A+p_{out}^{A|B}\eta_{10}^B)^2\nonumber\\&&+2(p_{in}^A\eta_{10}^A+p_{out}^{A|B}\eta_{10}^B)(p_{in}^A\eta_{11}^A+p_{out}^{A|B}\eta_{11}^B),
\label{eta_2}
\end{eqnarray}
\begin{eqnarray}
\eta_{01}^A=&&(p_{in}^A\eta_{01}^A+p_{out}^{A|B}\eta_{01}^B)^2\nonumber\\&&+2(p_{in}^A\eta_{01}^A+p_{out}^{A|B}\eta_{01}^B)(p_{in}^A\eta_{11}^A+p_{out}^{A|B}\eta_{11}^B).
\label{eta_3}
\end{eqnarray}

These consist a quadratic system of $6$ equations with $6$
variables. This system of equations is numerically solvable. The
solutions are continuous with respect to coefficients $p_{in}^A$,
$p_{out}^{B|A}$. The new equations of Eqs.
(\ref{eta_1},\ref{eta_2},\ref{eta_3}) form a generalization of the
system studied in \cite{jorgbook}.

The above procedure can also be easily generalized to system with
$q>2$ components leading to more terms on the righthand side of Eqs.
(\ref{eta_1},\ref{eta_2},\ref{eta_3}). In general, we can define an
abstract function $g$:

%
%
\begin{eqnarray*}
\{0,1,2\}^q\setminus\{(0,0,...0)\}&&\rightarrow\{0,1\}^q\setminus\{(0,0,...,0)\}\\
(a_1,a_2,...,a_q)&&\mapsto
\end{eqnarray*}
$\begin{cases}(a_1,a_2,...,a_q)&2\notin\{a_1,a_2,...,a_q\}\\
(\lfloor a_1/2\rfloor,\lfloor a_2/2\rfloor,...,\lfloor
a_q/2\rfloor)&2\in\{a_1,a_2,...,a_q\},
\end{cases}$\\
then for any
$\mathbf{a}=(a_1,a_2,...,a_q)\in\{0,1\}^q\setminus\{(0,0,...,0)\}$
and $1\leq i\leq q$, we have the equation
\begin{eqnarray}
\eta_\mathbf{a}^{A_i}=
&&\sum_{\substack{g(\mathbf{u}+\mathbf{v})=\mathbf{a}\\
\mathbf{u},\mathbf{v}\in\{0,1\}^q\setminus\{(0,0,...,0)\}}}(p_{in}^{A_j}\eta_{\mathbf{u}}^{A_i}+\sum_{j\neq
i}p_{out}^{A_i|A_j}\eta_\mathbf{u}^{A_j})\times\nonumber\\&&(p_{in}^{A_j}\eta_{\mathbf{v}}^{A_i}+\sum_{j\neq
i}p_{out}^{A_i|A_j}\eta_\mathbf{v}^{A_j}). \label{eq:cavity}
\end{eqnarray}

In the above equation (\eqnref{eq:cavity}), $\mathbf{a}$ denotes the
q-dimensional incoming message composed of $0$ and $1$s. We
introduce ($\mathbf{u}$,$\mathbf{v}$) to be any pair of vectors that
``sum up to'' a given vector $\mathbf{a}$, in the sense of
$g(\mathbf{u}+\mathbf{v})=\mathbf{a}$. We are able to numerically
evaluate the order parameter
$\eta(\mathbf{u})^{state}$ as a function of $p_{in}$.

From this, we can obtain the phase boundaries of the solvable region. Furthermore, to test
whether our simulation result matches the theory, we perform the
same accuracy test using our greedy algorithm on ER graphs with
$\langle k \rangle =16$ and four equal-sized clusters. Our result of Fig. 12 in
\cite{peter1} in Appendix A is consistent with the cavity inspired
result of Fig. 6.7 in \cite{jorgbook}. In both plots of the percentage
of correctly identified nodes in terms of $p_{in}/Z_{out}$, the
critical value of $p_{in}^{critical}$ and $Z_{out}^{critical}$ for
the accuracy drops are the same if we transfer $Z_{out}$ into
$p_{in}$ via the relation $p_{in}=\frac{16-Z_{out}}{16}$
(for the graphs considered therein with an average total coordination
number per node of $\langle k \rangle= 16$). In, e.g., Fig. 6.7 of
Ref. \cite{jorgbook}, the threshold value of $p_{in}$ is given by
$p_{in}^c\approx 45\%$. In Fig. 12 of \cite{peter1}, the critical
$Z_{out}$ obtained by our greedy algorithm is $Z_{out}\approx 9$,
which corresponds  to $p_{in}\approx
\frac{16-9}{16}=43\%\approx 45\%$.

\section*{Appendix B: Heat Bath Algorithm}
\label{HBA}

We extend the zero-temperature (greedy) algorithm of
\cite{peter2,peter1} to finite temperature via a heat bath
algorithm. This algorithm allows each node to become a member of one
community with probability set by a thermal distribution \cite{RB}.
The probability is
\begin{eqnarray}
p_{a\to b}=\frac{\exp(-\Delta E_{a\to b}/T)}{\sum_d \exp(-\Delta
E_{a\to d}/T)}. \label{eq:probability}
\end{eqnarray}
Here $\Delta E_{a\rightarrow b}$ is the change of energy for moving
this node from cluster $a$ to cluster $b$, and $d$ runs through all
connected clusters (neighbors) of this node (including the case that
$d=a$, i.e., this node remains in cluster $a$; and the case that $d$
is a newly added cluster, i.e., this node becomes a new sole-node
cluster).

The steps of our heat bath algorithm are as follows:

(1) \emph{Initialize the system}. Symmetrically initialize the
system by assigning each node to its own community. (i.e., $q_0=N$).
If the number of communities is constrained to some value $q$, we
instead randomly initialize the system into $q$ communities.

(2) \emph{Find the best cluster for node $i$}. Sequentially ``pick
up'' each node and scan its neighbor list (include its current
cluster and the newly added cluster). Calculate the energy change as
if it were moved to each connected cluster. Then calculate the
probability for an arbitrary node in cluster $i$ to be moved to a
connected cluster $b$ using \eqnref{eq:probability}. Then we use all
the probabilities for different j's to determine which cluster to be
moved to; i.e., generate a random number between $0$ and $1$, then
determine which probability range the random number is in, and move
the node from cluster $a$ to the selected cluster $b$.

(3) \emph{Repeat step 2 for all nodes in the system}. A node is
frozen for the current iteration once it has been considered for a
move.

(4) \emph{Merge clusters}. Allow for the merger of two communities together based on
the merge probability. Towards this end, we calculate the energy change as if the current
community is merged with its neighbors. We then use
\eqnref{eq:probability} to calculate merge probabilities.

(5) \emph{Repeat the above two steps}. Repeat step $2$ to $4$ until
the maximum number of iterations is reached.

(6) \emph{Repeat all the above steps for s trials}. Repeat step
$1$-$5$ for $s$ trials and select the lowest energy result as the
best solution. Each trial randomly permutes the order of nodes in
the symmetric initial state.

The new algorithm is similar to the earlier greedy algorithm \cite{peter2,peter1} except
for steps (2) and (4). The nodes are moved based on a random process.
Thus, the outcome may be sometimes sensitive to the initial random seed state.
As noted within the main text, when the system is within the easy phase, all seeds lead to the same
final outcome.
However, when the system is within the hard phase changing the random
seed may significantly alter the final result. In such a case, different initial
conditions enable the system to get stuck in different local
minima (each corresponding to a different partition of the system
into disparate communities). This is why we repeat the procedures $1-5$ for $s$
trials (usually $s$ is set to be $4$). The additional trials sample
different solutions evenly with the symmetric initialization, and it
will reduce the dependence on initial conditions. In the unsolvable
phase, for any finite number of trials $s$, the quality of the solutions
does not visibly change.

We should note that the new ``heat bath algorithm''  that we introduced above {\em is different from the
commonly used ``simulated annealing algorithm''}. (The latter is a generalization of
the ``Metropolis Monte Carlo'' procedure (MMC)
\cite{kirkpatrick})).

Within the conventional MMC procedure, the probability for an arbitrary node to be
moved in cluster $i$ to a connected cluster $j$ is given by
$\min(1,\exp(-\beta (E_b-E_a)))$.  This implies that a node $i$
in community $a$ will (with certainty)
be moved to cluster $b$ if the energy change is negative.
Such an algorithm precludes for  a lower energy move
(if such a later move will be found later on).
By contrast, within our ``heat bath algorithm'',  nodes
are not immediately moved to the first tried clusters if the
energy change is negative. We compute the probabilities
of connected clusters. Obviously, the cluster with the
largest energy decrease would have the largest probability
to be the ``candidate of absorption'' for node $i$.
Thus, in contrasting the commonly used MMC procedure and
our HBA, it seems be easier
to get to the lowest energy state of the studied system within our
algorithm. Our procedure allows nodes to explore more energy states in each step
and better equilibrate.

The results obtained at low temperature by our HBA are very close to
the results obtained by the zero temperature ``greedy algorithms''  \cite{peter2,peter1}.

\section*{Appendix C: Memory effect in $I_N$ versus noise plot}
\label{sec:memory_effect}

\begin{figure}[ht]
\begin{center}
\subfigure[$T=0.1$]{\includegraphics[width= 2.5in]{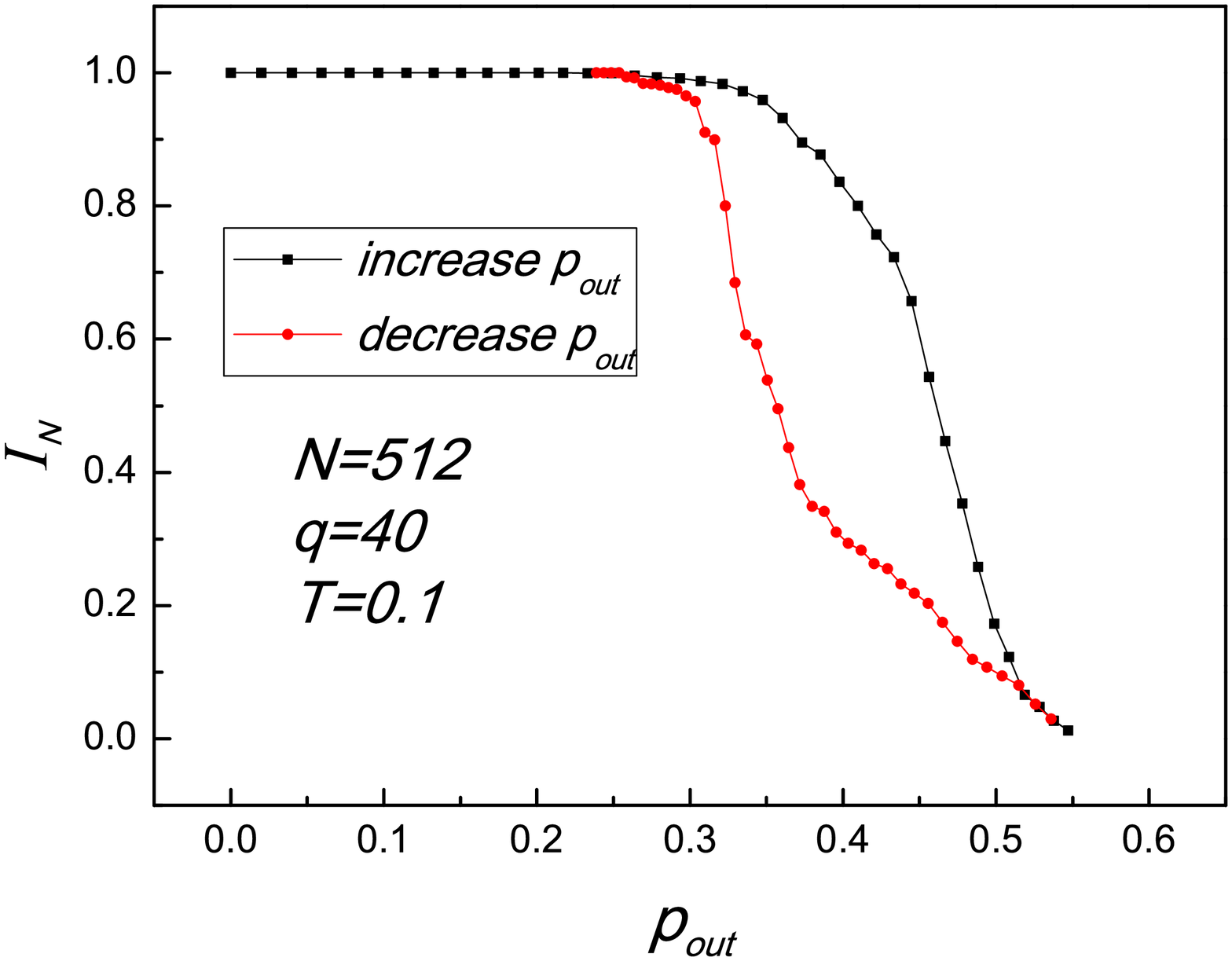}}
\subfigure[$T=1$]{\includegraphics[width= 2.5in]{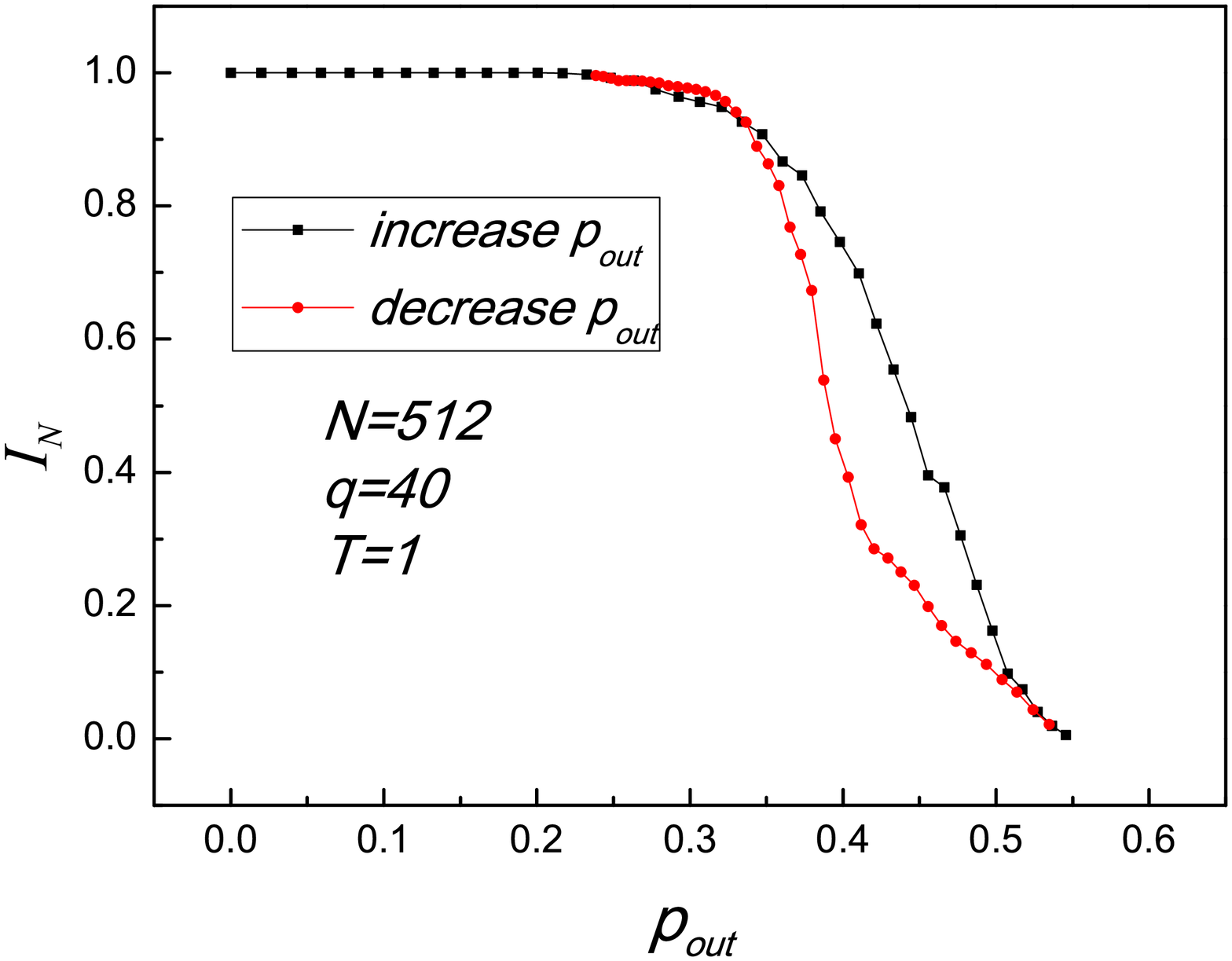}}
\subfigure[$T=2$]{\includegraphics[width=2.5in]{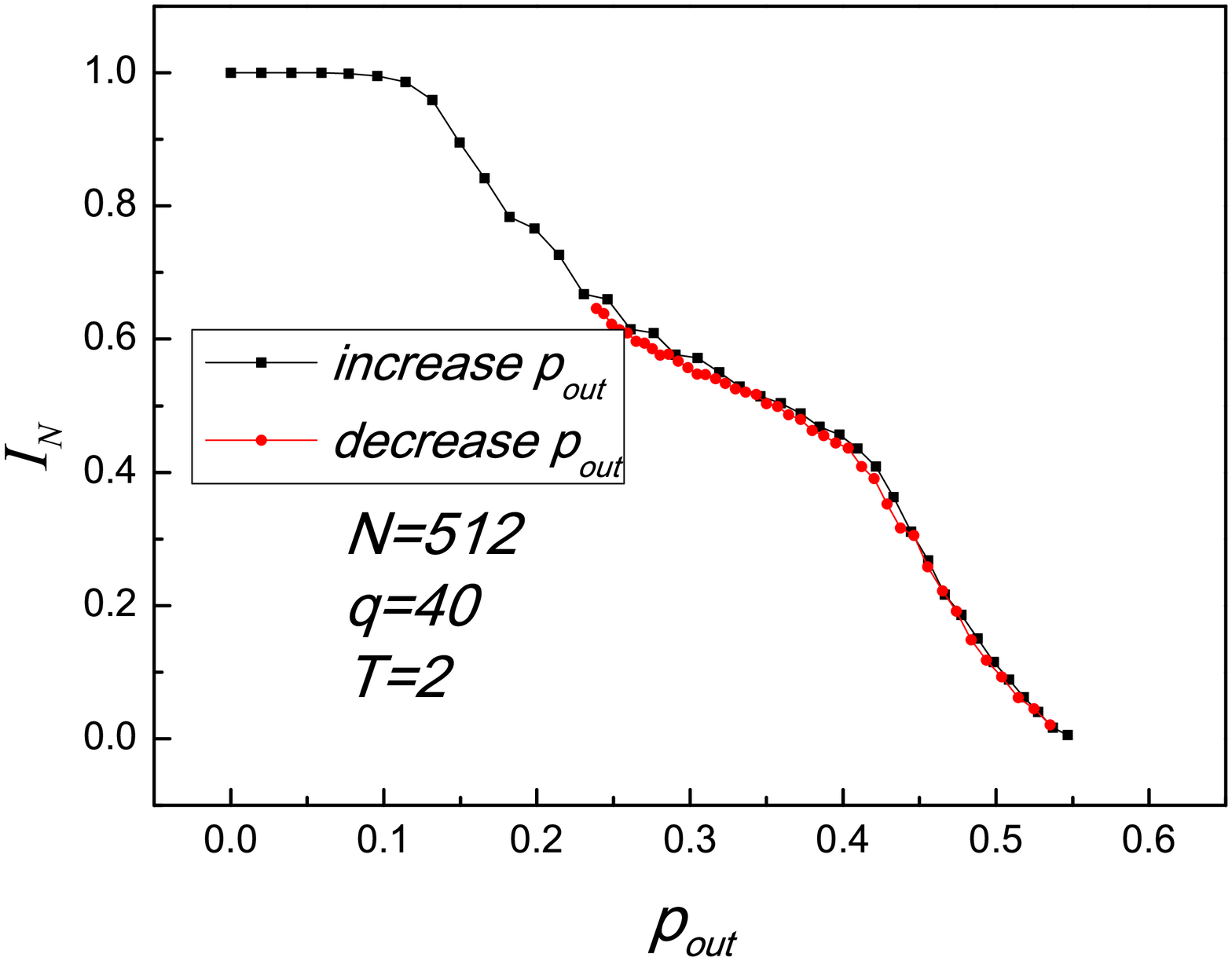}}
\caption{The plot of $I_N$ in terms of $p_{out}$ for system with
$N=512$, $q=40$. ($I_N$ is a normalized variant of mutual
information, for detailed explanation, see \secref{information}.)
From top to bottom, the temperature varies from $T=0.1$ to $T=2$.
Note that the curves in panel(a) and (b) show the effect of hysteresis at
low temperatures. Hysteresis disappears when the
temperature is sufficiently high, e.g., $T=2$ in panel (c). }\label{fig:NMI}
\end{center}
\end{figure}

In the main text of the article, we provided an example of a hysteresis by
decreasing and then
increasing {\em the temperature} of the system (see \figref{fig:EvsTemp}). However, examples
are not limited to this particular cycle  \cite{david}.
Other ways to see
the memory effect include varying {\em the noise level}.
This appendix is devoted to the study of the hysteresis
curves in such a case. That is, in this appendix we consider the effect
of adding external edges between disparate communities
(i.e., increasing $p_{out}$) and then removing these edges (i.e.,
decreasing $p_{out}$). We examine the accuracy of
solutions as a function of noise and see whether the two curves
coincide. The non-coincidence between the two processes will exhibit exactly
the same memory effect that we earlier
reported on by varying the temperature.

\figref{fig:NMI} shows the results of the above experiments
at three temperatures: $T=0.1$, $1$ and $2$.  In the $T=0.1, 1$ systems, the curves with increasing
$p_{out}$ and decreasing $p_{out}$ form hysteresis loops.
The hysteresis loop in temperature $T=1$ in panel (b) is
less significant than its counterpart for at $T=0.1$ in panel (a).
Upon further increase of the temperature, the hysteresis disappears
(as shown in panel (c)).

The plots in \figref{fig:NMI} have already exhibited
decreasing memory effects as the temperature increased. Thus, there must
exist a temperature beyond which the effect disappears. We investigated
the plots at temperatures $T=1.1, 1.2, ..., 1.9$ (not shown here). The
hysteresis loop disappeared at about $T=1.7$ in line with our other
reported results including the disappearance (at $T=1.6)$ of memory of initial
conditions to which we turn to next.

\section*{Appendix D: Memory effect in correlation functions with different initial conditions}
\label{sec:memory_effect}

\begin{figure}[ht]
\begin{center}
\subfigure[$T=0.2$]{\includegraphics[width= 2.5in]{CI02.eps}}
\subfigure[$T=1.6$]{\includegraphics[width=2.5in]{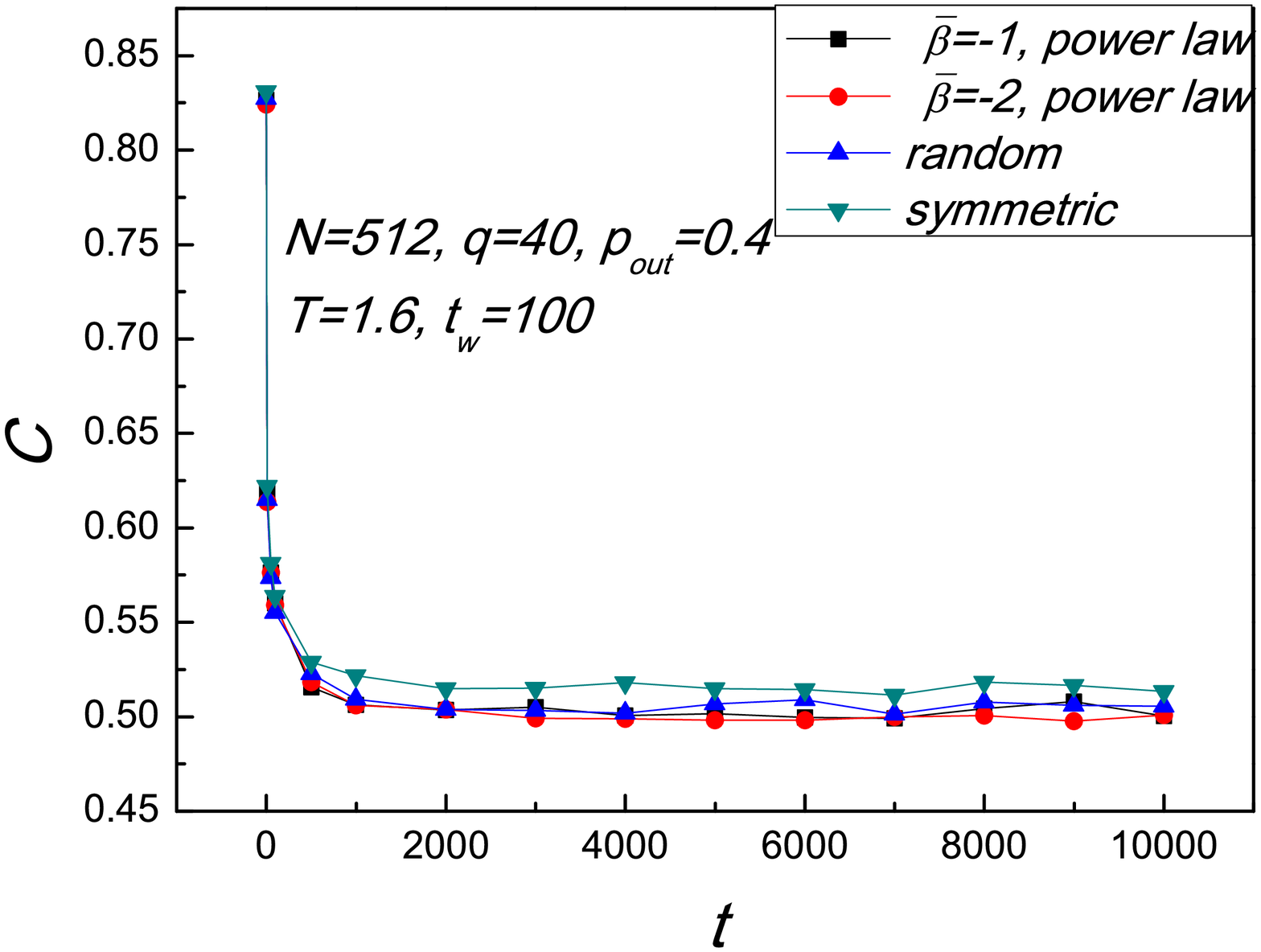}}
\subfigure[$T=2$]{\includegraphics[width=2.5in]{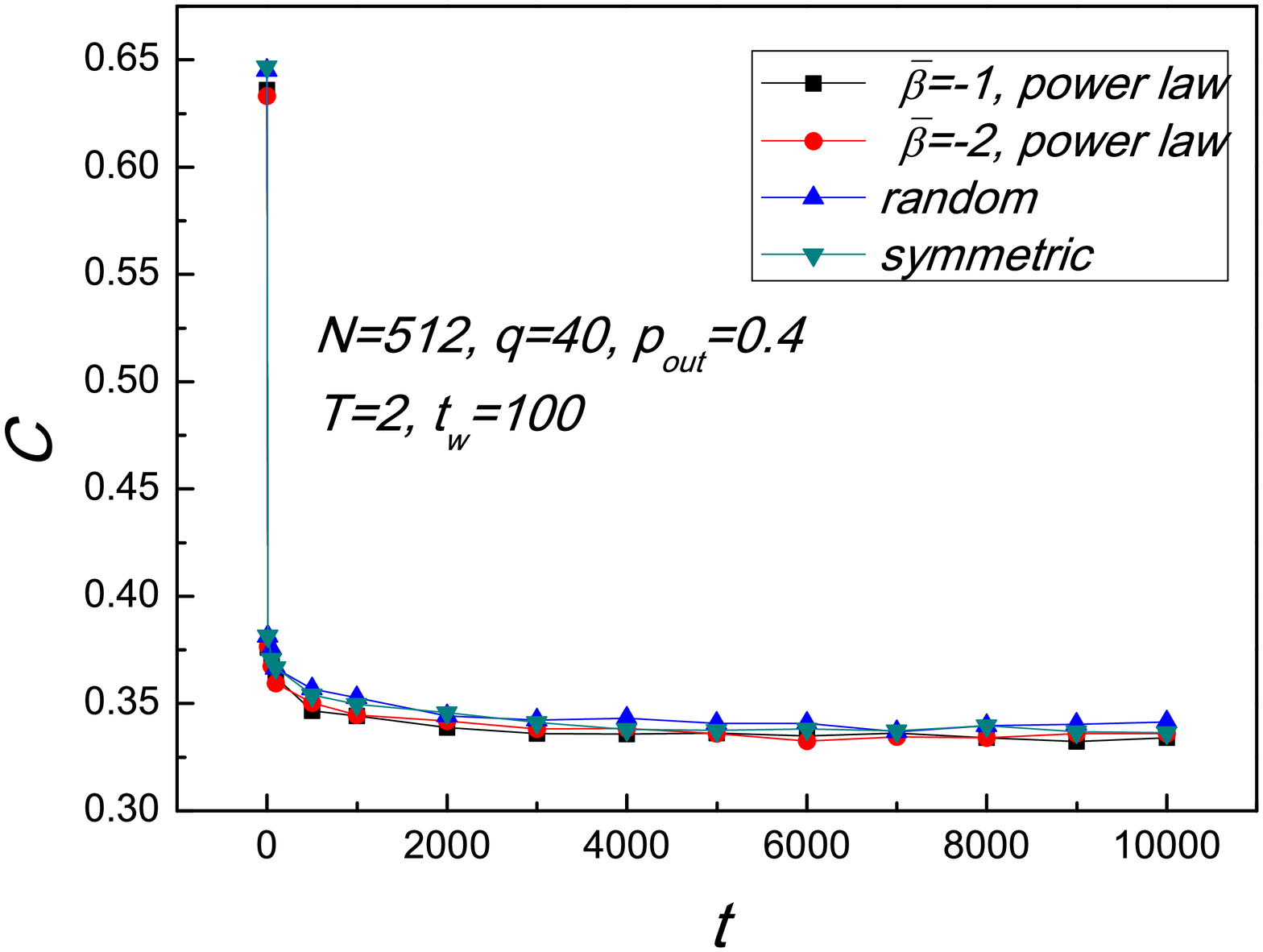}}
\caption{The autocorrelation function as a function of time for system
$N=512$, $q=40$, $p_{out}=0.4$ (above the transition
point in this system). The waiting time $t_w=100$ in all the panels. The four curves in
each panel represent four different initializations for the studied
system. Temperature varies from $T=0.2$ to $T=2$. At low
temperature, all the curves with different initializations separate
from each other even up to $t=10000$ (panel(a)). Then, as $T$
increases, all of the curves start moving towards (panel(b)), and finally overlap (panel(c)).
with each other. }\label{fig:CI}
\end{center}
\end{figure}

In this appendix, we report on the autocorrelation functions (\eqnref{eq:correlation})
for {\em three different types of initial configurations}. The conclusion of
this appendix is that the system may be sensitive to initial conditions.
The three initializations are
denoted as  symmetric, random and power law distribution.

$\bullet$ ``Symmetric'' initialization alludes to an initialization wherein each node forms its own community, so there are
$N$ communities in the beginning (as in step (1) of the algorithm outlined in
Appendix B).

$\bullet$ ``Random'' refers to randomly filling in
$q_0$ communities with nodes, where $q_0$ is a random number
generated between $2$ and $\frac{N}{2}$.

$\bullet$ In the ``Power law distribution'' $N$ nodes are
partitioned into different communities whose size adheres to a power
law distribution (Prob $\sim n^{-{\overline{\beta}}}$) with a
negative exponent ${\overline{\beta}}$. In the cases displayed, we
set  $\overline{\beta}=-1, -2$.

The maximal community size in the true solution is set to be
$50$ and the minimal community size is $8$.

\figref{fig:CI} vividly illustrates that all the curves with
different initializations separate, at low temperatures, from each
other even up to times of size $t=10000$. The curve with symmetric
initialization lies on the bottom in panel (a). However, as
temperature increases, all of the curves veer towards each another.
The symmetric curve moves form the bottom to the top at a
temperature $T=1.6$ as shown in panel (b). As temperature increases
furthermore ($T=2$), all of the curves overlap in panel (c).

At a temperature of $T=1.6$, systems with different initial configurations
start to overlapping. Beyond this temperature, there is
no remaining memory of the initial conditions. Furthermore, the relative position of the curves become
different, which is another indication for the lose of memory. The spin
temperature at which we found the hysteresis loop to disappear
in \figref{fig:NMI},  $T=1.7$, nearly coincides with the
temperature found here.

In \figref{fig:CI}, the relative positions for the ``random''
and ``power law distribution'' do not persist: their positions
change irregularly as temperature varies. This indicates that these
two are similar to each other--  there is no essential difference
between them. However, the curve of symmetric initialization lies
below the other two until the temperature rises up to $T=1.6$, which
happens in all the waiting times that we tested ($t_w=100$, $t_w=10$ and
$t_w=1000$ (not shown here)). This suggests that the symmetric
initialization differs, in an essential way, from the other two initializations.

\section*{Appendix E: Finite size effects} \label{first_transition}

In this appendix, we examine the zero temperature transition at $p_{out}= p_{1}$
for systems with different system sizes $N$ and community numbers $q$.

\begin{figure}[ht]
\begin{center}
\subfigure[$p_1$ vs $q$]{\includegraphics[width= 3in]{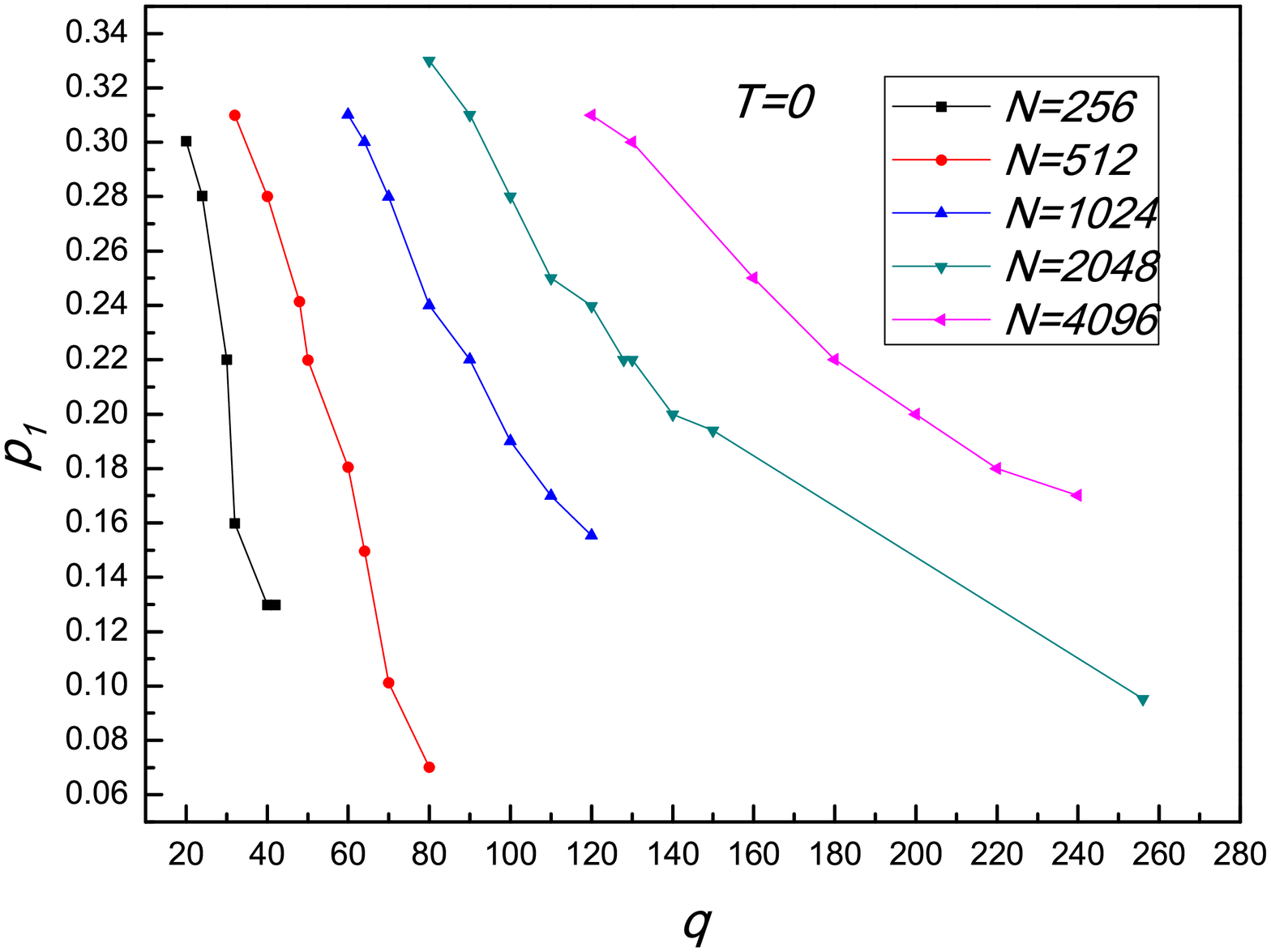}}
\subfigure[$p_1$ vs $N$]{\includegraphics[width= 3in]{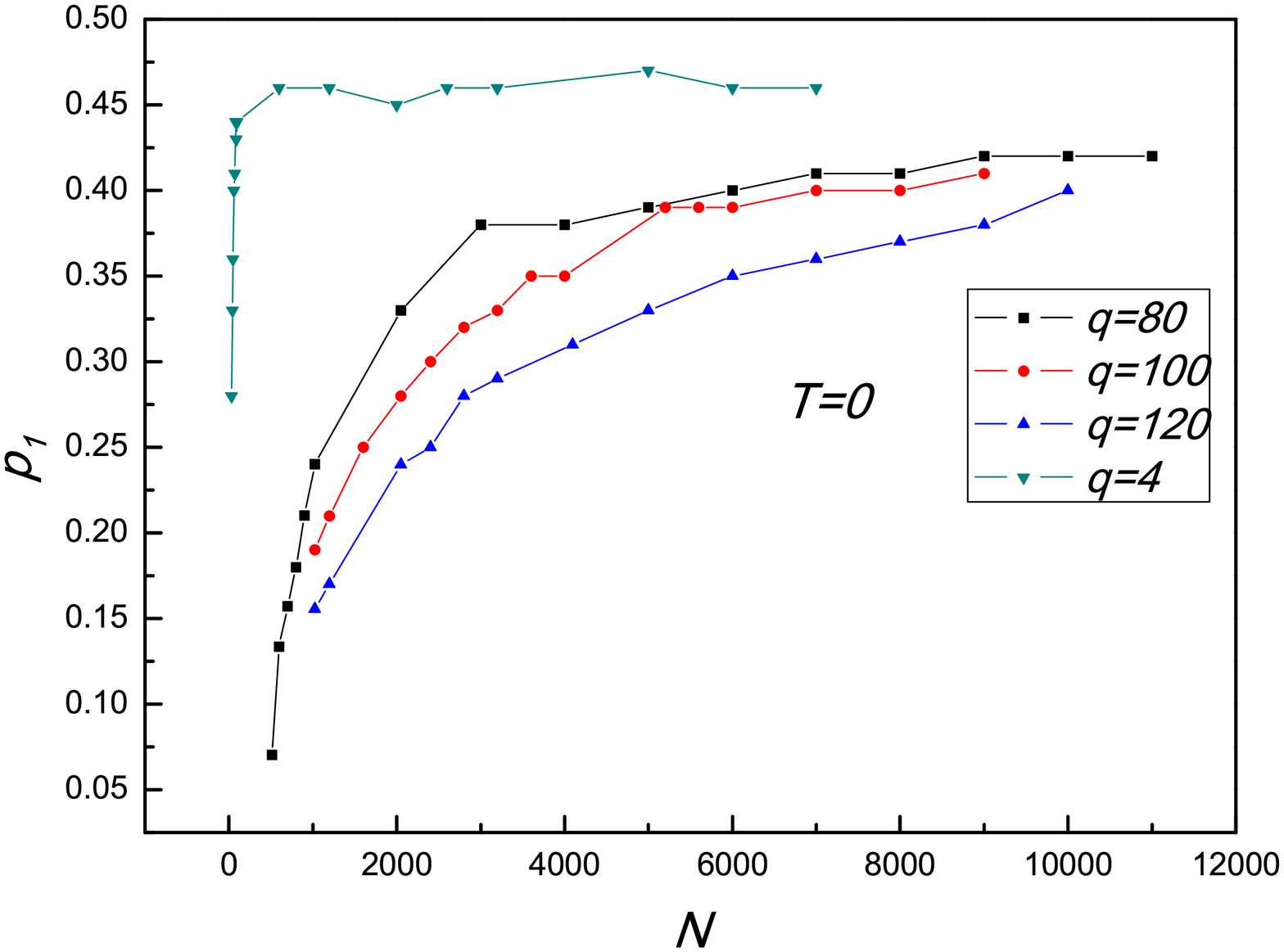}}
\caption{The first transition point $p_1$ as a function
of $q$ (panel(a)) and as a function of $N$ (panel(b)) at zero
temperature.}\label{fig:pc}
\end{center}
\end{figure}

\myfig{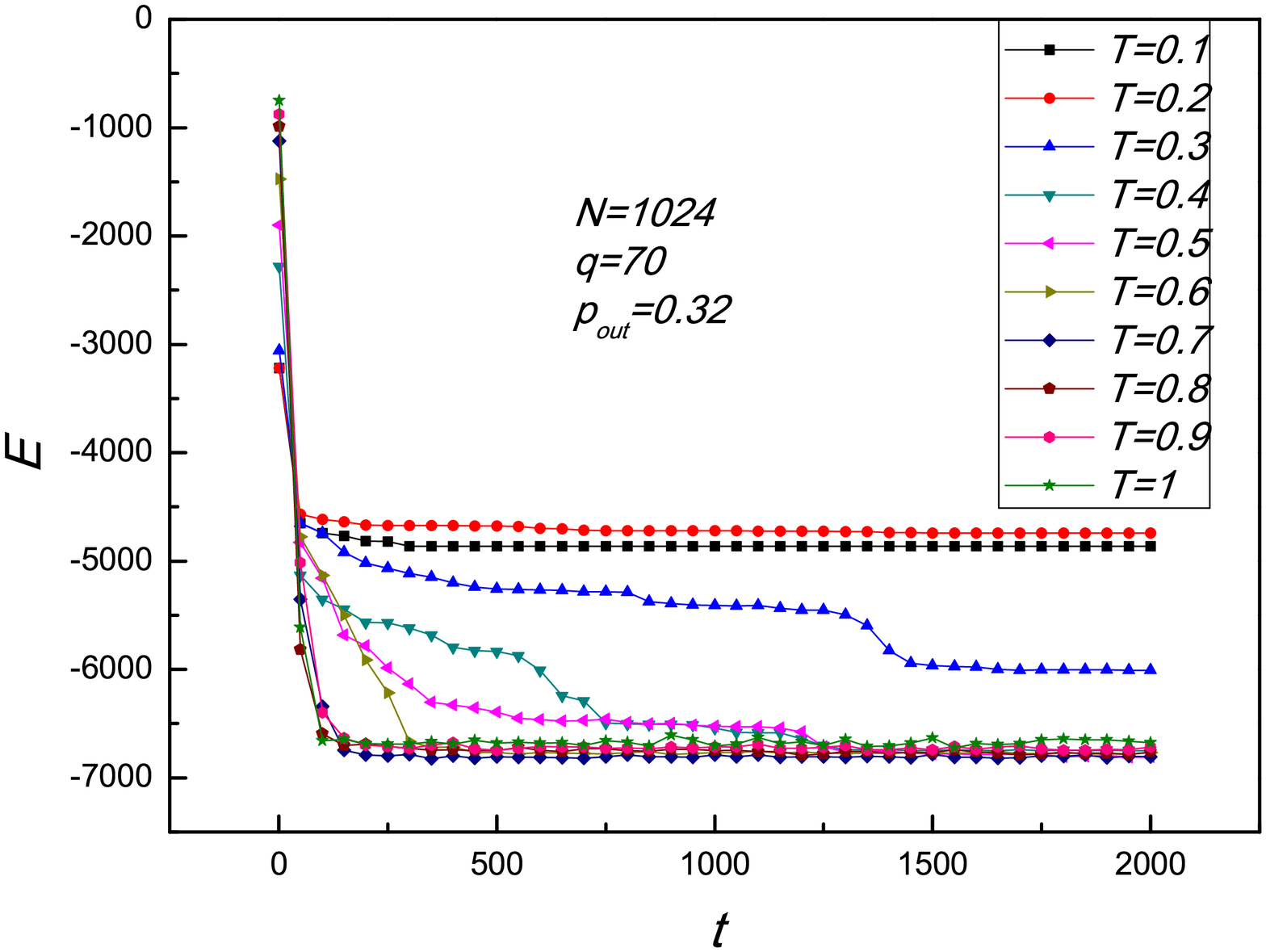}{The plot of energy versus time for the system
$N=1024$, $q=70$. We fix the noise as $p_{out}=0.32$ which is above
the first transition point $p_1=0.3$ (in the ``hard'' region). Note
that the energy curves with different temperatures have a
``crossover'' at about $t=1250$. Before that, the curve with low
temperature is always above the one with high temperature. After
that, except at temperatures of $T=0.1$, $0.2$ or $0.3$, the curves
of the low temperature systems dip below those of the higher
temperature ones. The ``crossover'' property shown here is a sign of
transition from non-equilibrium to equilibrium.
}{fig:Etime}{0.9\linewidth}{}

In \figref{fig:pc}, we display, at zero temperature, the first phase transition point $p_1$
(we remind the reader that this nose levels marks the first transition point encountered as $p_{out}$ is
increased) as a function of the community number $q$ (panel (a)) and the system size
$N$ (panel (b)). From the numerical results that we obtained, we find that
$p_1$ relates linearly $1/q$.  As seen in \figref{fig:pc} panel (a), the value of the
first phase transition point $p_1$ in each curve approaches zero as
$q$ increases. This is consistent with what is expected: for a fixed
system size, increasing the number of spin flavors $q$
introduces a multitude of possible states and the system
becomes progressively disordered.

This may also
be made analytical via a $(1/q)$ type expansion wherein
the partition of the Potts model is expanded in terms of
correlations (of having two ($\sigma_{i}= \sigma_{j}$) and then three etc.) connected
spins be of the same flavor. The resulting terms in such an expansion
illustrate that increasing $q$ emulates (not too surprisingly increasing the
temperature $T$). For a system at large $N$ (i.e., a system in thermodynamic
limit), increasing $q$ renders the system progressively less ordered.
Thus, in situations such as that of an increasing number
of communities $q$ that scales linearly with the system
size $N$ (such that the average community size remains
constant), the transitions become less well
defined as $N \to \infty$.  In the fitting
form below of Eq.(\ref{p1Nq}), the
saturation of the system phase diagram
for large $N$ and the relatively quick drop
in the sensitivity of our results to finite size
effects becomes apparent.

On the other hand, from panel (b) in \figref{fig:pc}, for a fixed $q$, when $N$ is small,
$p_1$ first increases with a very steep slow and thence increases very
slowly with a nearly plateau behavior. We can interpret these data by the
function $p_1=a(q)+bN^{-1}$, where $a(q)$ is a constant for each $q$
(e.g., $a=0.45$ for $q=4$, and $a(q=80) =0.4$).
Combining both panels, we
can present $p_1$ in the examined range as a two-variable function $N$ and $q$,
\begin{eqnarray}
p_1\propto \frac{1}{q} (a(q)+\frac{1}{N}).
\label{p1Nq}
\end{eqnarray}
Thus, as alluded to above, finite size effects drop
and features of the system phase diagram (as evidenced
by $p_{1}$ above) saturate for large $N$.
Thus, in considering limits such $N \to \infty$
while holding the average community size $n = N/q$
fixed, we essentially increase $q$ for a system
in the thermodynamic limit. 

\section*{Appendix F: Equilibration times}
\label{sec:Et}

In equilibrium, the energy is (of course) constant.
The system energy is set by its temperature. In this appendix
we investigate, at different temperatures, the evolution of the system from an initial high energy
states. In the particular results that we provide below,
the energy of the disparate systems at low temperatures
would, at short times, naively seem to violate thermodynamic expectations.
Systems with lower temperature can have higher energies
than those at higher temperature. The origin of this
and similar effects is that significant time may be required
to achieve thermodynamic equilibrium. Within the low temperature
unsolvable phase the system is out of equilibrium. In the hard
phase, equilibrium is achieved yet it requires long times.

\begin{figure}[]
\begin{center}
\subfigure[$\vec{h}_i=0.5(1,0,0)$]{\includegraphics[width=
1.4in]{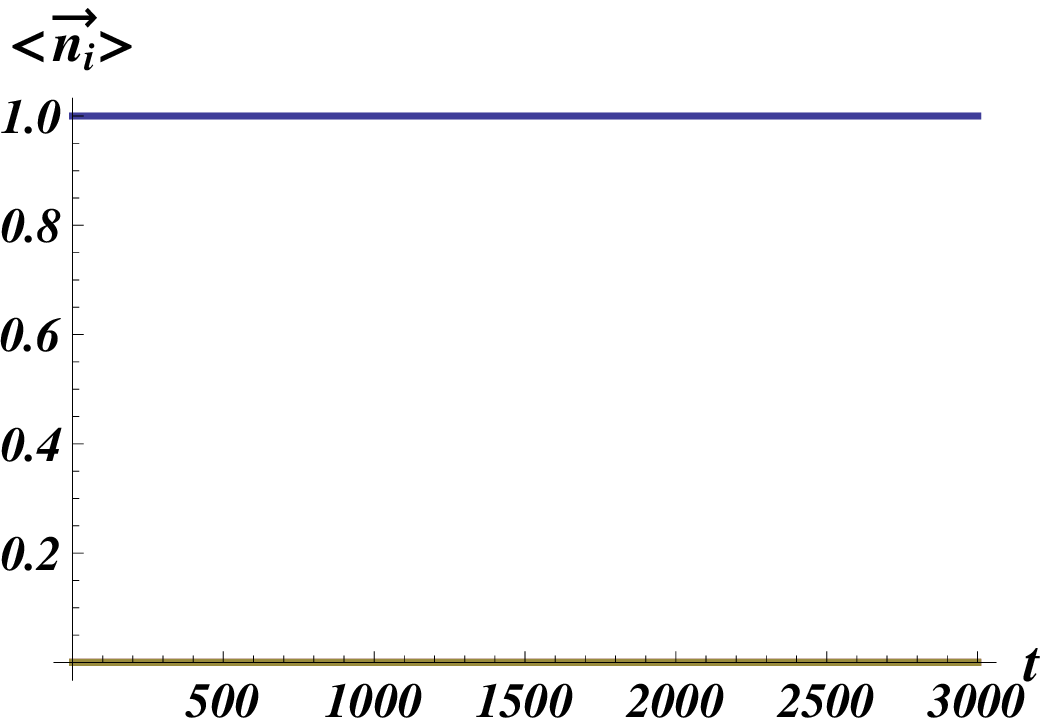}}
\subfigure[$\vec{h}_i=0.5(-\frac{1}{3},\frac{2\sqrt{2}}{3},0)$]{\includegraphics[width=
1.5in]{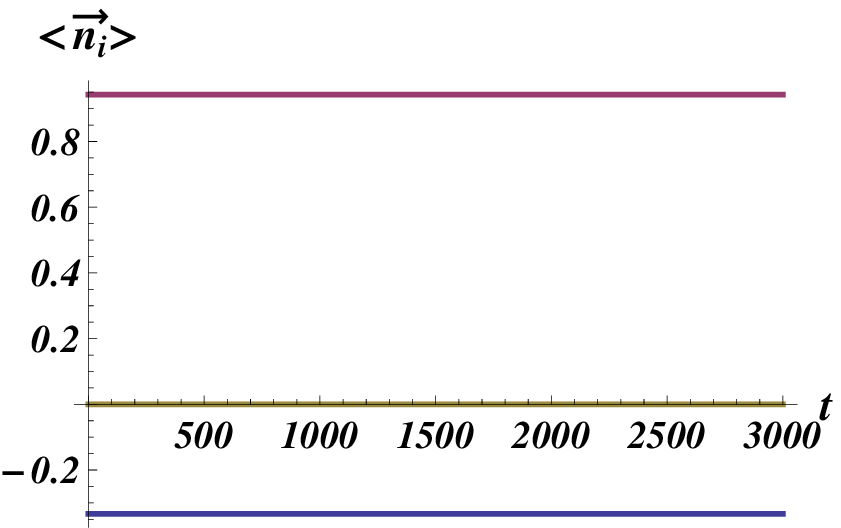}}
\subfigure[$\vec{h}_i=0.5(-\frac{1}{3},-\frac{\sqrt{2}}{3},\frac{\sqrt{2}}{\sqrt{3}})$]{\includegraphics[width=
1.5in]{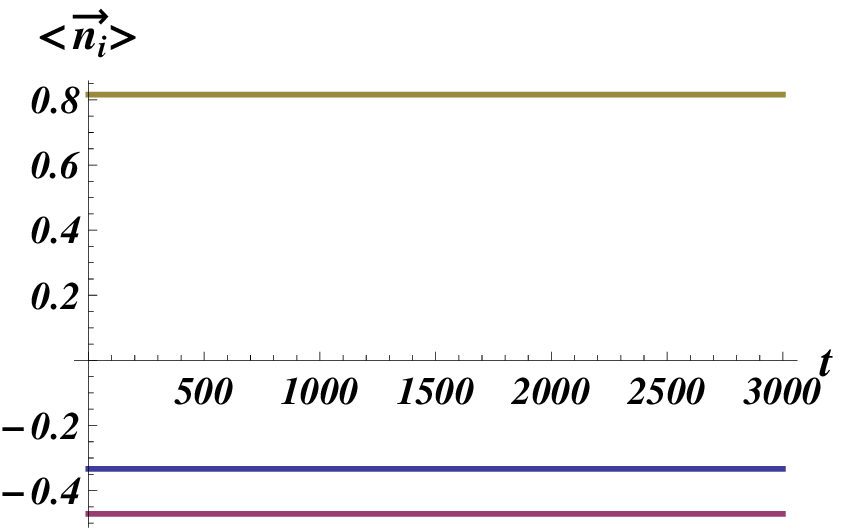}}
\subfigure[$\vec{h}_i=0.5(-\frac{1}{3},-\frac{\sqrt{2}}{3},-\frac{\sqrt{2}}{\sqrt{3}})$]{\includegraphics[width=
1.5in]{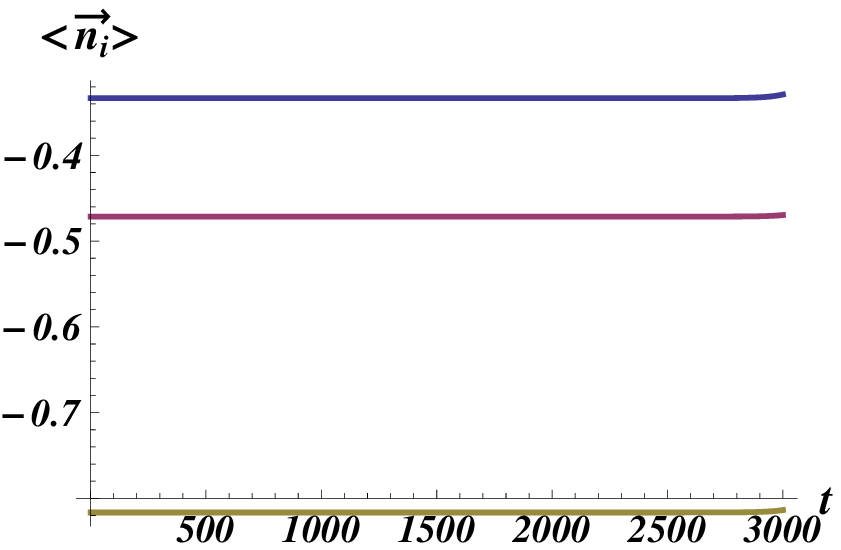}} \caption{The node trajectories in the presence of
the weak perturbing field for system of size $N=24$ with $q=4$
communities with a noise level of $p_{out}=0.1$ at a temperature of
$T=0.01$. As discussed earlier, in this system $\langle \vec{n}_{i}
\rangle$ (for any node $i$) is a three component vector. Each
Cartesian component is labeled by a different color (shade) in the
above figure. The field $\vec{h}_i$ is chosen to be the same as that
of the preset cluster membership for node $i$, i.e., $\vec{n}_i$.
The averages $\langle\vec{n}_i\rangle$ in panels (a) to (d) indicate
the node location under applied fields $\vec{h}_i$ (below each
plot). These fields bias the node trajectories towards the solution
of the system.}\label{fig:bias}
\end{center}
\end{figure}

We now present our results. We set the system size (number of nodes) to be $N=1024$ with $q=70$
communities and a value of the noise given by $p_{out} =0.32$. As
such, with this value of $p_{out}$ which is larger than the
threshold value of $p_{1} = 0.3$ for this system, the system is in
the ``hard'' phase. We examine the system evolution with the
algorithm time steps in \figref{fig:Etime}. In this plot, the system
has a ``crossover'' at about $t=1250$. Prior to that time, the energy
always decreases as $T$ increases. This reflects the fact
times below $t=1250$ are not long enough for the system
to equilibrate. After that, except for the cases of $T=0.1$, $0.2$ or $0.3$,
the energy turns to increase as $T$ increases. Thus, $t=1250$
constitutes sufficient time for equilibration except a few systems
at very low temperature (that require yet longer times).
This ``crossover'' property for system is a
sign of the restoration of equilibrium at sufficiently long times.

All the curves show a decrease of the energy with time until a
plateau in reached. When time is not sufficiently long, the system is
not ergodic and out of equilibrium. As seen in \figref{fig:Etime},  times $t>2000$ are required for
lowest temperature systems (e.g., $T=0.1$, $T=0.2$) to equilibrate.

\section*{Appendix G: Nodes' trajectory after applying the perturbation field}
\label{sec:perturbation}

As mentioned earlier in the text, effective fields may direct the continuous dynamical system
of Section (\ref{dee}) towards correct non-trivial solutions. In this brief appendix,
we outline how this is achieved and provide some results.

The dynamical equation for a node moving under the effective field
is
\begin{eqnarray}
\frac{d\vec{\eta}_i}{dt} =-\vec{f}_i=
         \frac{\delta H_{eff}}{\delta \vec{\eta}_i}\bigg|_{\vec{h}_i}\nonumber\\
=\beta^{-1}\sum_j \frac{1}{2}(\beta A')_{ij}^{-1} \vec{\eta}_j
         -\beta^{-1}\frac{\sum_{\vec{n}_i} \vec{n}_i e^{\vec{n}_i(\vec{\eta}_i+\beta\vec{h}_i)}
              }{\sum_{\vec{n}_i}
             e^{\vec{n}_i(\vec{\eta}_i+\beta\vec{h}_i)}} \label{eq:de}
\end{eqnarray}

Similarly to Section (\ref{dee}) yet now with general applied fields, we have
\begin{eqnarray}
\langle\vec{n}_i\rangle|_{\vec{h}_i}=\frac{\delta\ln
Z\{\vec{h}_i\}}{\delta \beta\vec{h}_i}\bigg|_{\vec{h}_i} =
\frac{\sum_{\vec{n}_i}\vec{n}_i
e^{\vec{n}_i(\vec{\eta}_i+\beta\vec{h}_i)}}{\sum_{\vec{n}_i}
e^{\vec{n}_i(\vec{\eta}_i+\beta\vec{h}_i)}}.\label{eq:derivative}
\end{eqnarray}

As shown in \figref{fig:bias}, if we choose the perturbation field
to favor a preset community membership for each node, i.e., let
$\vec{h}_i=\alpha\vec{n}_i$, where $\alpha$ is a small constant
value, then within the solvable phase the nodes  will be biased
towards the corresponding particular partition of the system.


\begin{thebibliography}{99}
\bibitem{fortunato}
S. Fortunato,  {\it Community detection in graphs}, Physics Reports
{\bf 486}, 75-174 (2010).

\bibitem{rosvall}
M. Rosvall and C. Bergstrom,  {\it Maps of random walks on complex
networks reveal community structure}, Proc Natl Acad Sci USA {\bf
105}, 1118-1123 (2008).

\bibitem{radicchi}
F. Radicchi, C. Castellano, F. Cecconi, V. Loreto, and D. Parisi,
{\it Defining and identifying communities in networks}, Proc Natl
Acad Sci USA {\bf 101}, 2658-2663 (2004).

\bibitem{ramasco}
A. Lancichinetti, F. Radicchi and J. Ramasco, {\it Statistical
significance of communities in networks}, Phys. Rev. E {\bf 81},
046110 (2010).

\bibitem{blondel}
V. Blondel, J. Guillaume, R. Lambiotte and E. Lefebvre, {\it Fast
unfolding of community hierarchies in large networks},  J. Stat.
Mech. P10008 (2008).





\bibitem{gregory}
S. Gregory, {\it Finding overlapping communities in networks by
label propagation},  New J. Phys. {\bf 12}, 103018 (2010).
\bibitem{duch}
J. Duch and A. Arenas, {\it Community detection in complex networks
using extremal optimization}, Phys. Rev. E {\bf 72}, 027104 (2005).

\bibitem{mod}
M. E. J. Newman and M. Girvan, {\it Finding and evaluating community
structure in networks}, Phys. Rev. E {\bf 69}, 026113 (2004);
M. Girvan and M. E. J. Newman,
{\it Community structure in social and biological networks}, Proc.
Natl. Acad. Sci. USA {\bf 99}, 7821-7826 (2002).

\bibitem{peter2}
P. Ronhovde and Z. Nussinov, {\it Local resolution-limit-free Potts
model for community detection}, Phys. Rev. E {\bf 81}, 046114
(2010).

\bibitem{peter1}
P. Ronhovde and Z. Nussinov, {\it Multiresolution community
detection for megascale networks by information-based replica
correlations}, Phys. Rev. E {\bf 80}, 016109 (2009).

\bibitem{RB}
J. Reichardt and S. Bornholdt, {\it Statistical mechanics of
community detection},  Phys. Rev. E {\bf 74}, 016110 (2006).

\bibitem{gudkov}
V. Gudkov, V. Montealegre, S. Nussinov, and Z. Nussinov, {\it
Community detection in complex networks by dynamical simplex
evolution}, Phys. Rev. E  {\bf 78}, 016113 (2008).

\bibitem{Hastings} M. Hastings, {\it Community detection as an inference problem}, Phys. Rev. E {\bf 74}, 035102 (R) (2006).

\bibitem{jsima}
J. \v{S}\'{i}ma, S. Schaeffer, {\it On the NP-Completeness of some
graph cluster measures},
 Proceedings of the Thirty-second
International Conference on Current Trends in Theory and Practice of
Computer Science (Sofsem 06), in: Lecture Notes in Computer Science,
Vol. {\bf 3831}, p. 530 (2006).
\bibitem{bayati}
M. Bayati, C. Borgs, A. Braunstein, J. Chayes, A. Ramezanpour, and
R. Zecchina, {\it Statistical Mechanics of Steiner Trees}, Phys.
Rev. Lett. {\bf 101}, 037208 (2008).
\bibitem{jie}
J. Zhou and H. Zhou, {\it Ground-state entropy of the random
vertex-cover problem}, Phys. Rev. E {\bf 79}, 020103(R) (2009).
\bibitem{hidetoshi}
H. Nishimori and K. Y. M. Wong, {\it Statistical mechanics of image
restoration and error-correcting codes}, Phys. Rev. E {\bf 60}, 132
(1999).

\bibitem{zecchina}
R. Monasson, R. Zecchina, S. Kirkpatrick, B. Selman, and L.
Troyansky, {\it Determining computational complexity from
characteristic 'phase transitions'},
 Nature {\bf 400}, 133-137 (1999).

\bibitem{mezard}
M. M\'{e}zard, G. Parisi, and R. Zecchina, {\it Analytic and
algorithmic solution of random satisfiability problems}, Science
{\bf 297}, 812 (2002).

\bibitem{florent}
F. Krzakala and L. Zdeborov\'{a}, {\it Phase transitions and
computational difficulty in random constraint satisfaction
Problems}, Journal of Physics: Conference Series {\bf 95}, 012012
(2008).


\bibitem{taghogg}
T. Hogg, B. Huberman, and C. Williams, {\it Phase transitions and
the search problem}, Artificial Intelligence {\bf 81}, 1-15 (1996).



\bibitem{goodMC}
B. Good, Y. de Montjoye, and A. Clauset, {\it Performance of
modularity maximization in practical contexts}, Phys. Rev. E {\bf
81}, 046106 (2010).







\bibitem{cavity1}
M. M\'{e}zard, G. Parisi, and M. Virasoro,
 {\it Spin Glass Theory and Beyond},
World Scientific Lecture Notes in Physics, vol. 9 (1986).


\bibitem{cavity2}
A. Braunstein, M. M\'{e}zard, and R. Zecchina, {\it Survey
Propagation: An algorithm for satisfiability},  Random Structures
and Algorithms {\bf 27}, 201-226 (2005).




\bibitem{gallager}
R. Gallager, {\it Low Density Parity Check Codes}, Monograph, M.I.T.
Press (1963).

\bibitem{bp} J. Pearl,  {\it Reverend Bayes on inference engines: A distributed hierarchical approach}, Proceedings of the Second National Conference on Artificial Intelligence. AAAI-82: Pittsburgh, PA. Menlo Park, California: AAAI
Press,133 (1982).

\bibitem{image}
D. Hu, P. Ronhovde, and Z. Nussinov, {\it A Replica Inference
Approach to Unsupervised Multi-Scale Image Segmentation},
http://arxiv.org/abs/1106.5793, (2011).


\bibitem{peter3}
P. Ronhovde, S. Chakrabarty, D. Hu, M. Sahu, K. Kelton, and N. Mauro, K.
Sahu, Z. Nussinov, {\it Detecting hidden spatial and spatio-temporal
structures in glasses and complex physical systems by
multiresolution network clustering}, http://arxiv.org/abs/1102.1519
(2011).

\bibitem{peter4}
P. Ronhovde, S. Chakrabarty, M. Sahu, K. Sahu, K. Kelton, and N.
Mauro, and Z. Nussinov, {\it Detection of hidden structures on all
scales in amorphous materials and complex physical systems: basic
notions and applications to networks, lattice systems, and glasses},
http://arxiv.org/pdf/1101.0008 (2011).



\bibitem{parisi}
M. M\'{e}zard and G. Parisi, {\it The cavity method at zero
temperature}, Journal of Statistical Physics {\bf 111}, Nos. 1/2
(2003).
\bibitem{RB2}
J. Reichardt and M. Leone, {\it (Un)detectable Cluster Structure in
Sparse Networks}, Phys. Rev. Lett. {\bf 101}, 078701 (2008).
\bibitem{jorgbook}
J. Reichardt, {\it Structure in Complex Networks}, Lecture Notes in
Physics Vol. {\bf 766}, Springer-Verlag, Berlin, (2009).
\bibitem{Allah}
A. Allahverdyan, G. Ver Steeg and A. Galstyan, {\it Community
detection with and without prior information},
 EuroPhysics Letters {\bf 90}, 18002 (2010).


\bibitem{lfr_bench}
A. Lancichinetti, S. Fortunato, and F. Radicchi, {\it Benchmark
graphs for testing community detection algorithms}, Phys. Rev. E
{\bf 78}, 046110 (2008).

\bibitem{res_lim}
S. Fortunato and M. Barth\'{e}lemy, {\it Resolution limit in
community detection}, Proc. Natl. Aca. Sci. U.S.A. {\bf 104}, 36
(2007);

J. M. Kumpula, J. Saram\"{a}ki, K. Kaski, and J. Kert\'{e}sz, {\it
Limited resolution in complex network community detection with Potts
model approach}, Euro. Phys. J. B {\bf 56}, 41 (2007).

\bibitem{villain}
J. Villain, R. Bidaux, J. P. Carton and R. Conte, {\it Order as an
Effect of Disorder}, J. Phys. (Paris) 41, no.11, 1263¨C1272 (1980).

\bibitem{shender}
E. F. Shender, {\it Antiferromagnetic Garnets with Fluctuationally
Interacting Sublattices}, Sov. Phys. JETP 56 178¨C184 (1982).

\bibitem{henley}
C. L. Henley, {\it Ordering Due to Disorder in a Frustrated Vector
Antiferromagnet}, Phys. Rev. Lett. 62 2056¨C2059 (1989).

\bibitem{biskup}
Z. Nussinov, M. Biskup, L. Chayes, and J. van den Brink, {\it
Orbital order in classical models of transition-metal compounds},
Europhysics Letters, 67, 990-996 (2004).

\bibitem{jonason}
K. Jonason, E. Vincent, J. Hammann, J. Bouchaud, and P. Nordblad,
{\it Memory and Chaos Effects in Spin Glasses}, Phys. Rev. Lett.
{\bf 81}, 3243 (1998).

\bibitem{jonason2}
K. Jonason, P. Nordblad, E. Vincent, J. Hammann, and J. Bouchaud,
{\it Memory interference effects in spin glasses}, Eur. Phys. J. B
{\bf 13}, 99 (2000).

\bibitem{square_lattice}
Setting, in Eq.(\ref{eq:ourpotts}), $A_{ij} = \delta_{|i-j|,1}$ with
$i$ and $j$ sites of a square lattice and $|i-j|$ the distance
between the sites (with the lattice constant set to unity)  yields a
q-state Potts model on a square lattice. For an unweighted graph,
for which ($a_{ij}=b_{ij}=1$), \eqnref{eq:ourpotts} becomes the
standard Potts model on a square lattice: $H=-\frac{1}{2}\sum_{i\neq
j}A_{ij}'\delta(\sigma_i,\sigma_j)$, with $A_{ij}'\equiv
(A_{ij}-\gamma(1-A_{ij}))$.

\bibitem{david}
D. Foster, J. Foster, M. Paczuski, and P. Grassberger, {\it
Clustering Phase Transition and Hysteresis: Pitfalls in Constructing
Network Ensembles}, Phys. Rev. E {\bf 81}, 046115 (2010).

\bibitem{stariolo}
D. Stariolo, M. Montemurro, and F. Tamarit, {\it Aging dynamics of
$\pm J$ Edwards-Anderson spin glasses}, Eur. Phys. J. B {\bf 32},
361 (2003).

\bibitem{young}
L. Berthier and A. Young, {\it Aging dynamics of the Heisenberg spin
glass}, Phys. Rev. B {\bf 69}, 184423 (2004).

\bibitem{brandes} U. Brandes, D. Delling, M. Gaertler, R. Gorke, M. Hoefer,
Z. Nikoloski, and D. Wagner, {\it Maximizing Modularity is hard},
http://arxiv.org/pdf/physics/0608255 (2006).

\bibitem{cook} S. Cook, {\it The complexity of theorem proving procedures},
Proceedings, Third Annual ACM Symposium on the
Theory of Computing, ACM, New York, 151-158 (1971)

\bibitem{wolynes} V. Lubchenko and P. Wolynes, {\it Theory of Structural Glasses and Supercooled Liquids}, Annu Rev Phys Chem {\bf 58},
235-266 (2006).

\bibitem{science} F. Ricci-Tersenghi, {\it Being Glassy Without Being Hard to
Solve}, Science {\bf 330}, 1639 (2010).




%






\bibitem{arenas2}
A. Arenas, A. D\'{i}az-Guilera, and C. P\'{e}rez-Vicente, {\it
Synchronization Reveals Topological Scales in Complex Networks},
Phys. Rev. Lett. {\bf 96}, 114102 (2006).



\bibitem{damp}
Similar expressions and analysis appear for an inertial system
(wherein the force is given by $(d^{2} \vec{\eta}_{i}/dt^{2})$). The
damping accentuates the difference between converging and diverging
trajectories as seen on the experimental time scale as discussed in
the text.

\bibitem{fisher}
D. Fisher and D. Huse, {\it Equilibrium behavior of the spin-glass
ordered phase}, Phys. Rev. B {\bf 38}, 386 (1998).
\bibitem{bray}
A. Bray and M. Moore, {\it Chaotic Nature of the Spin-Glass Phase},
Phys. Rev. Lett. {\bf 58}, 57 (1987).
\bibitem{middleton}
C. Thomas, D. Huse, and A. Middleton, {\it Chaos and universality in
two-dimensional Ising spin glasses},
http://arxiv.org/abs/1012.3444v1 (2010)

\bibitem{us} D. Hu, P. Ronhovde, and Z. Nussinov, {\it Phase transition in the community detection problem: spin-glass type
and dynamic perspectives},  http://arxiv.org/abs/1008.2699 (2010).


\bibitem{zdeborova}
A. Decelle, F. Krzakala, C. Moore, and L. Zdeborov\'{a}, {\it Phase
transition in the detection of modules in sparse networks},
http://arxiv.org/abs/1102.1182 (2011).


\bibitem{kirkpatrick}
S. Kirkpatrick, C.D.Gelatt and M.P.Vecchi {\it Optimization by
Simulated Annealing},  Science. {\bf 220}: 4598 (1983).

















\end{thebibliography}
\end{document}